\documentclass[iop]{emulateapj}
\usepackage{amsmath}
\usepackage{epstopdf}
\usepackage{multirow}
\usepackage{mdwlist}
\newcommand{\kms}{km s$^{-1}$}
\newcommand{\whz}{W Hz$^{-1}$}
\newcommand{\as}{$^{\prime\prime}$}
\shorttitle{Bent-Tailed Radio Sources in the ATLAS-CDFS}
\shortauthors{Dehghan et al.}

\begin{document}

\title{Bent-Tailed Radio Sources in the Australia Telescope Large Area Survey of the Chandra Deep Field-South}
\author{S. Dehghan, M. Johnston-Hollitt}
\affil{School of Chemical \& Physical Sciences, Victoria University of Wellington, PO Box 600, Wellington, 6140, New Zealand}
\email{siamak.dehghan@vuw.ac.nz}
\author{T. M. O. Franzen, R. P. Norris}
\affil{Australia Telescope National Facility, CSIRO Astronomy \& Space Science, PO Box 76, Epping, NSW 1710, Australia}
\author{N. A. Miller}
\affil{Stevenson University, Department of Mathematics and Physical Sciences, Stevenson, MD 21153, USA}

\begin{abstract}
Using the 1.4 GHz Australia Telescope Large Area Survey (ATLAS), supplemented with the 1.4 GHz Very Large Array images, we undertook a search for bent-tailed (BT) radio galaxies in the Chandra Deep Field-South (CDFS). Here we present a catalog of 56 detections, which include 45 bent-tailed sources, four diffuse low-surface-brightness objects (one relic, two halos, and one unclassified object), and a further seven complex, multi-component sources. We report BT sources with rest-frame powers in the range $10^{22} \leq$ $\textrm{P}_{1.4 \textrm{ GHz}} \leq 10^{26}$ \whz, redshifts up to 2 and linear extents from tens of kpc up to about one Mpc. This is the first systematic study of such sources down to such low powers and high redshifts and demonstrates the complementary nature of searches in deep, limited area surveys as compared to shallower, large surveys. Of the sources presented here one is the most distant bent-tailed source yet detected at a redshift of 2.1688. Two of the sources are found to be associated with known clusters: a wide-angle tail source in Abell 3141 and a putative radio relic which appears at the infall region between the galaxy group MZ 00108 and the galaxy cluster AMPCC 40. Further observations are required to confirm the relic detection, which if successful would demonstrate this to be the least powerful relic yet seen with $\textrm{P}_{1.4 \textrm{ GHz}} = 9 \times 10^{22}$ \whz. Using these data we predict future 1.4 GHz all-sky surveys with a resolution of $\sim$ 10 arcseconds and sensitivity of 10 $\mu$Jy will detect of the order of 560,000 extended low-surface-brightness radio sources of which 440,000 will have a bent-tailed morphology. \end{abstract}

\keywords{surveys; radio continuum: galaxies; galaxies: active; galaxies: jets}

\section{Introduction}


Radio galaxies have long been known to display a wide range of structures, from the classic double lobed sources, to more complex twisted morphologies of tailed radio galaxies. For the last 40 years, radio galaxies have been divided based on morphology and power into Fanaroff-Riley Class I (FRI) and Class II (FRII) sources \citep{fr74}. FRIIs are powerful radio galaxies with strong, linear jets which terminate in hotspots surrounded by large radio lobes, whereas fainter core-dominates sources with less efficient jets are known as FRIs. The boundary between these two classes is often stated to be a function of environment \citep{ro76} as demonstrated by the fact that the FRI/FRII transition in more massive galaxies happens at higher luminosity \citep{lo96}. At the transition luminosity for these two classes are the so-called `Wide-Angle Tail' (WAT) and `Narrow-Angle Tail' (NAT) radio galaxies. Typically found in dense environments, WATs and NATs are morphologically FRI radio sources but with luminosities which place them close to the FRII transition. WATs are usually associated with central cluster galaxies possessing a pair of well-collimated jets with low opening angles ($\leq 60^{\circ}$) which persist for up to tens of kpc before flaring out into long, bent plumes. Conversely, NATs, also known as Head-Tail (HT) galaxies \citep{mpk72}, have plumes which are bent to such a degree that their whole radio structure lies on one side of the optical host galaxy. Projection effects make the distinction between WATs and NATs somewhat arbitrary\footnote{A WAT seen edge on will appear as a NAT.} and the term precludes some of the more twisted and complex morphologies that have recently been observed. Recent literature has adopted the term `Bent-Tailed' (BT) galaxies to encompass all sources in which the radio lobes and jets are not aligned linearly with the core galaxy.

BT radio galaxies are found almost exclusively in high density regions of the local Universe (z $<0.1$), i.e., galaxy groups and clusters \citep{mjs09}. BT radio galaxies' peculiar morphology is commonly believed to be the result of the ram pressure due to the relative movement of the galaxy through the intra-cluster/group medium \citep{gg72,mpk72}. An alternative explanation of the curved radio structure is the buoyancy forces due to deviations in the density of the intra-galactic medium, which is in an interaction with the intra-cluster medium \citep{cm75}.

In the more distant Universe, BT radio galaxies are believed to be the signature of overdensities in large-scale structure, with associations between BTs and groups and clusters occurring out to at least z $\sim$ 1 \citep{bgh03}. Furthermore, there is a growing evidence that such associations persist out to the limits of both cluster and BT radio galaxy detections at z $\sim$ 2 \citep{djm11}, making BTs excellent tools to trace the large-scale structures in the distant Universe \citep{bgh03,mss10,nab13}.

\begin{figure*}
\centering
{\hspace{-0.45cm}\includegraphics[width=7in, trim= 0 0 0 0, clip=true]{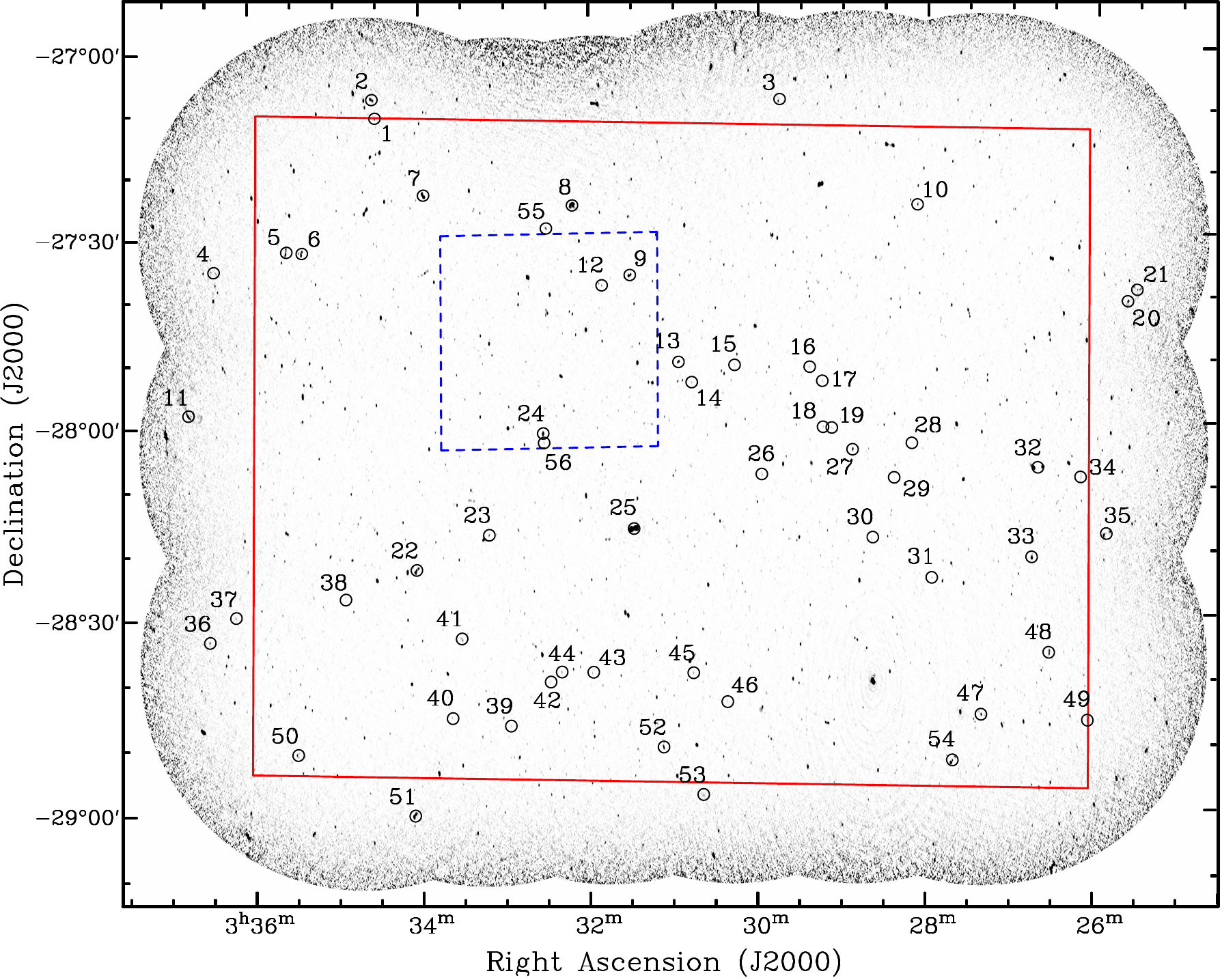}}
\caption{Map of the ATLAS-CDFS field (third data release) showing the location of detected halos, relics, complex and bent-tail radio sources. The red box shows the extent of the ATLAS-CDFS first data release. The blue dashed-line frame represents the extent of the final VLA-ECDFS image. ID numbers correspond to Figure \ref{fig:bts} and Table \ref{tab:prop}.}
\label{fig:map}
\end{figure*}

In order to assess the usability of BT sources as tracers of over-densities, systematic studies of such sources as a function of environment are required. This in turn requires multi-wavelength data to both detect radio sources and then undertake a full assessment of the environment in which they reside. The first such effort in this direction was undertaken by \citet{bgh01} on a sample of 40 large BTs within a 724 deg$^2$ area of the sky and with bright optical counterparts ($m_R \leq 19$). A more recent work has been to examine the environment of a sample of BT sources drawn from 9000 deg$^2$ of the Faint Images of the Radio Sky at Twenty Centimeters \citetext{FIRST, \citealp{bwh95}} survey \citep{wb11}. \citet{wb11} found that $\sim$ 80\% and 67\% from a visually-selected sample of BTs (up to z $\sim0.5$) are located within clusters or groups, and rich clusters, respectively. Using surveys such as FIRST allows detection of BT sources over a wide area with good resolution ($\sim$ 5\as), but with a relatively shallow sensitivity in the radio ($\sim$ 150 $\mu$Jy) meaning that exploration of the population is typically confined to redshifts less than 0.5 \citetext{\citealp{wb11} report only 9 BT sources with z between 0.5 and 0.68, and none beyond z = 0.68} and sources with radio powers in excess of $\textrm{P}_{1.4 \textrm{ GHz}} \geq 4 \times 10^{23}$ W Hz$^{-1}$ with the vast majority of sources having powers above $\textrm{P}_{1.4 \textrm{ GHz}} \geq 10^{24}$ W Hz$^{-1}$. 

An alternative approach is to undertake a study of a smaller area with greater radio sensitivity. Such an approach will produce complementary data as deep radio surveys are able to probe the BT population to higher redshifts. Additionally, focus on smaller areas allows selection of regions with considerably more multi-wavelength coverage, necessary for exploring a wider range of environmental parameters.

\begin{figure*}
\centering
{\includegraphics[width=1.72in, trim= 0 20 0 0, clip=true]{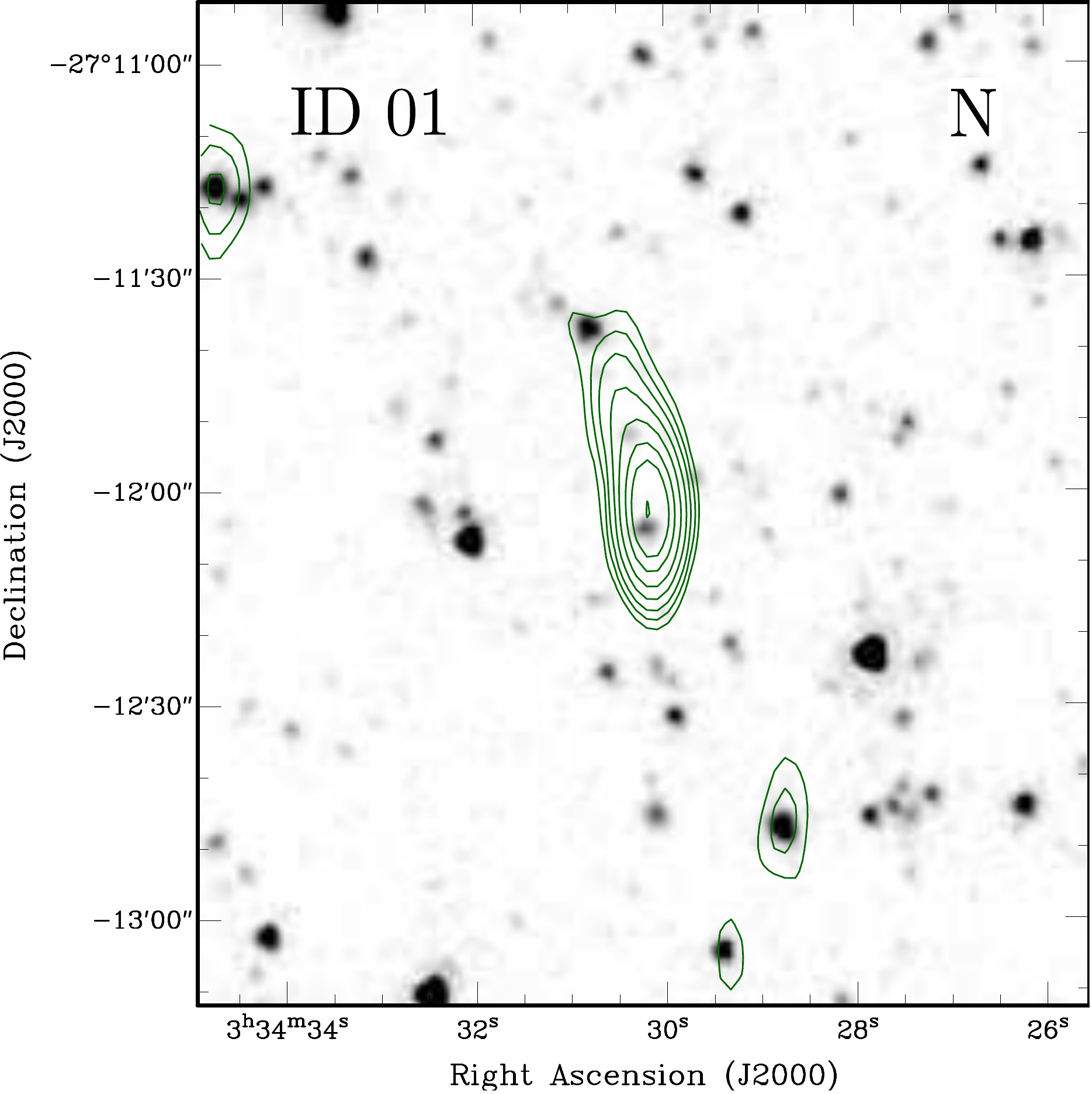}}
{\includegraphics[width=1.624in, trim= 30 20 0 0, clip=true]{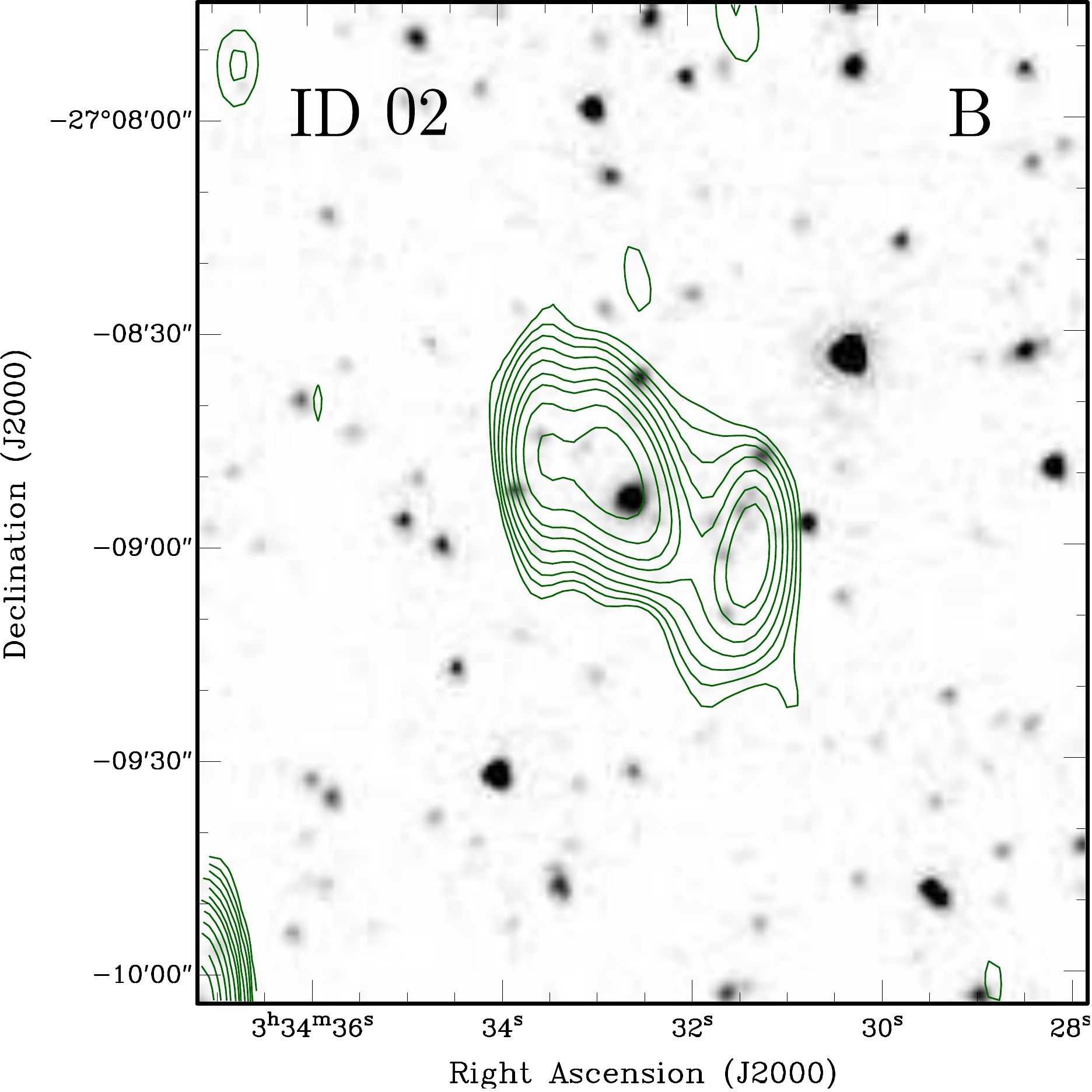}}
{\includegraphics[width=1.624in, trim= 30 20 0 0, clip=true]{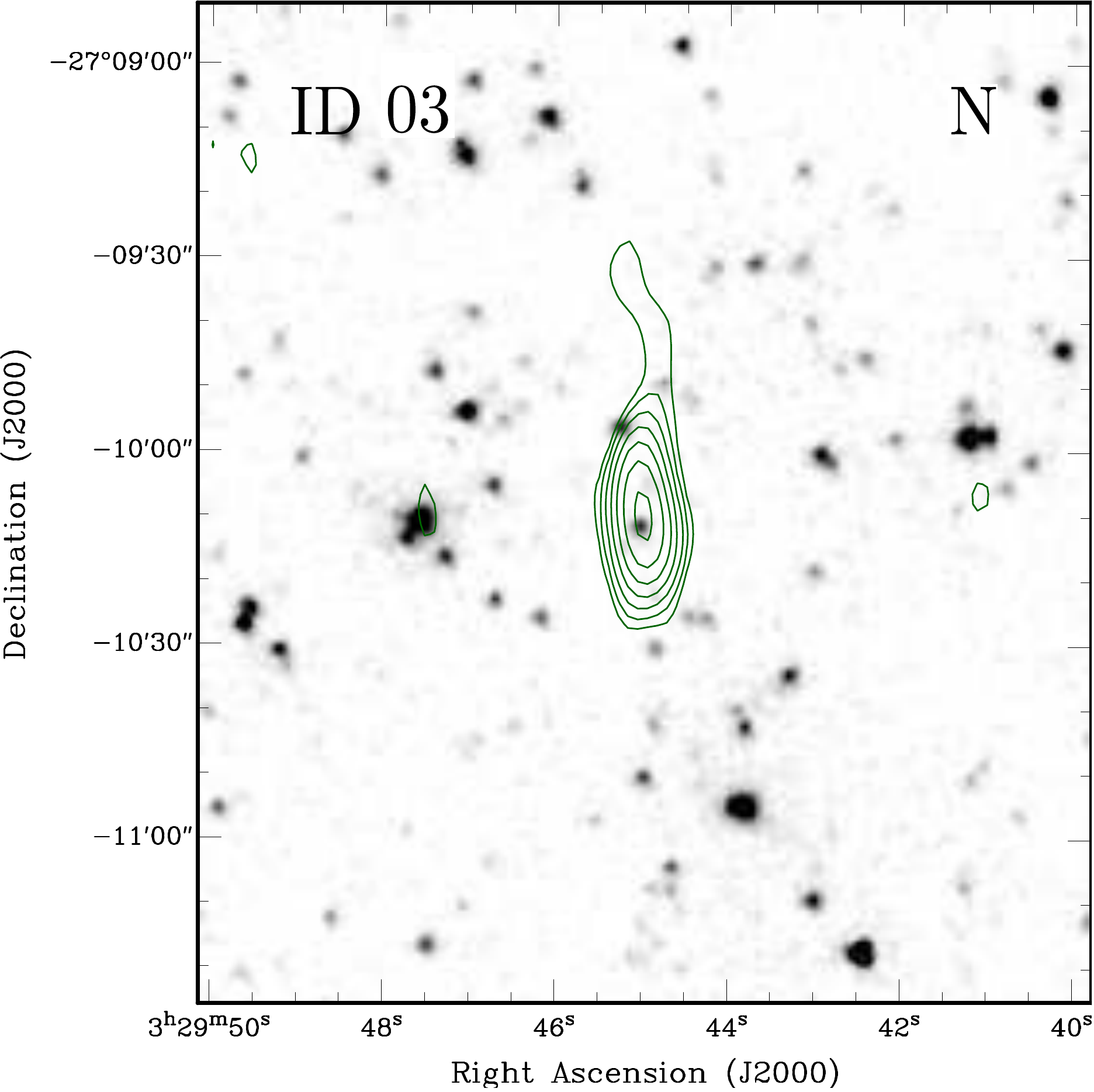}}
{\includegraphics[width=1.624in, trim= 30 20 0 0, clip=true]{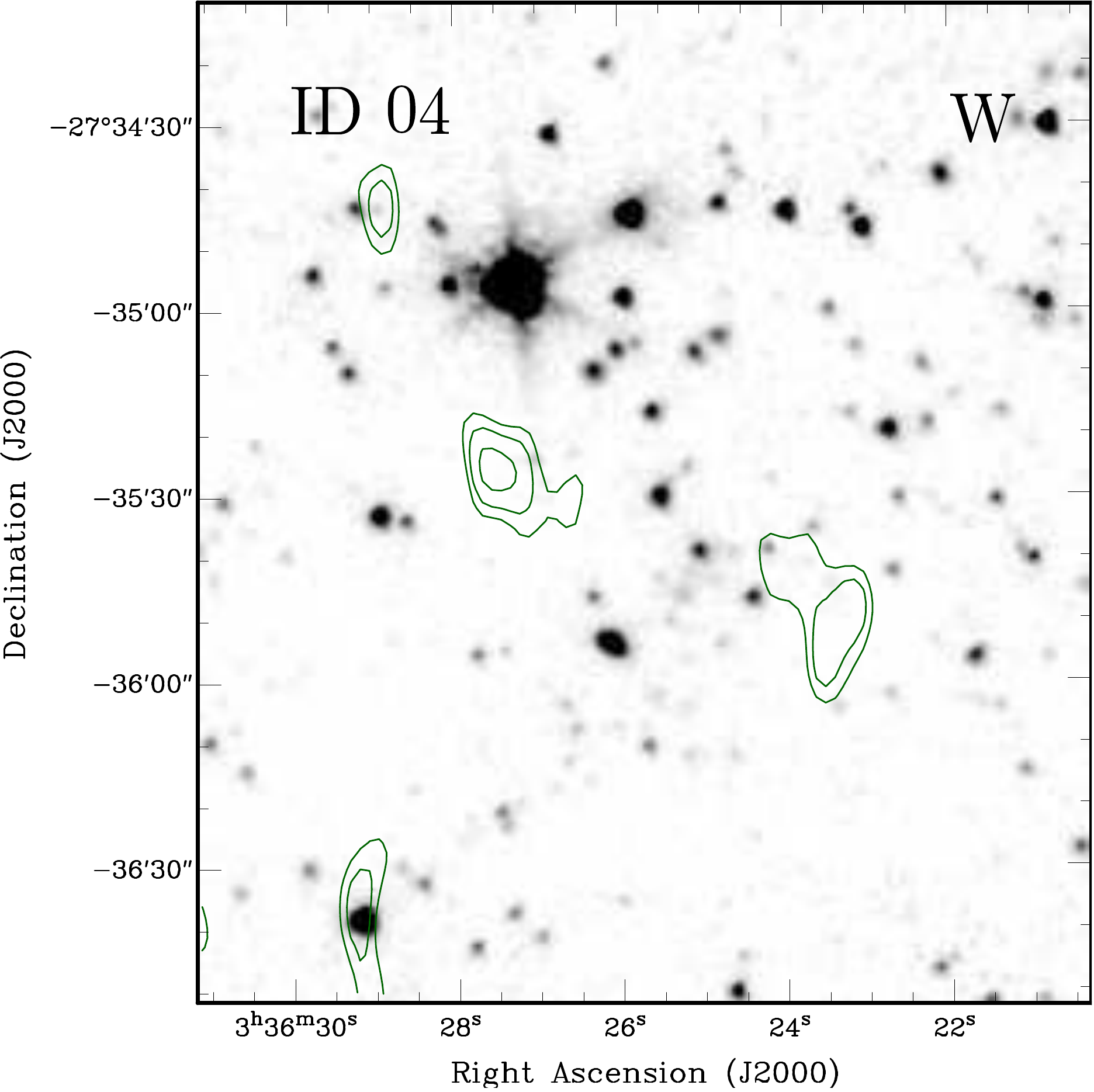}}
{\includegraphics[width=1.72in, trim= 0 20 0 0, clip=true]{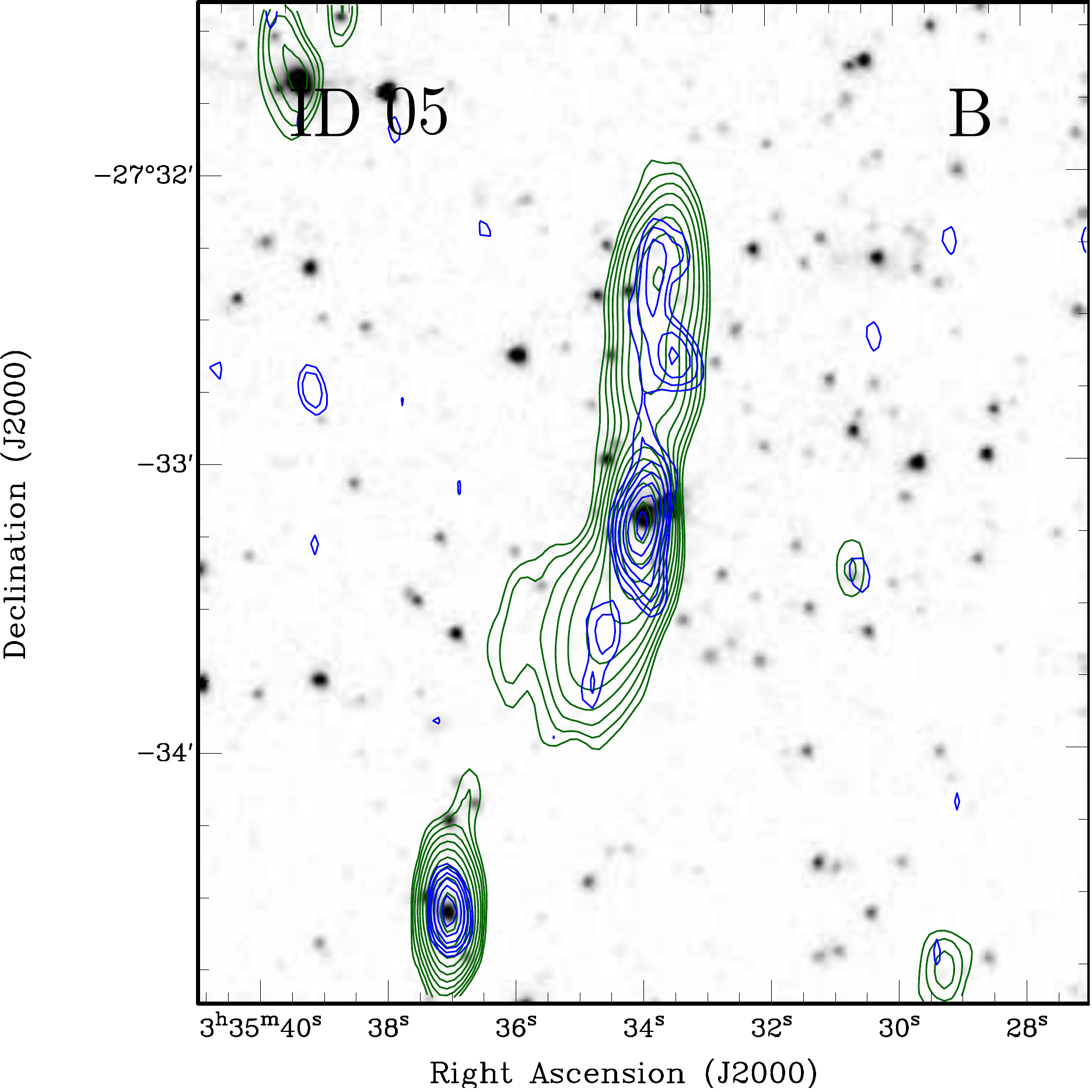}}
{\includegraphics[width=1.624in, trim= 30 20 0 0, clip=true]{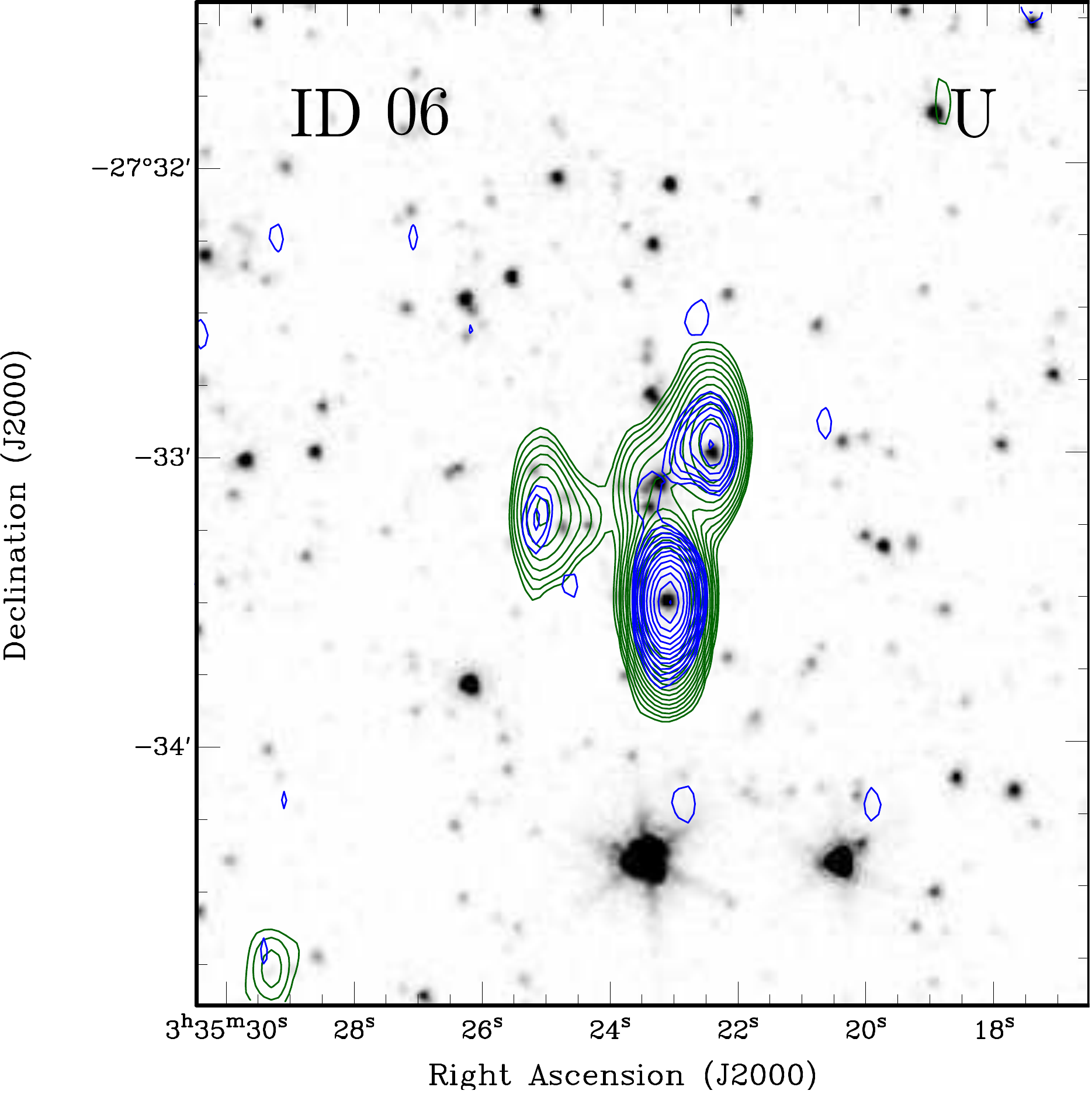}}
{\includegraphics[width=1.624in, trim= 30 20 0 0, clip=true]{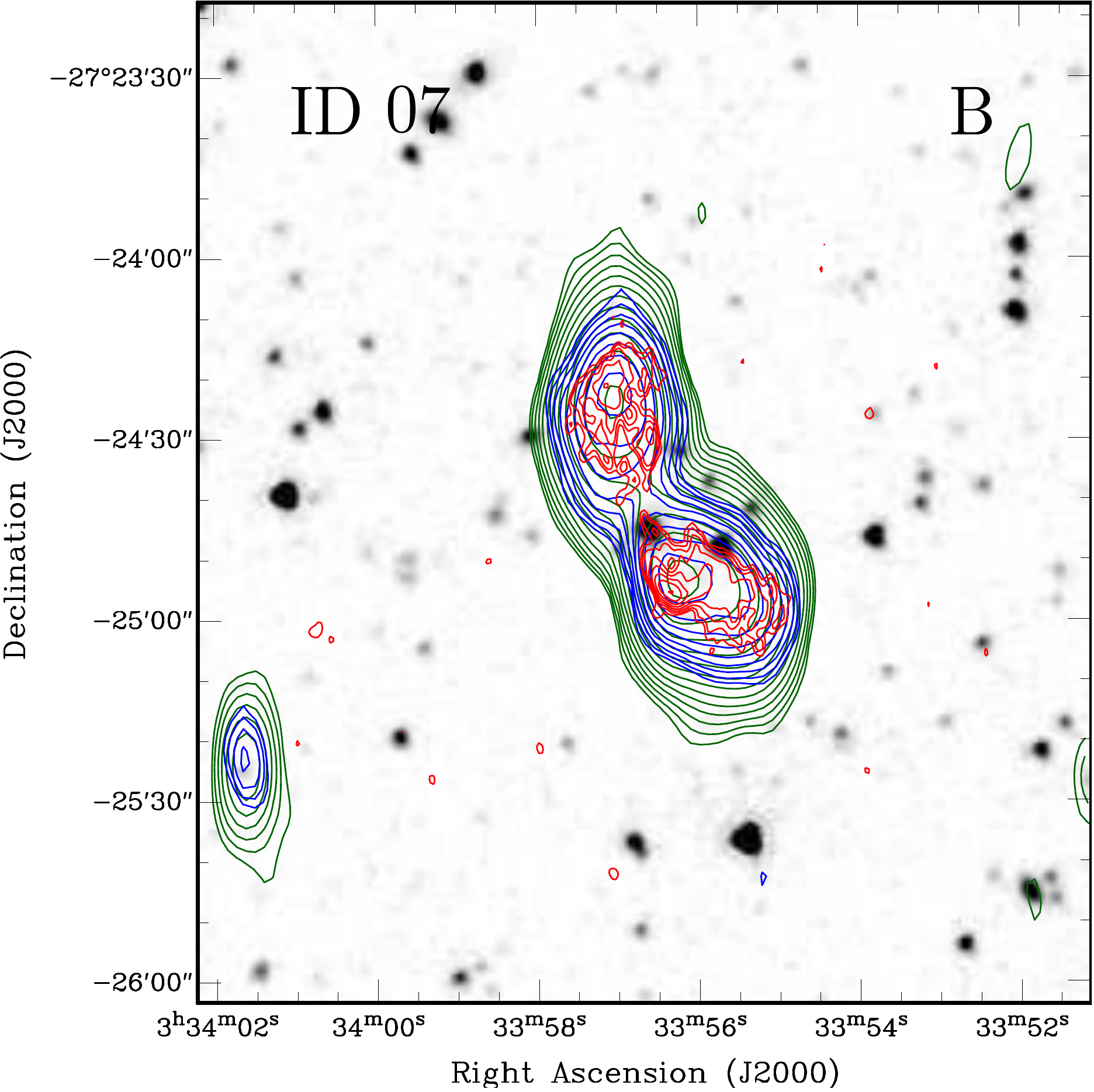}}
{\includegraphics[width=1.624in, trim= 30 20 0 0, clip=true]{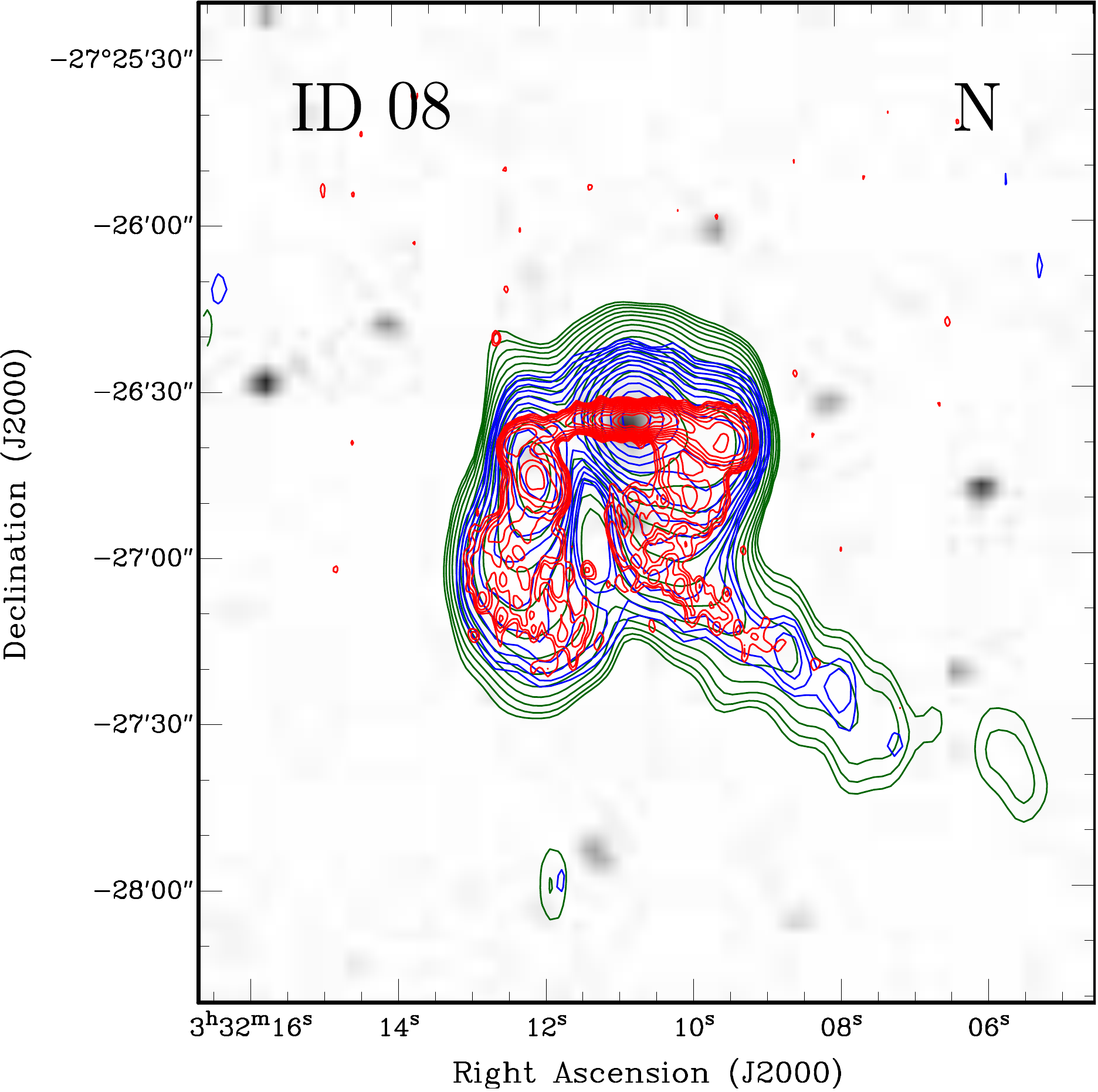}}
{\includegraphics[width=1.72in, trim= 0 20 0 0, clip=true]{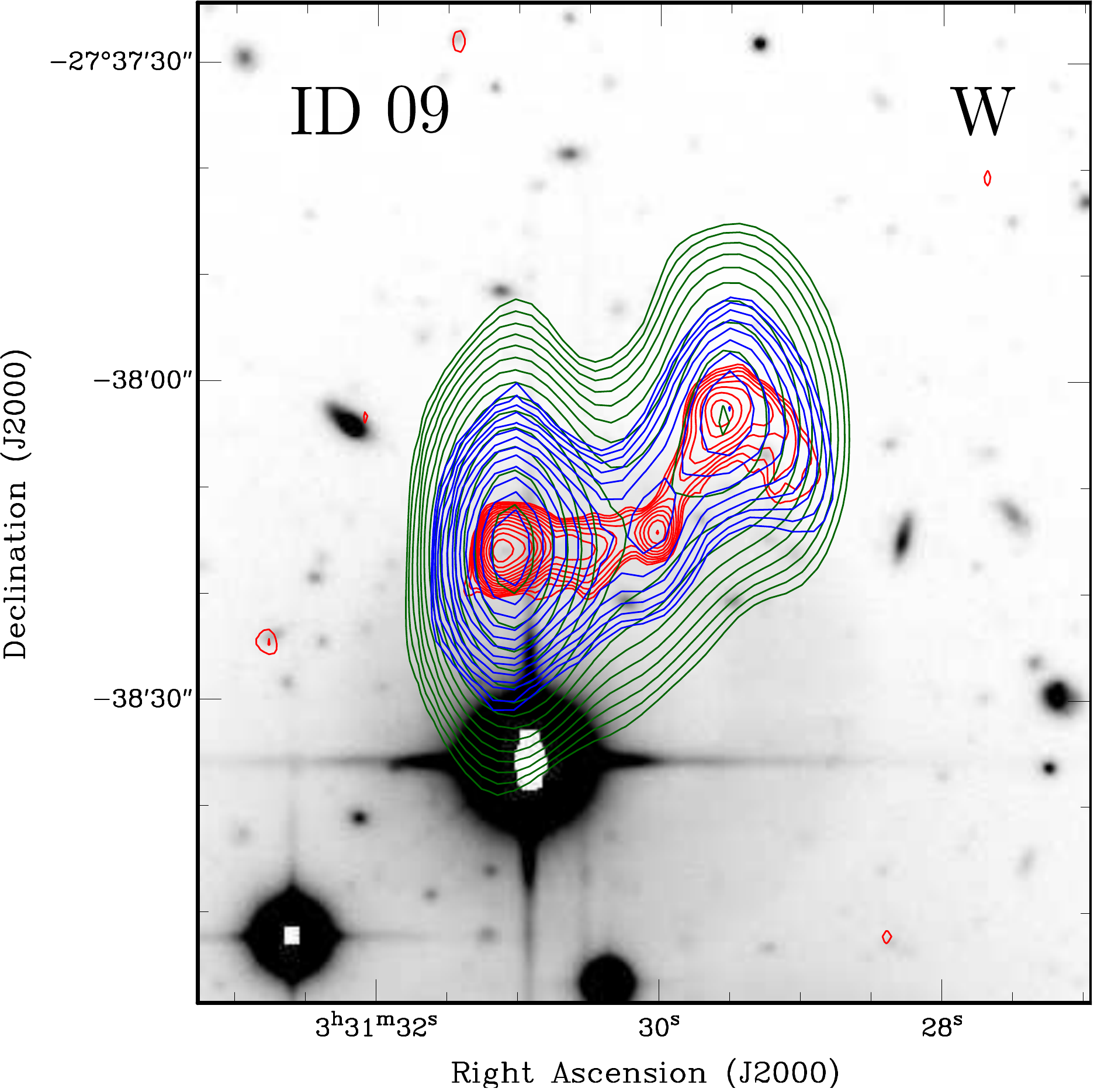}}
{\includegraphics[width=1.624in, trim= 30 20 0 0, clip=true]{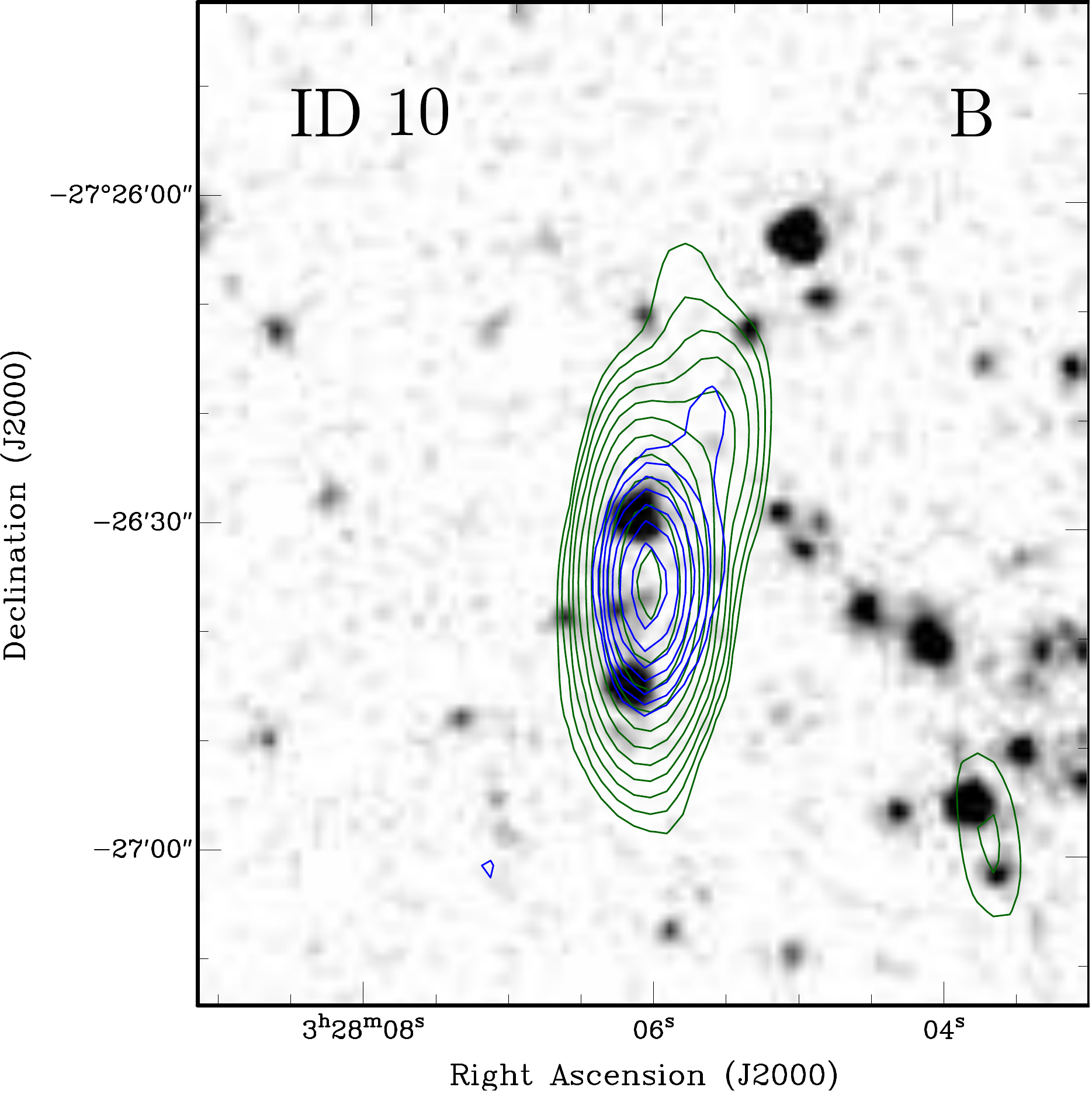}}
{\includegraphics[width=1.624in, trim= 30 20 0 0, clip=true]{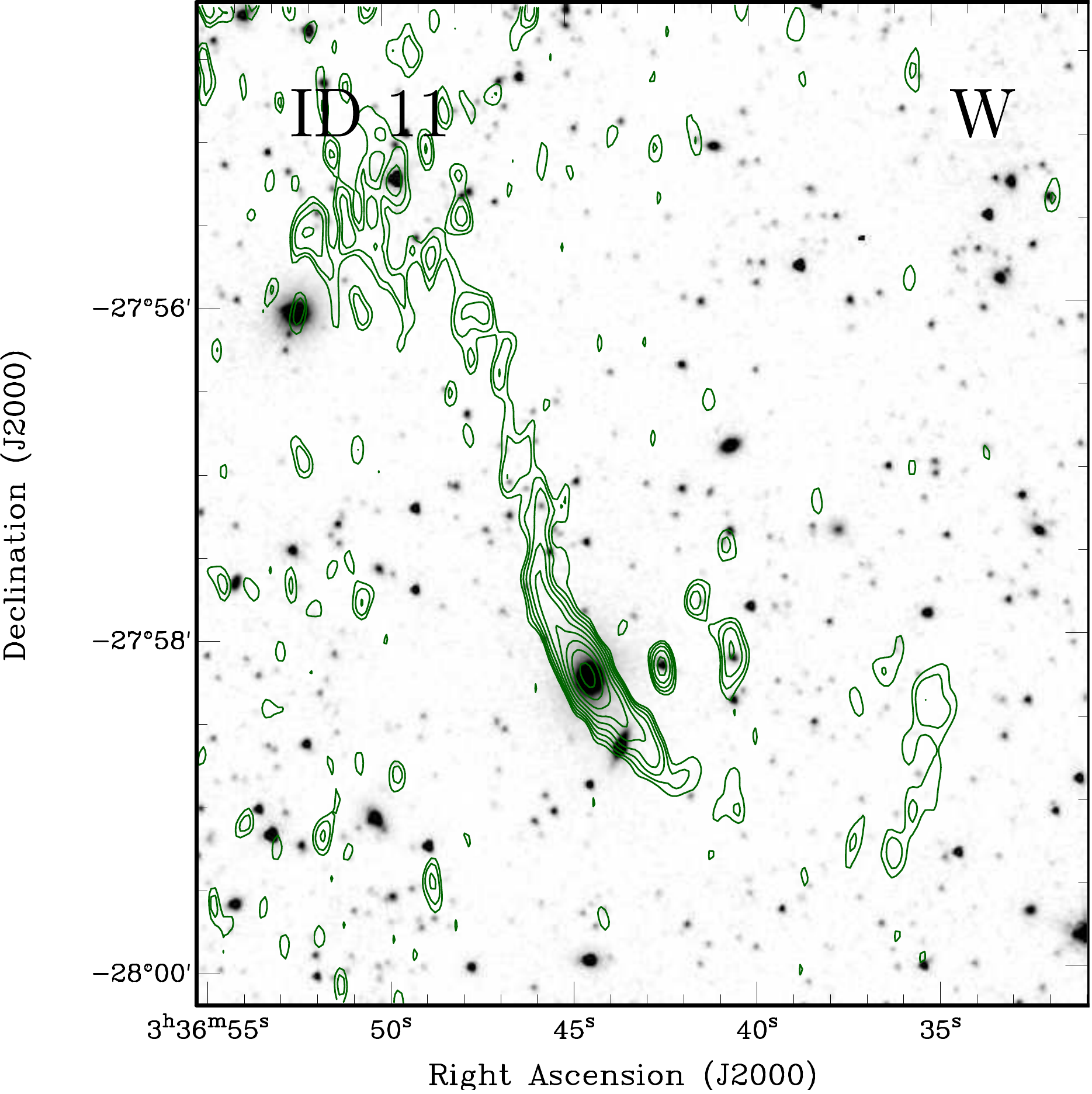}}
{\includegraphics[width=1.624in, trim= 30 20 0 0, clip=true]{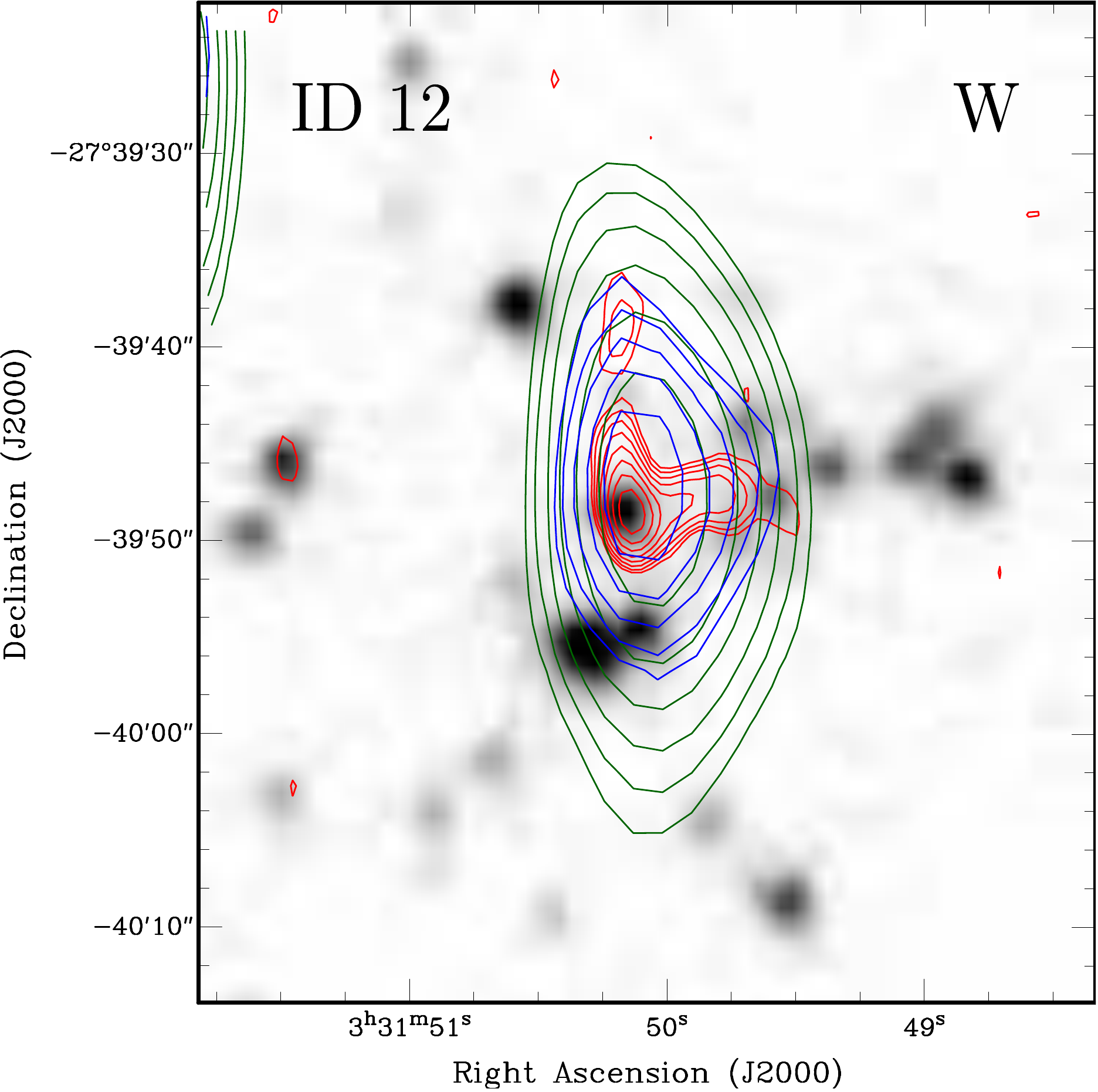}}
{\includegraphics[width=1.72in, trim= 0 20 0 0, clip=true]{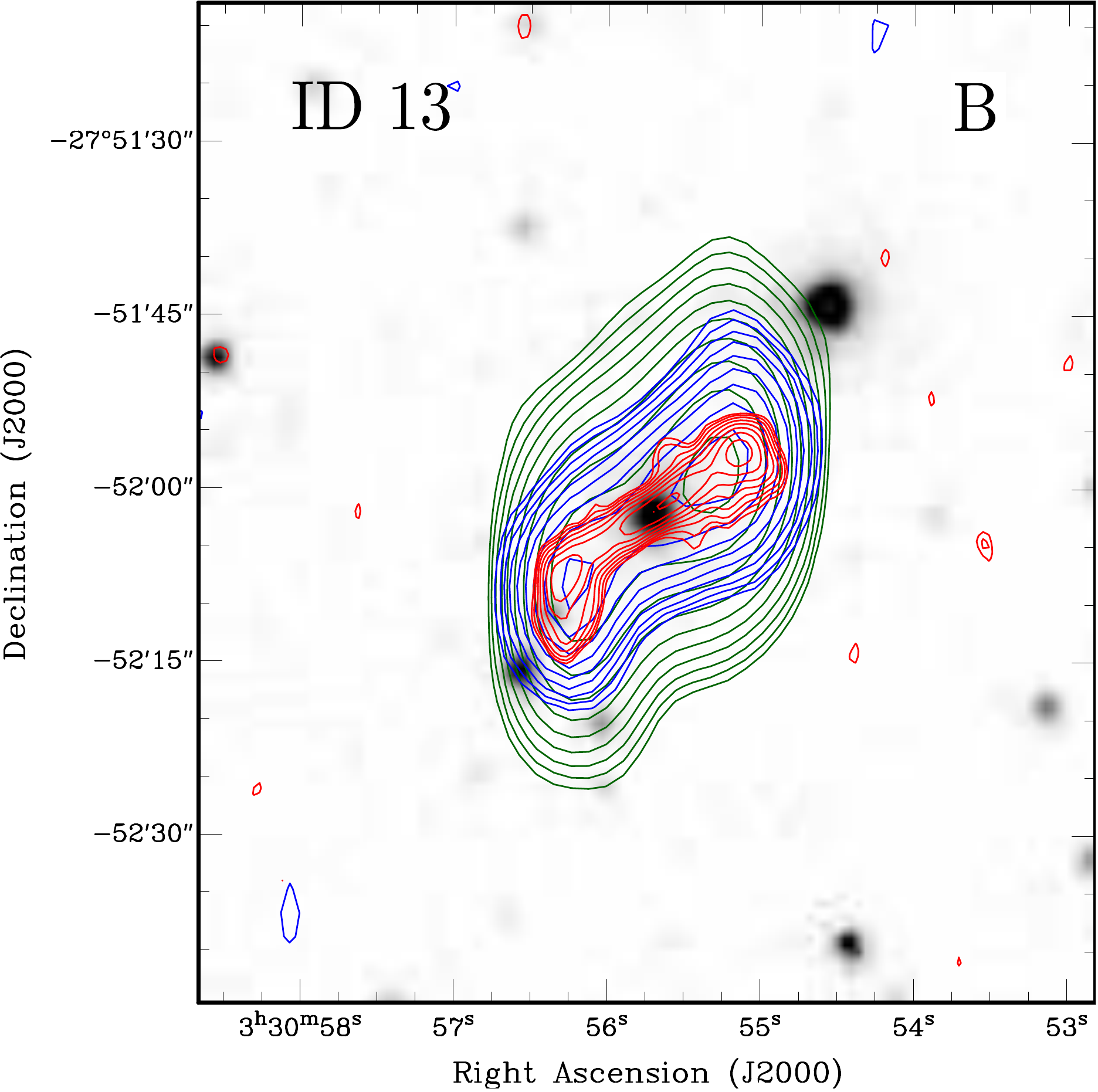}}
{\includegraphics[width=1.624in, trim= 30 20 0 0, clip=true]{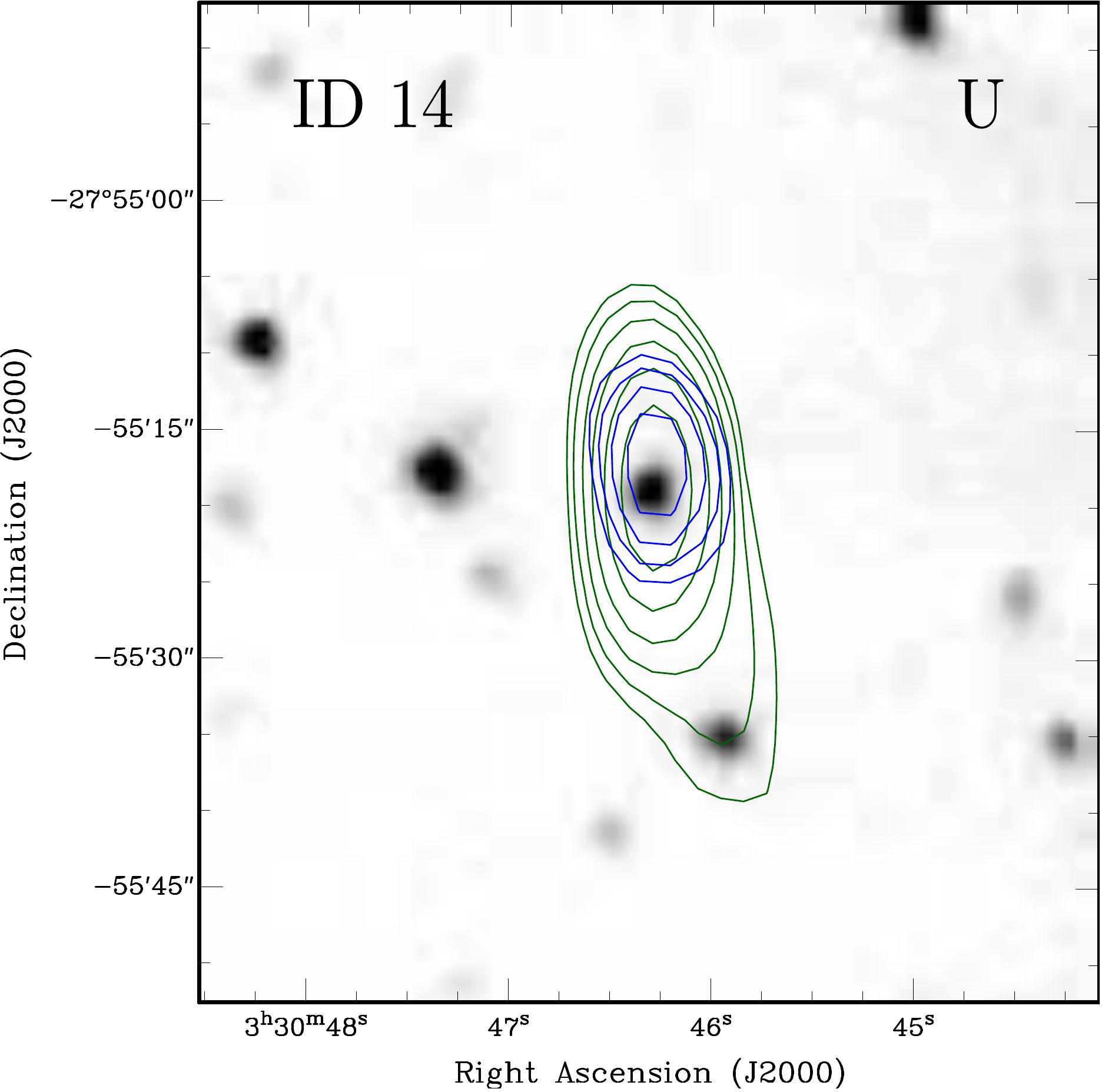}}
{\includegraphics[width=1.624in, trim= 30 20 0 0, clip=true]{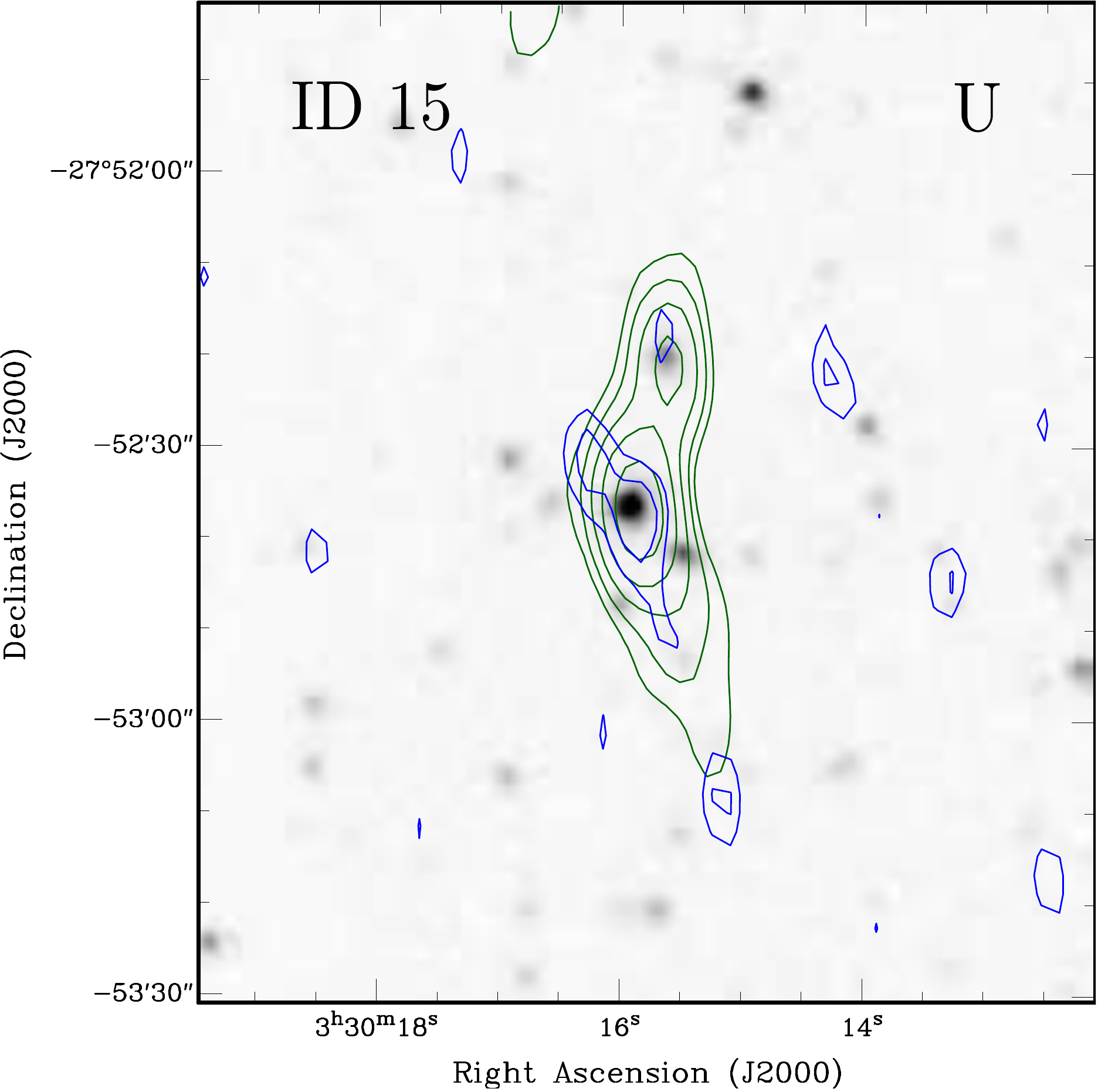}}
{\includegraphics[width=1.624in, trim= 30 20 0 0, clip=true]{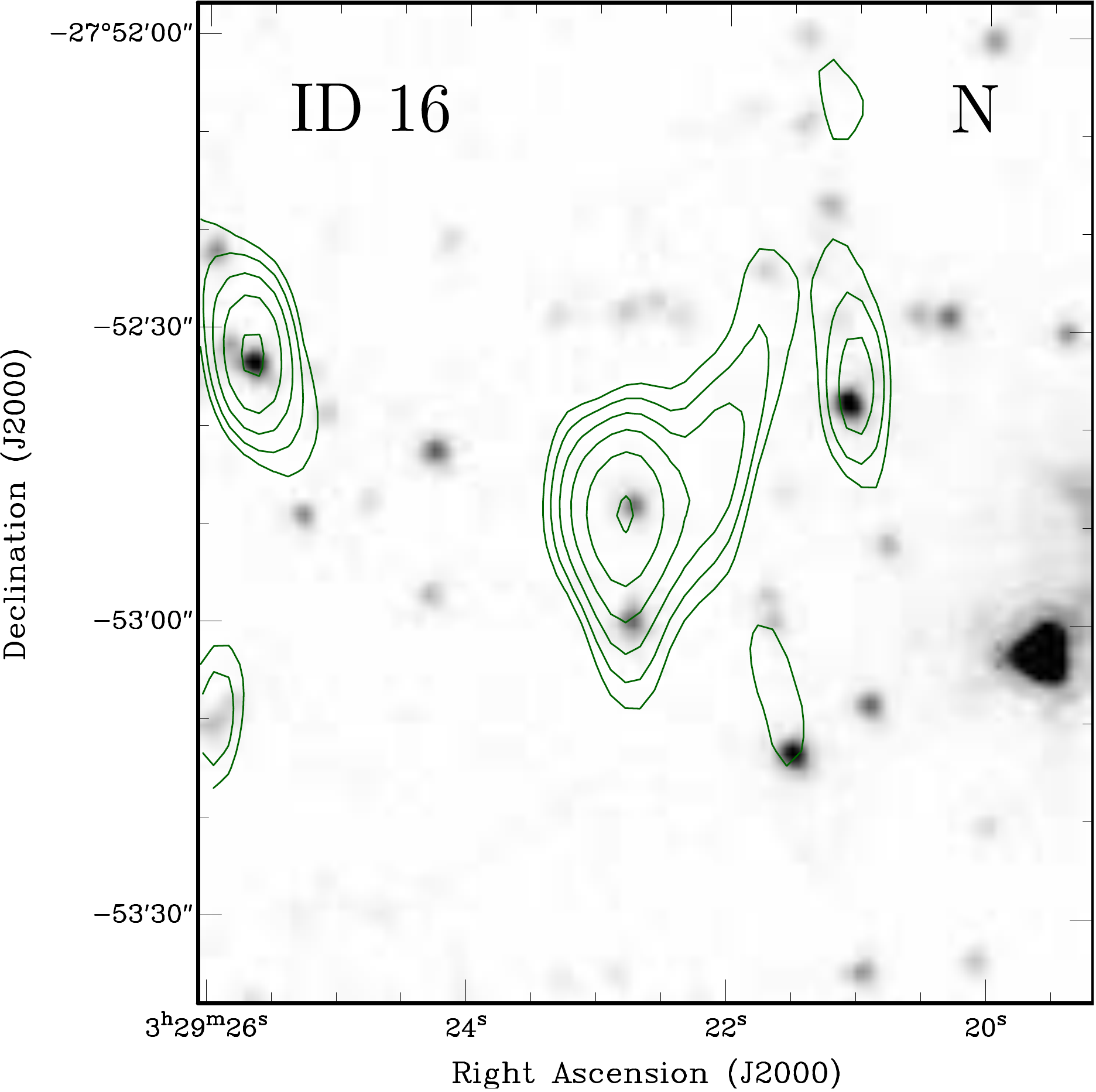}\vspace*{.02cm}}
{\includegraphics[width=1.72in, trim= 0 0 0 0, clip=true]{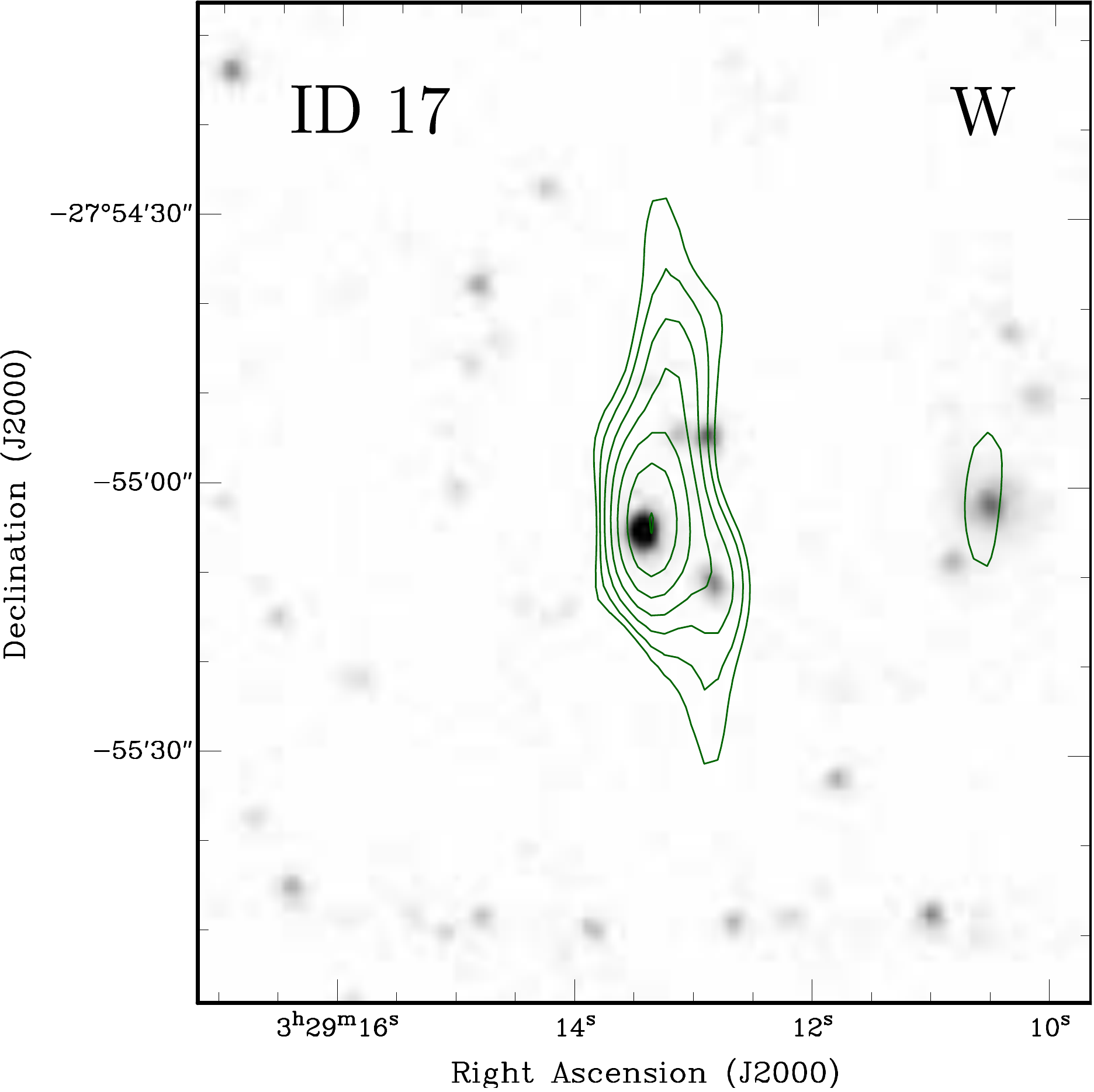}}
{\includegraphics[width=1.624in, trim= 30 0 0 0, clip=true]{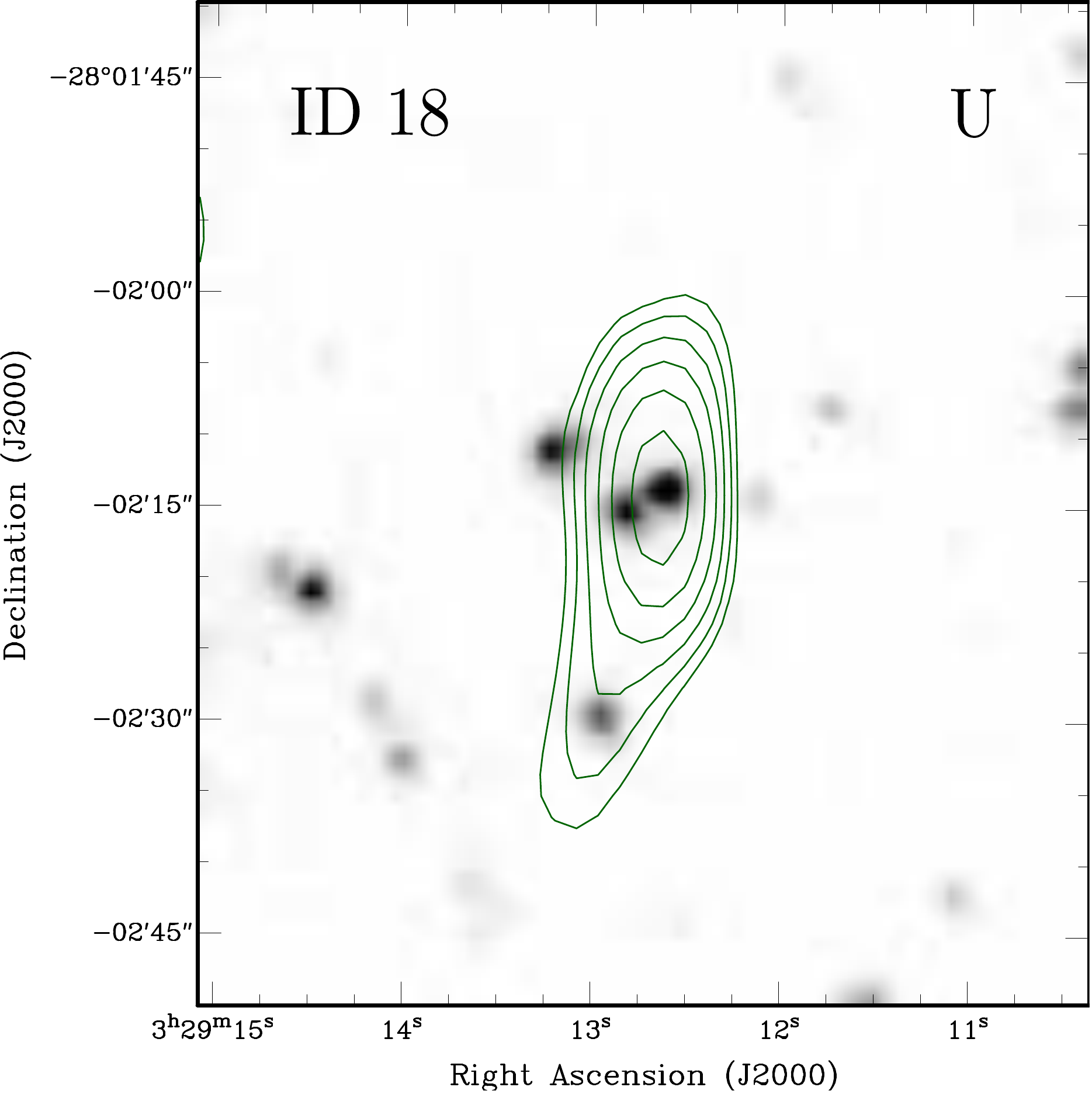}}
{\includegraphics[width=1.624in, trim= 30 0 0 0, clip=true]{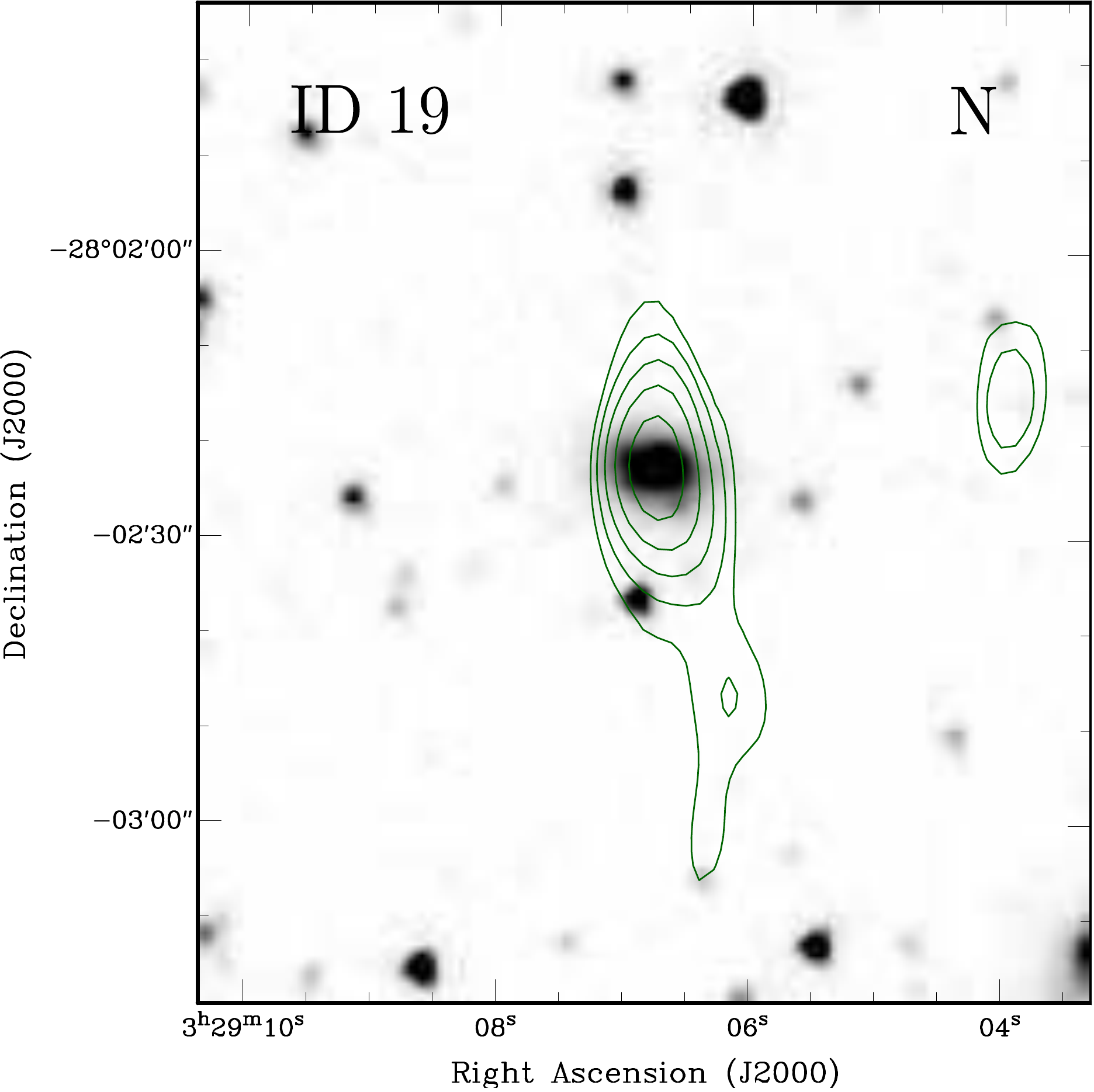}}
{\includegraphics[width=1.624in, trim= 30 0 0 0, clip=true]{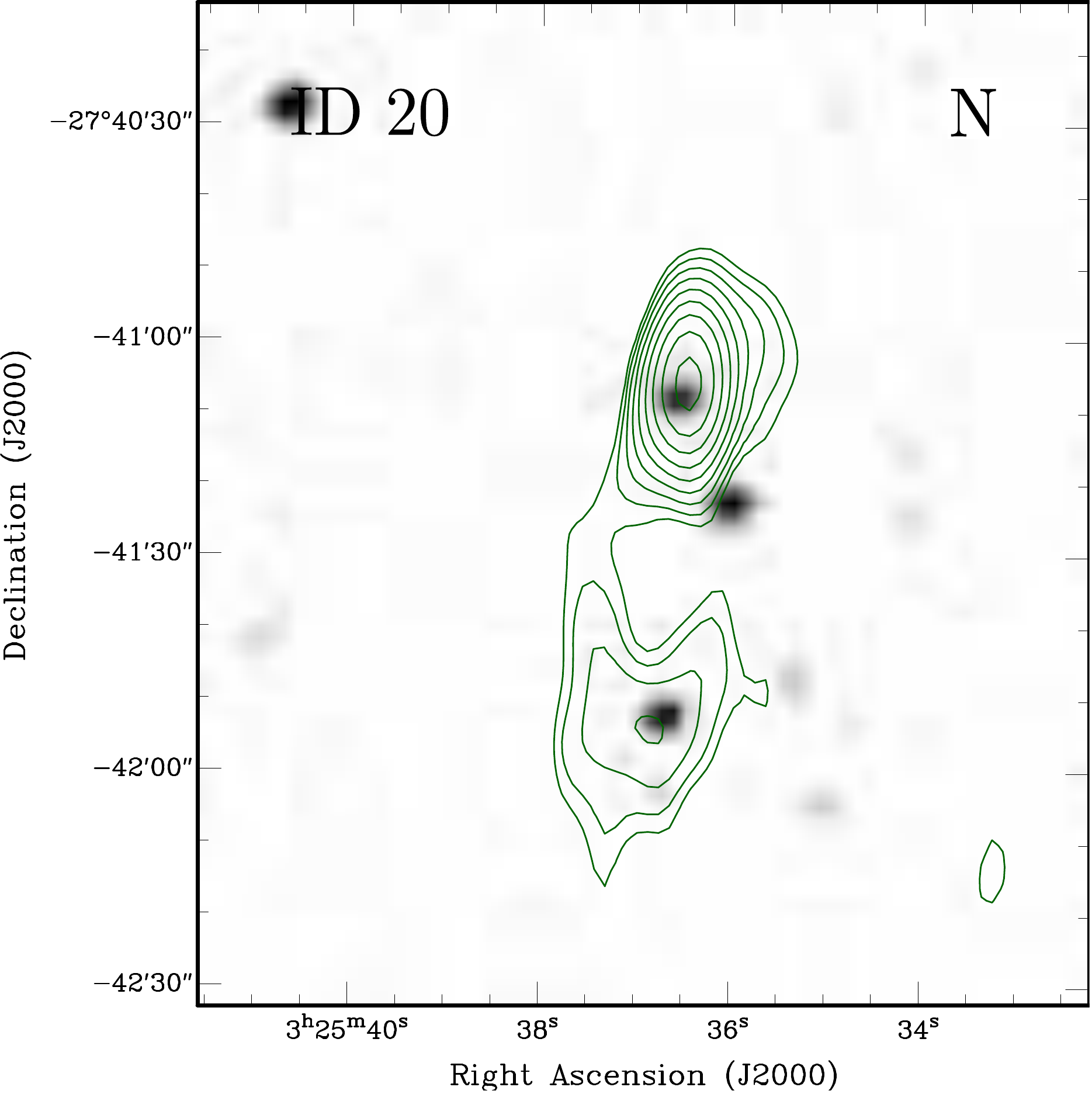}}

\caption{Detected sources in the ATLAS-CDFS field. Radio structure of the detected sources based on 1.4 GHz ATLAS first and third data releases along with VLA data, respectively shown with blue, green, and red contours, overlaid on optical images extracted from the SWIRE survey or GaBoDS. Radio contours start at three times the local rms value in the image and increase at intervals of $\sqrt{2}$. The IDs shown in the top left of each sub-image refer to Table \ref{tab:prop}. Labels shown in the top right of the sub-images represent the morphological classification of the sources; B, W, N, R, and U correspond to BT, WAT, NAT, radio relic, and unclassified or ambiguously classified sources, respectively.}
\label{fig:bts}
\end{figure*}

\begin{figure*}
\figurenum{2}
\centering
{\includegraphics[width=1.72in, trim= 0 20 0 0, clip=true]{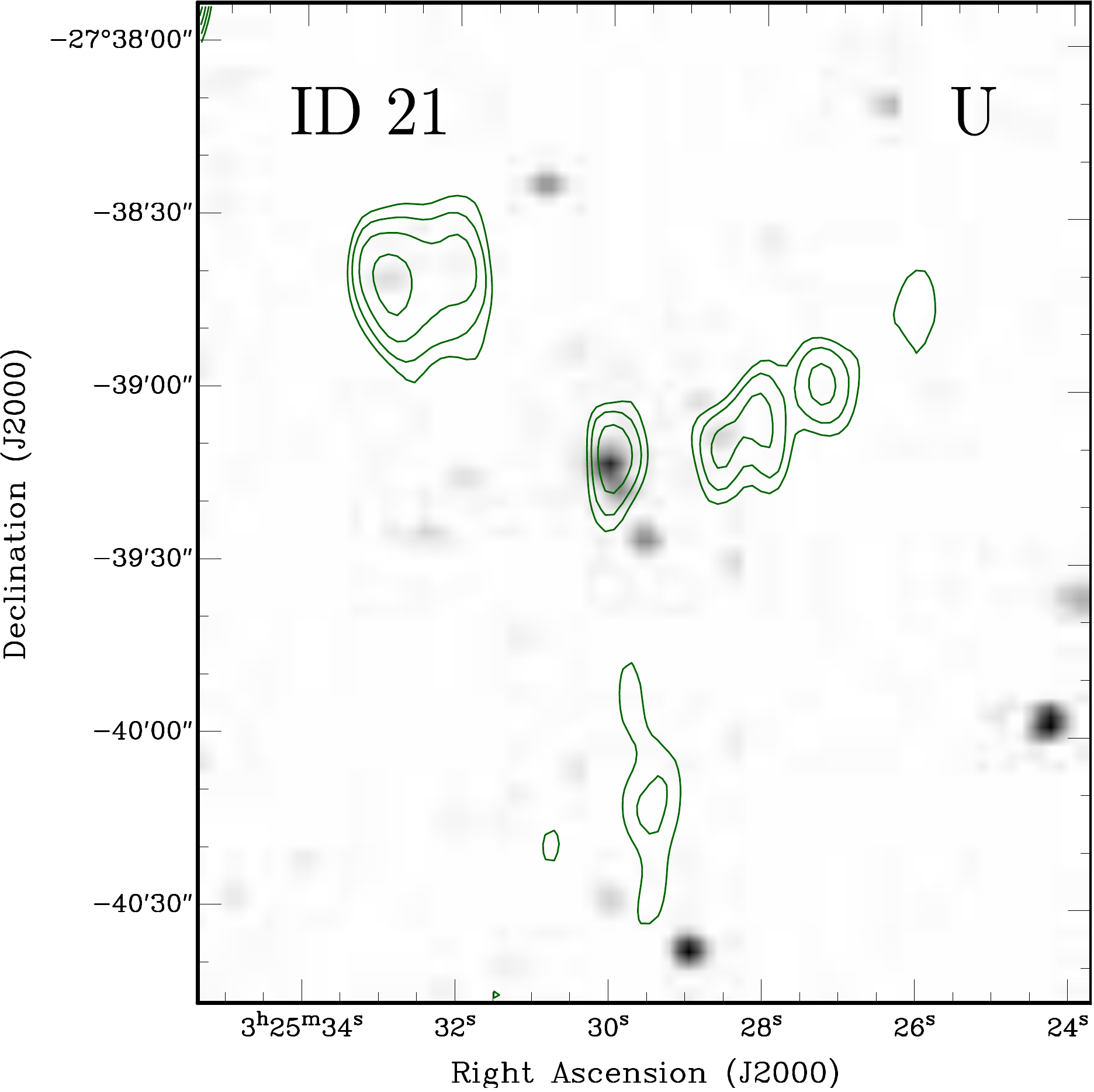}}
{\includegraphics[width=1.624in, trim= 30 20 0 0, clip=true]{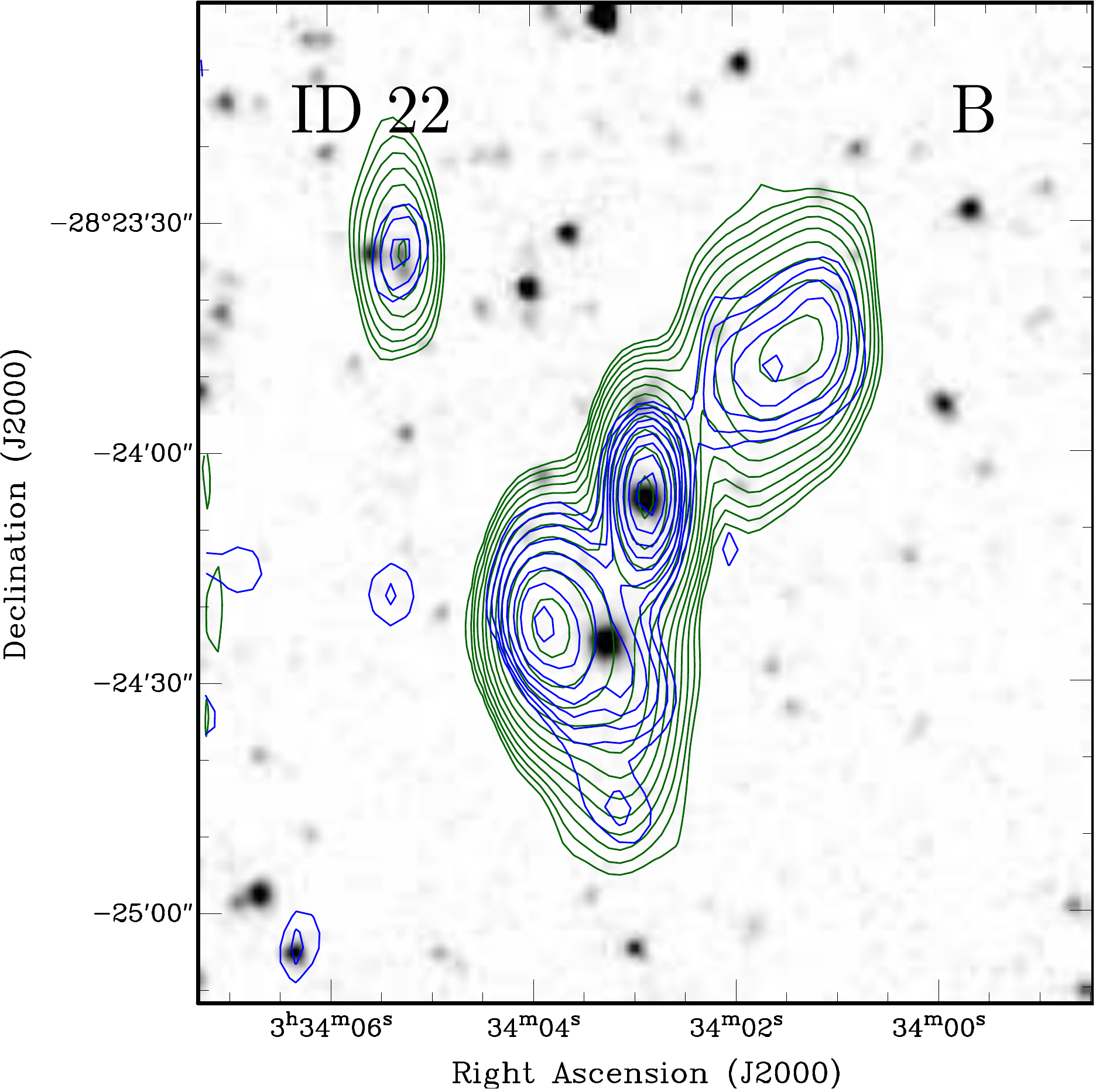}}
{\includegraphics[width=1.624in, trim= 30 20 0 0, clip=true]{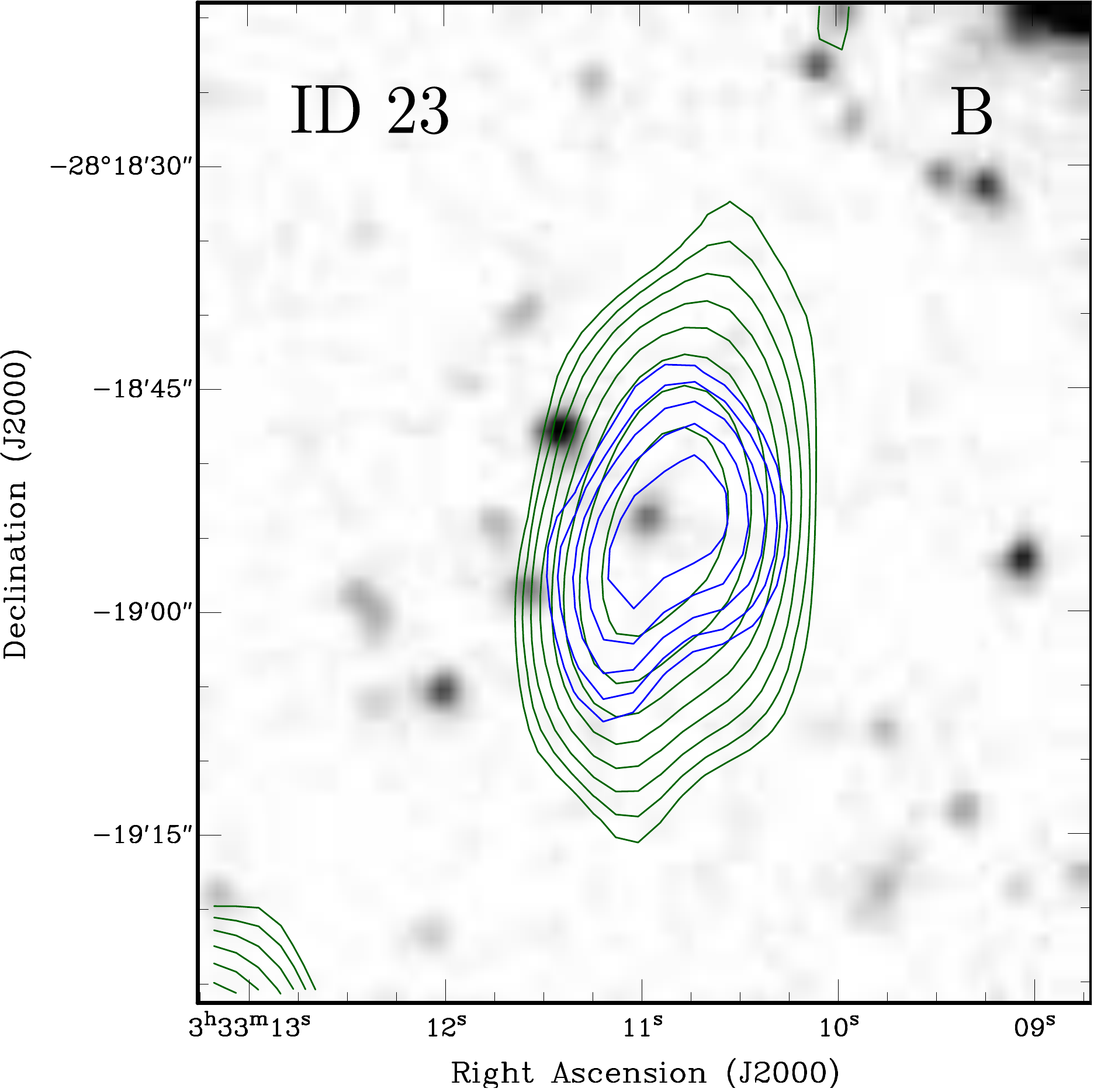}}
{\includegraphics[width=1.624in, trim= 30 20 0 0, clip=true]{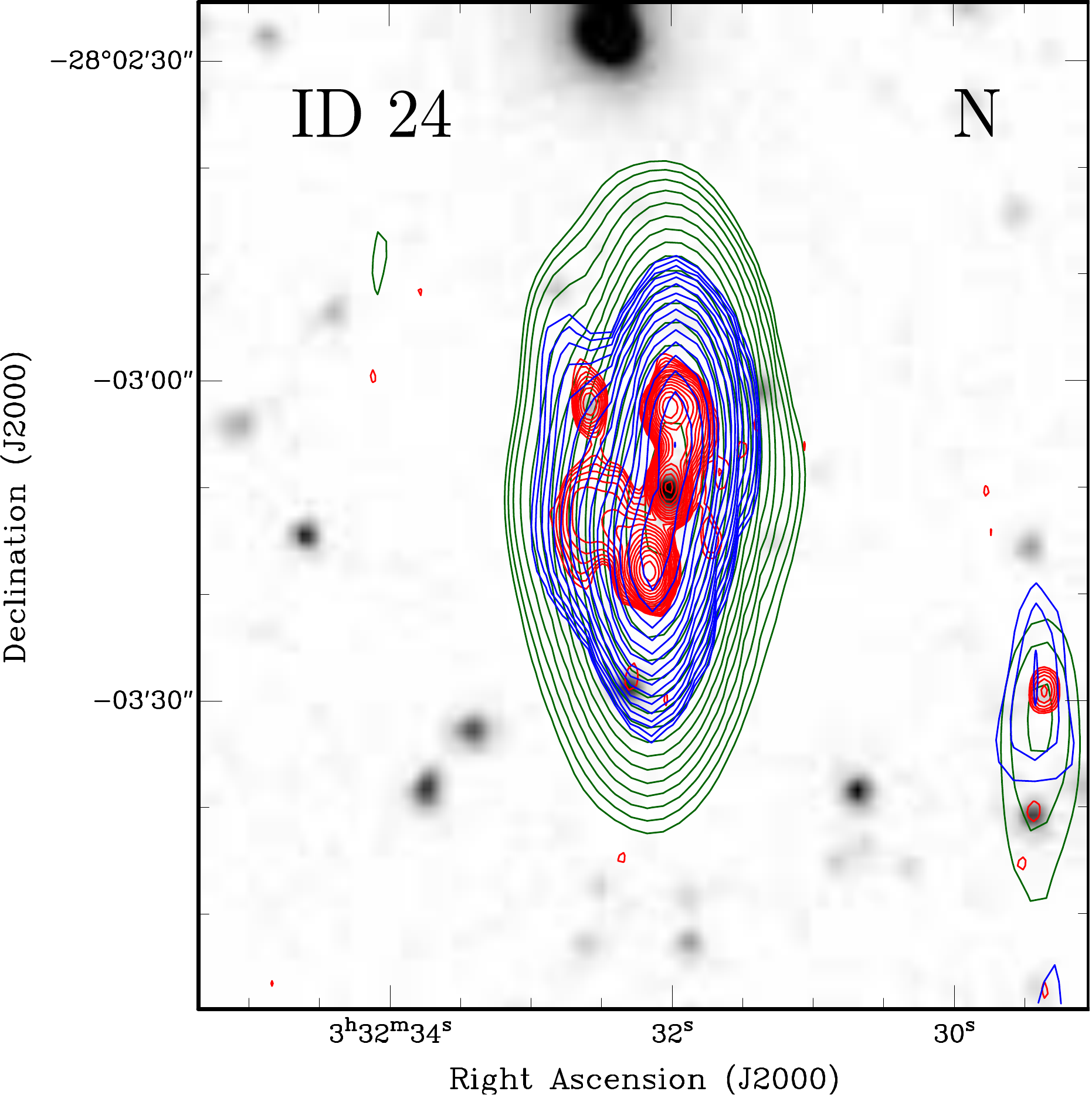}}
{\includegraphics[width=1.72in, trim= 0 20 0 0, clip=true]{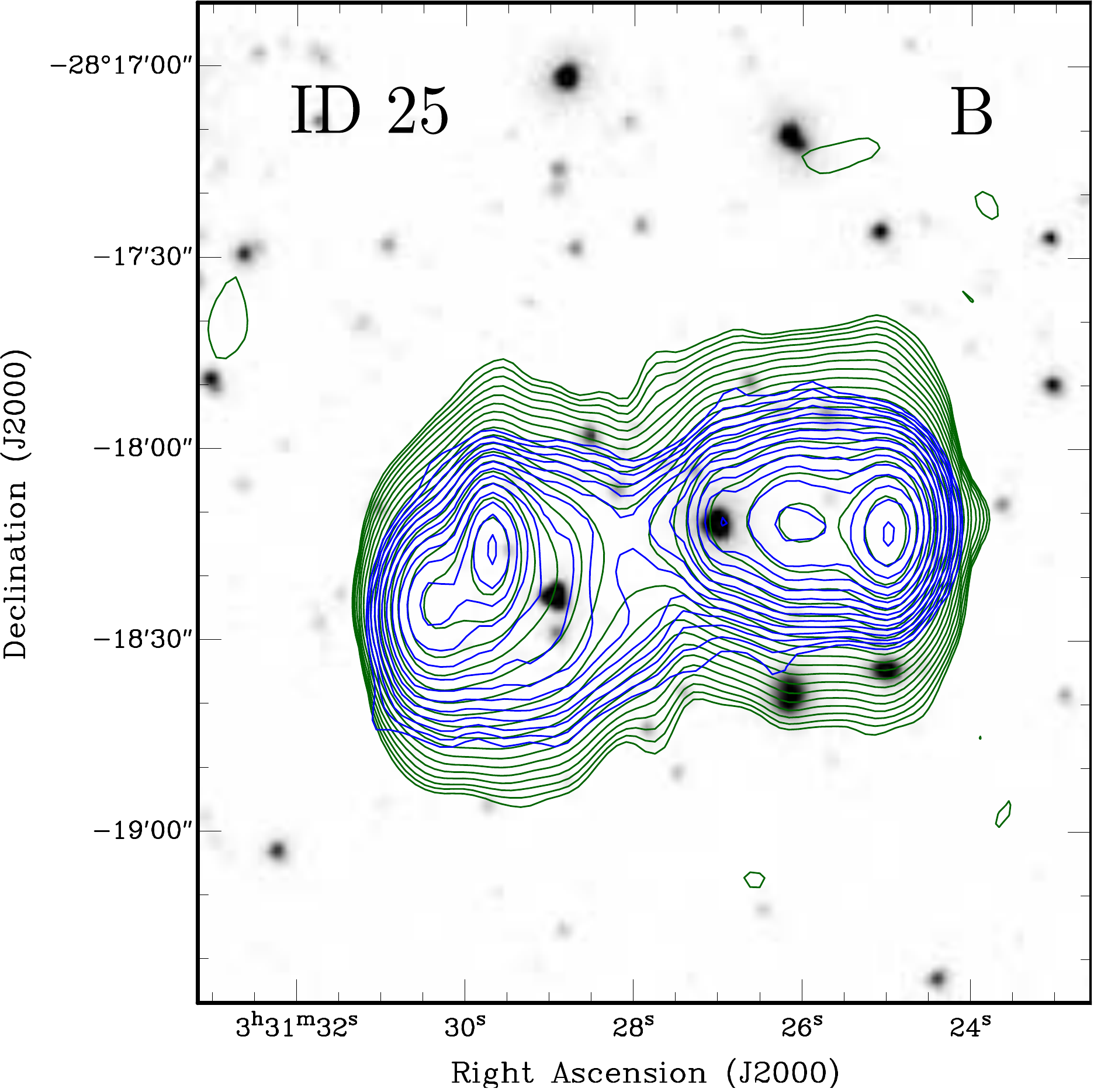}}
{\includegraphics[width=1.624in, trim= 30 20 0 0, clip=true]{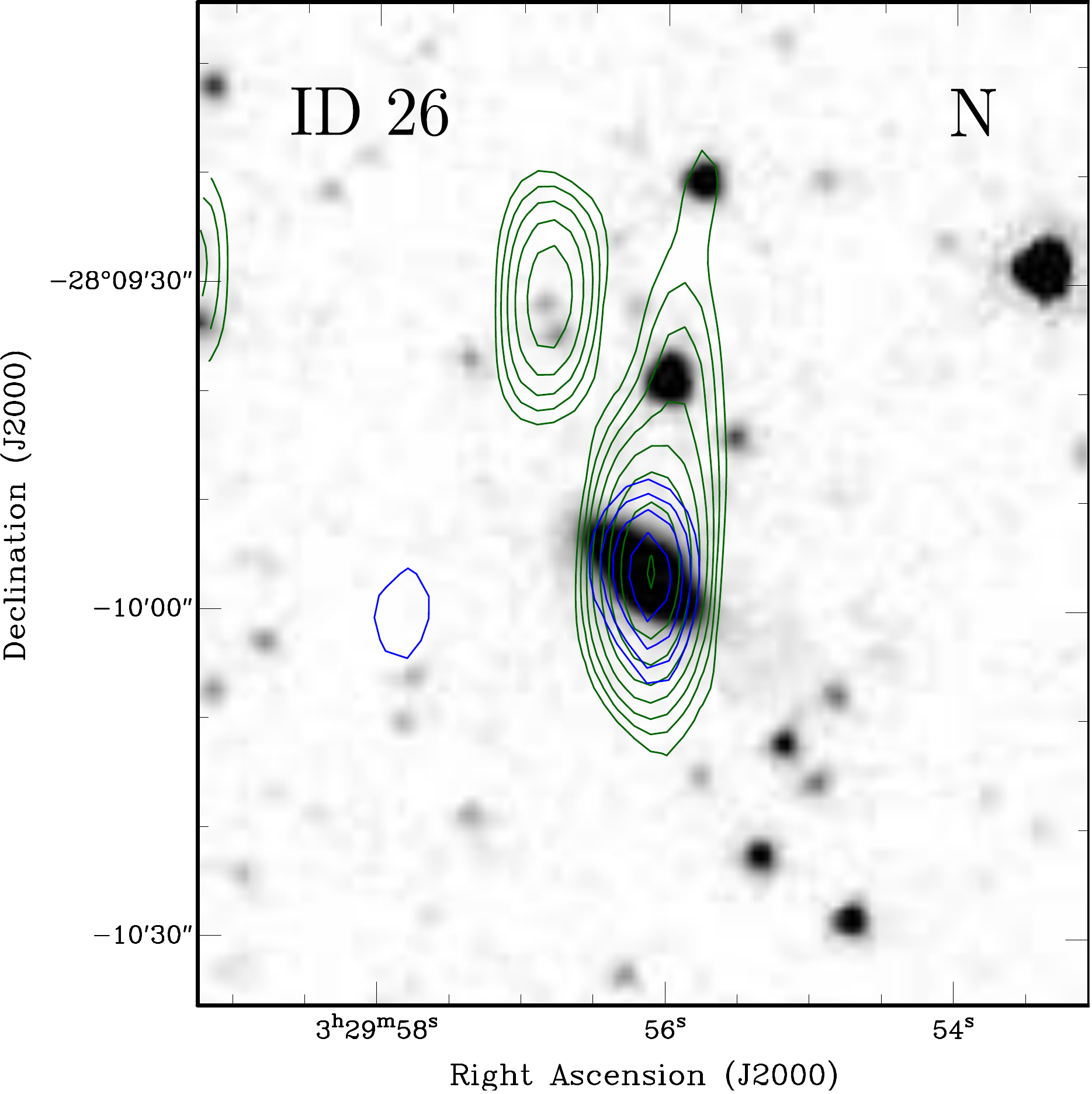}}
{\includegraphics[width=1.624in, trim= 30 20 0 0, clip=true]{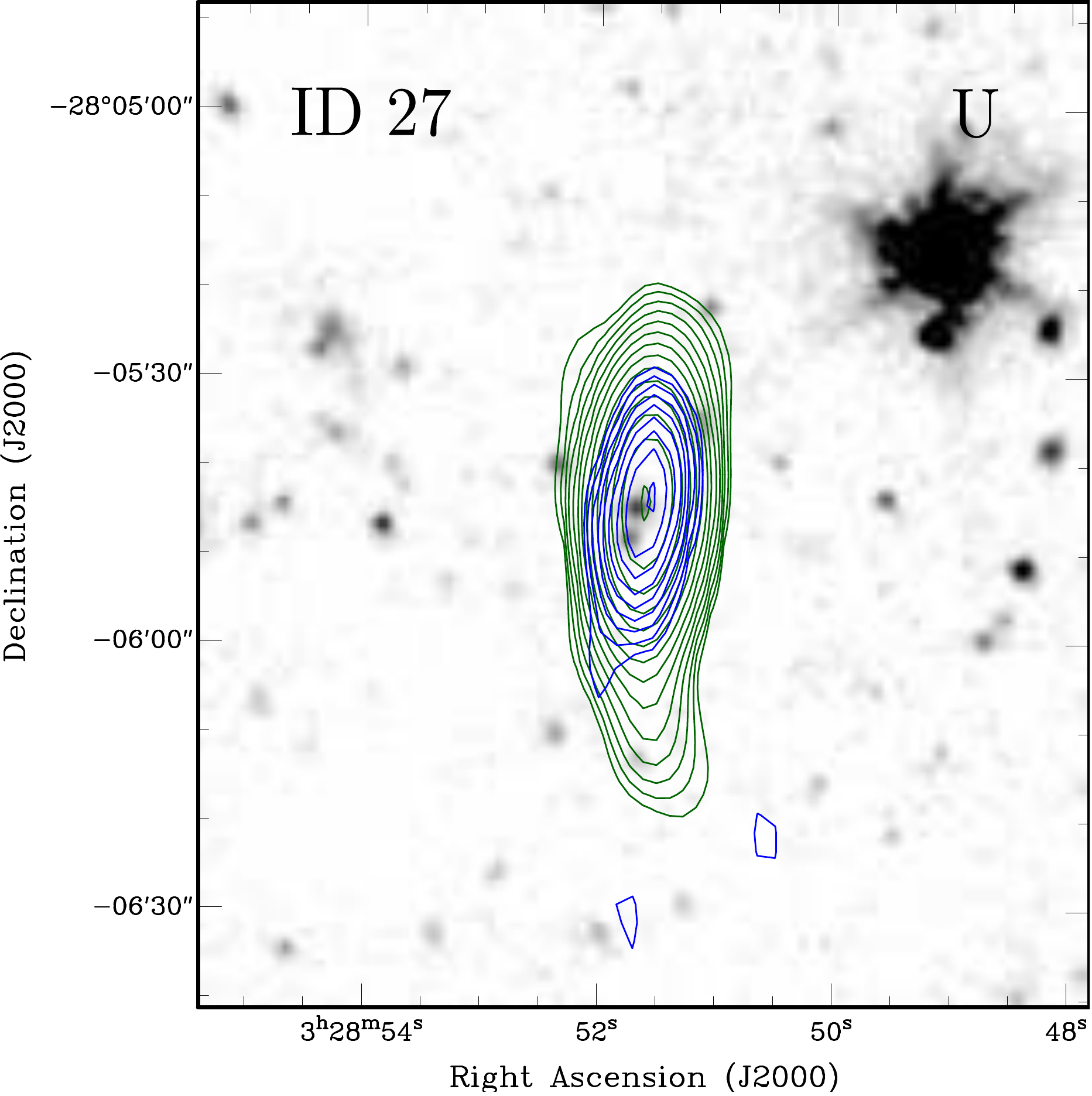}}
{\includegraphics[width=1.624in, trim= 30 20 0 0, clip=true]{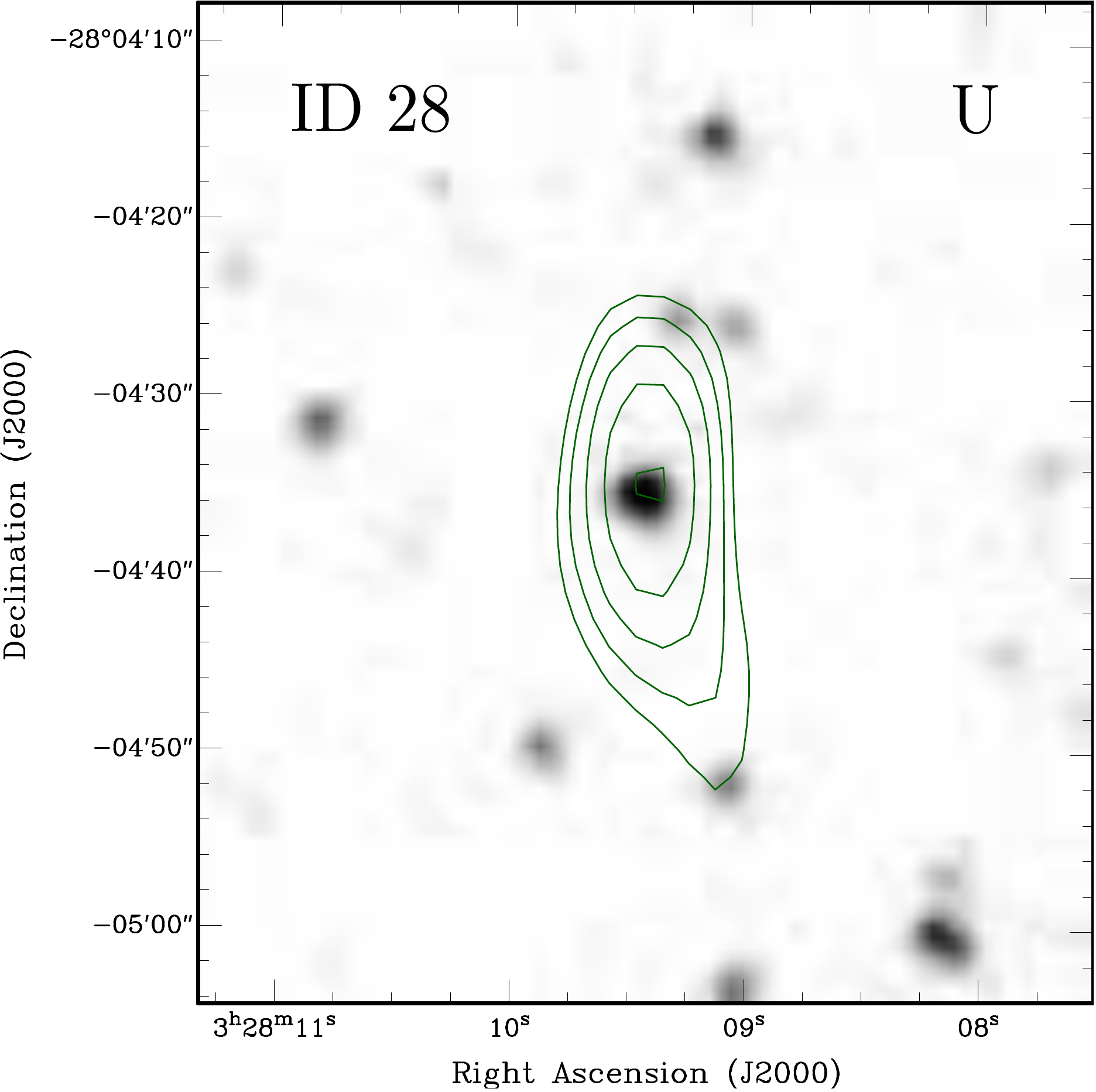}}
{\includegraphics[width=1.72in, trim= 0 20 0 0, clip=true]{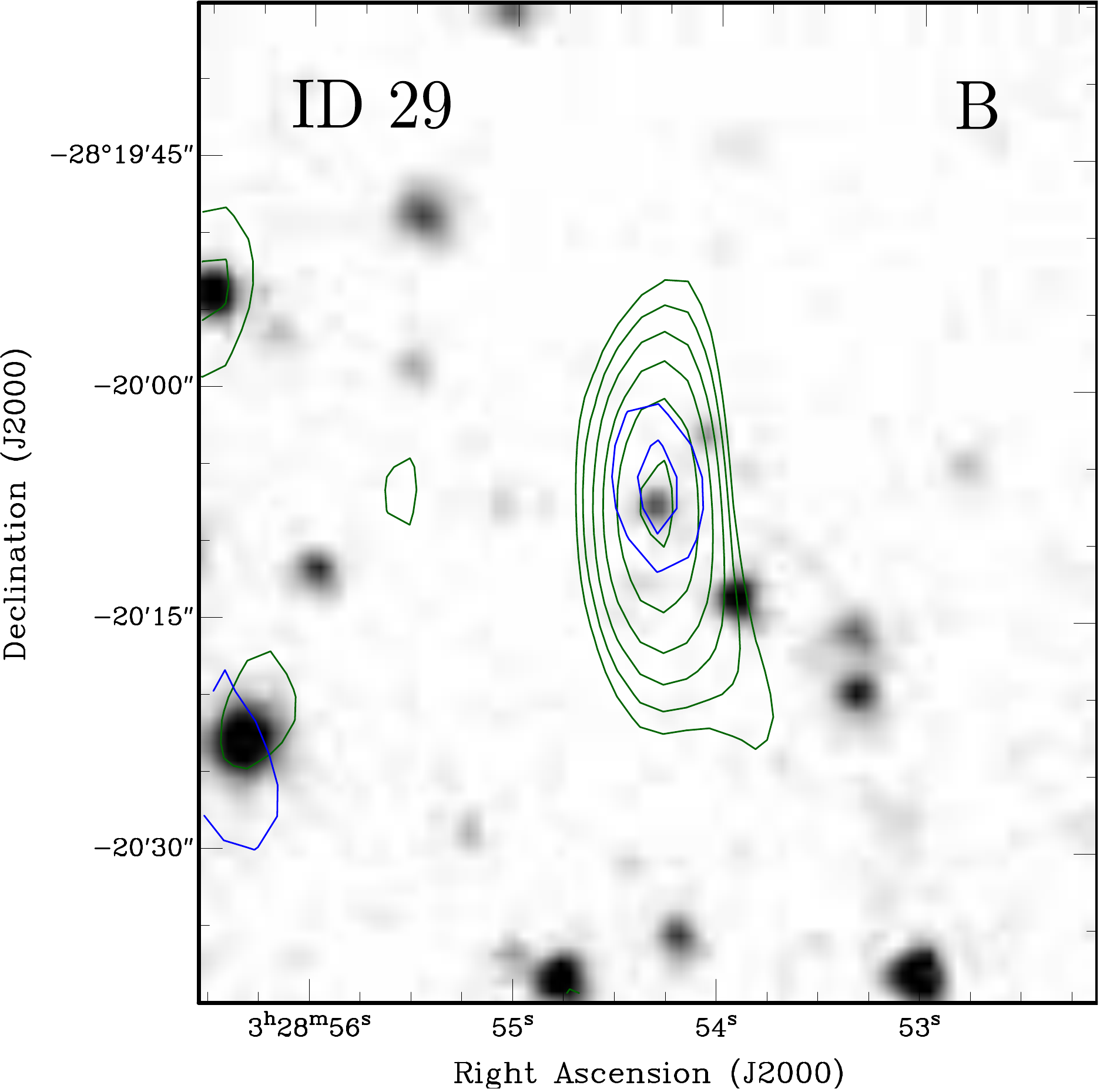}}
{\includegraphics[width=1.624in, trim= 30 20 0 0, clip=true]{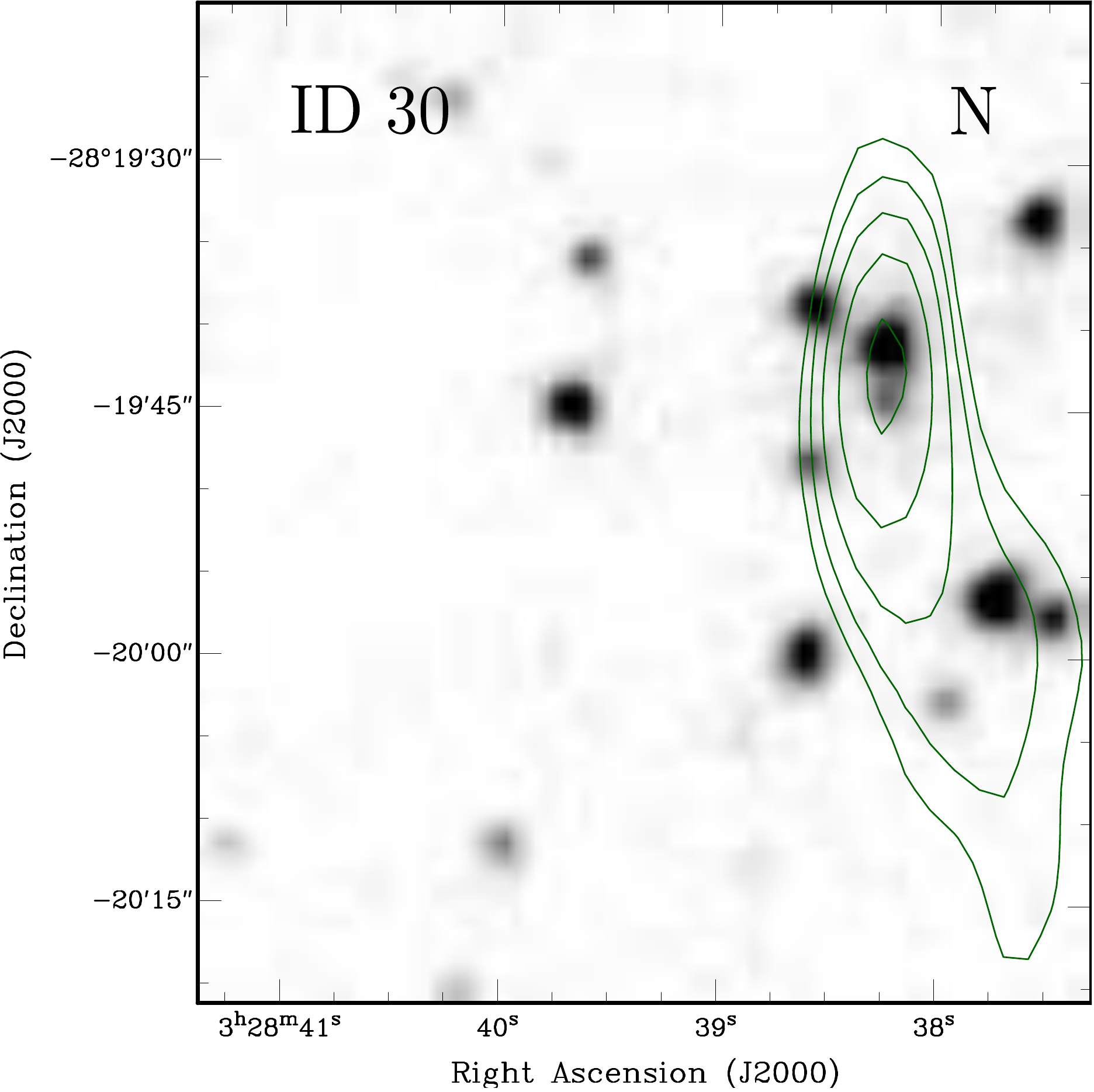}}
{\includegraphics[width=1.624in, trim= 30 20 0 0, clip=true]{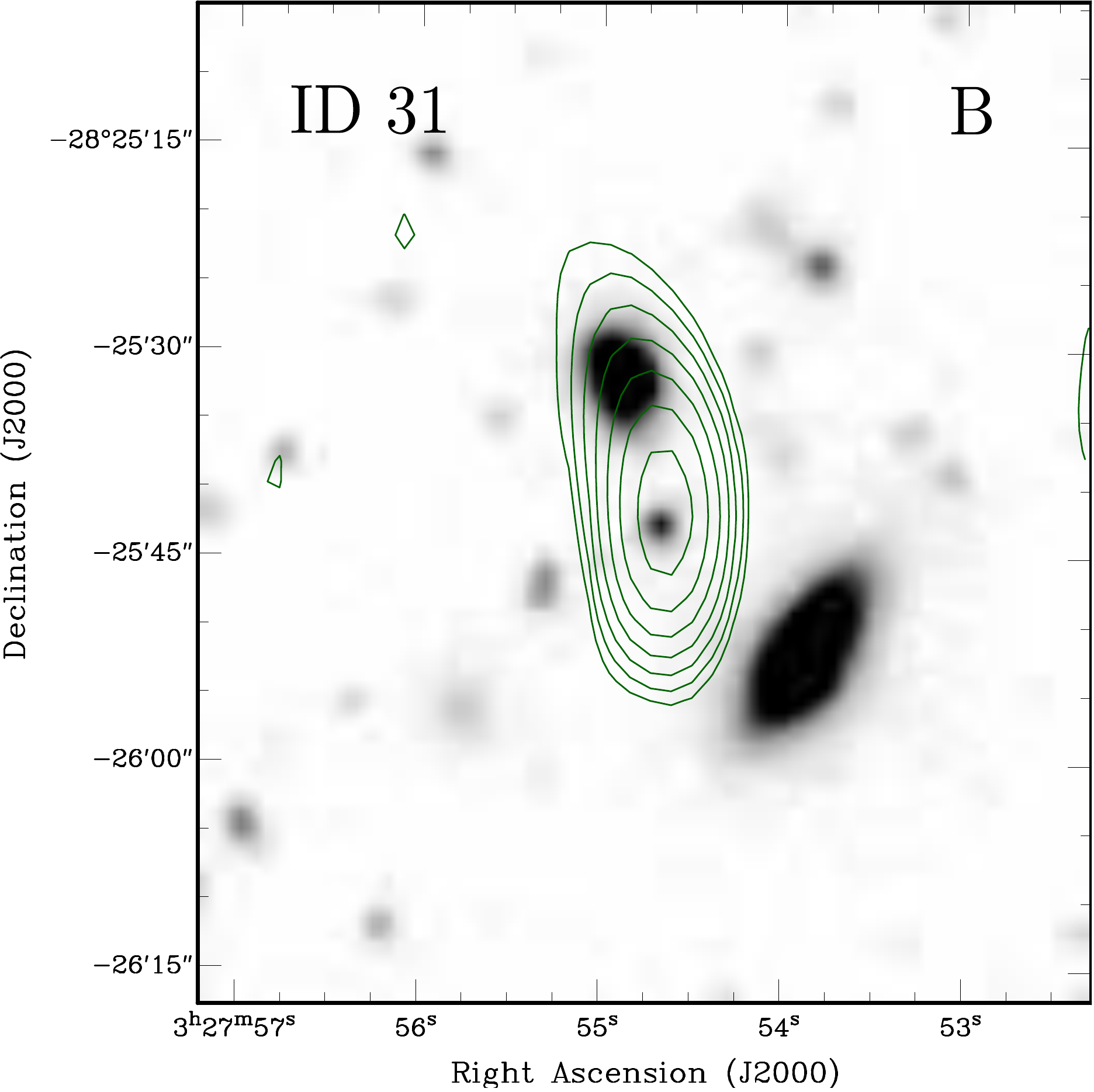}}
{\includegraphics[width=1.624in, trim= 30 20 0 0, clip=true]{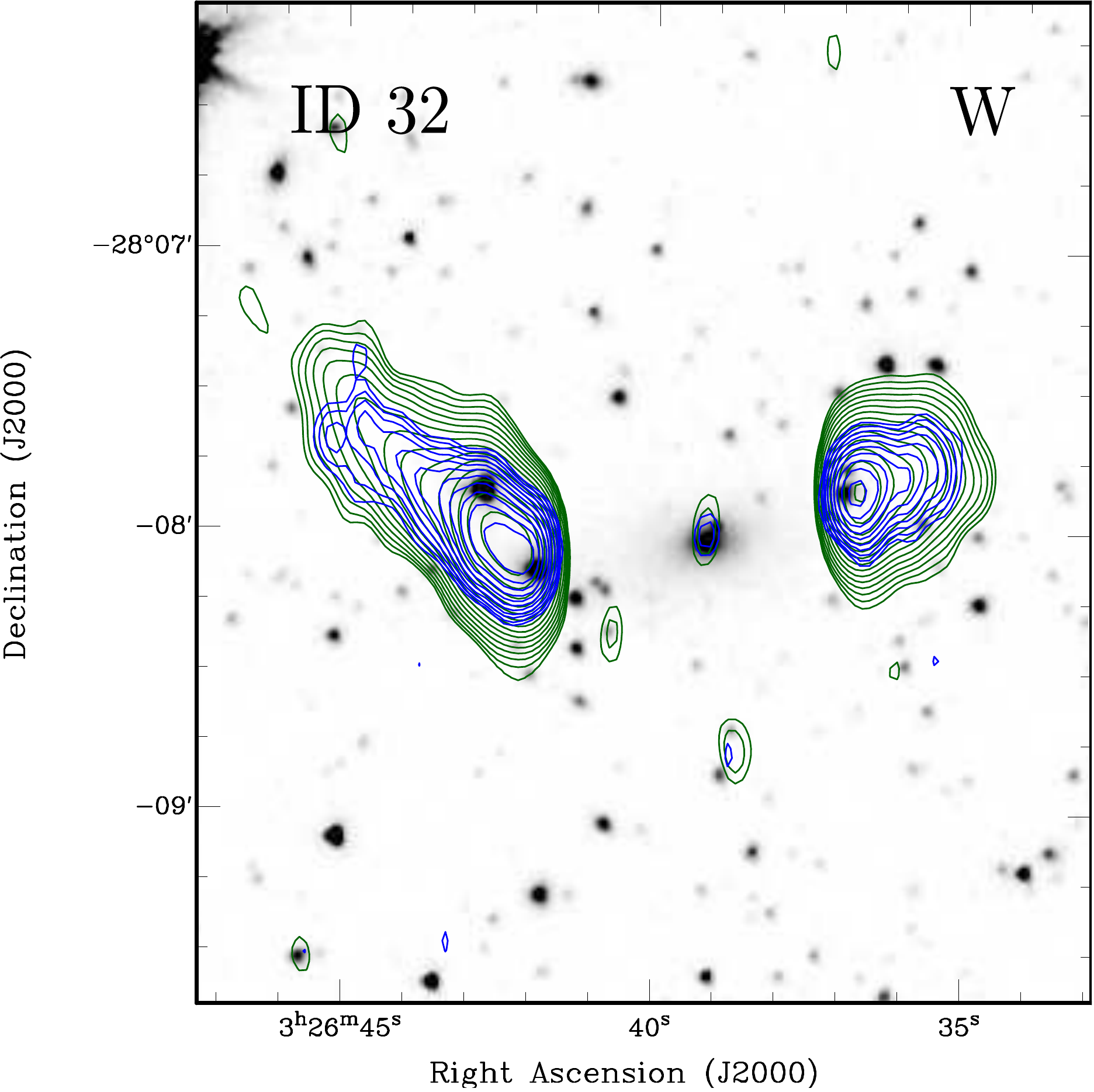}}
{\includegraphics[width=1.72in, trim= 0 20 0 0, clip=true]{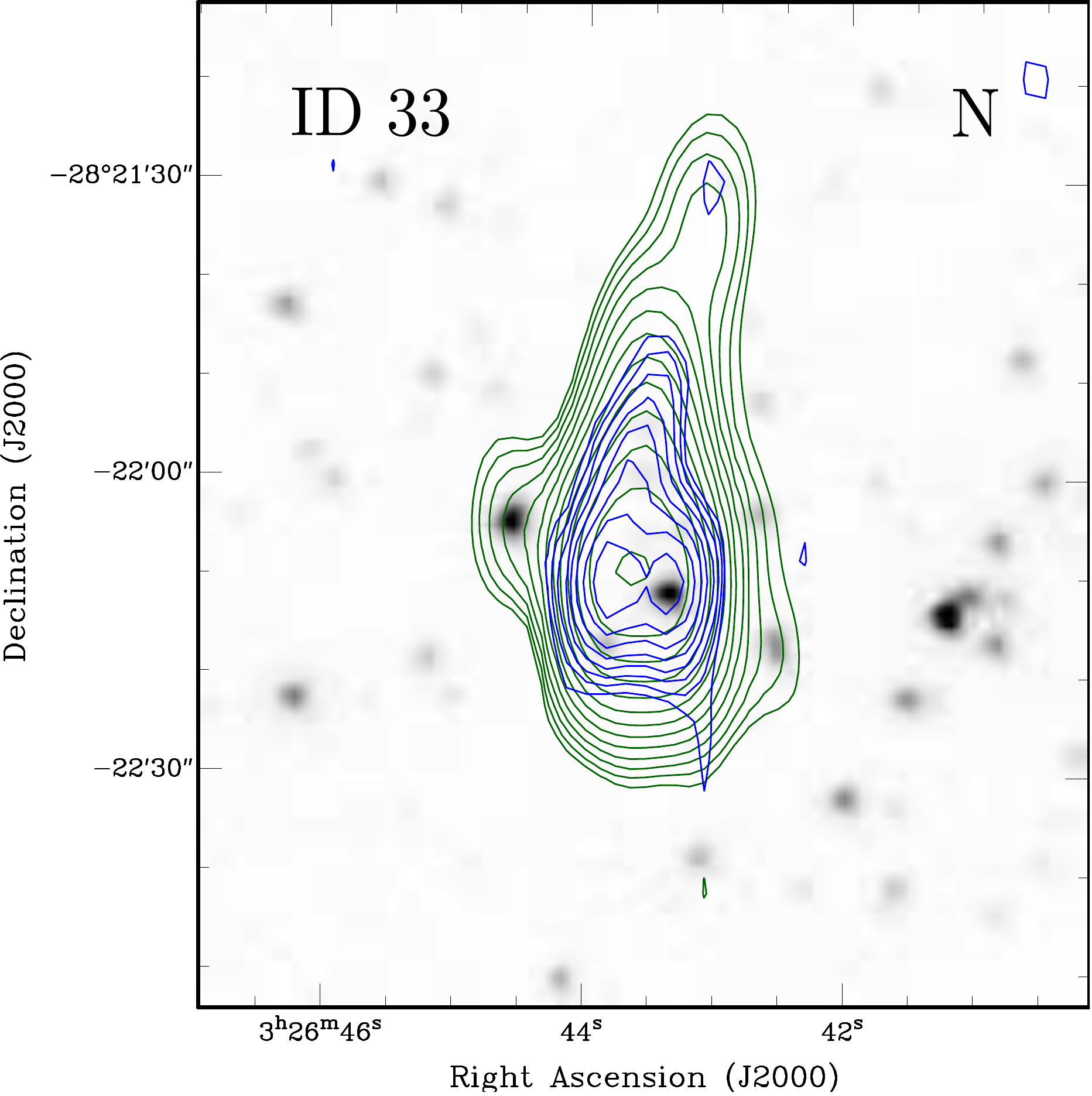}}
{\includegraphics[width=1.624in, trim= 30 20 0 0, clip=true]{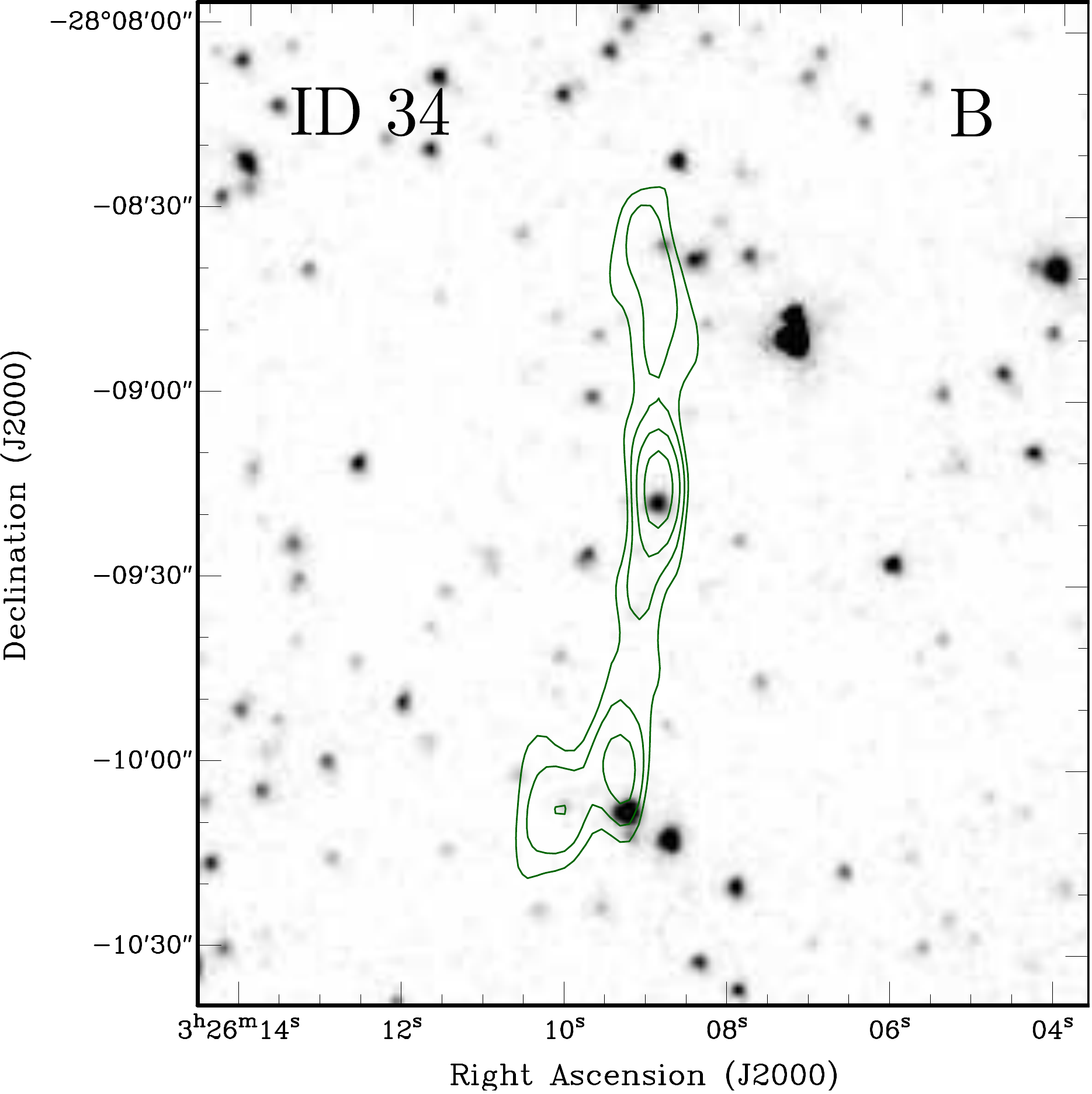}}
{\includegraphics[width=1.624in, trim= 30 20 0 0, clip=true]{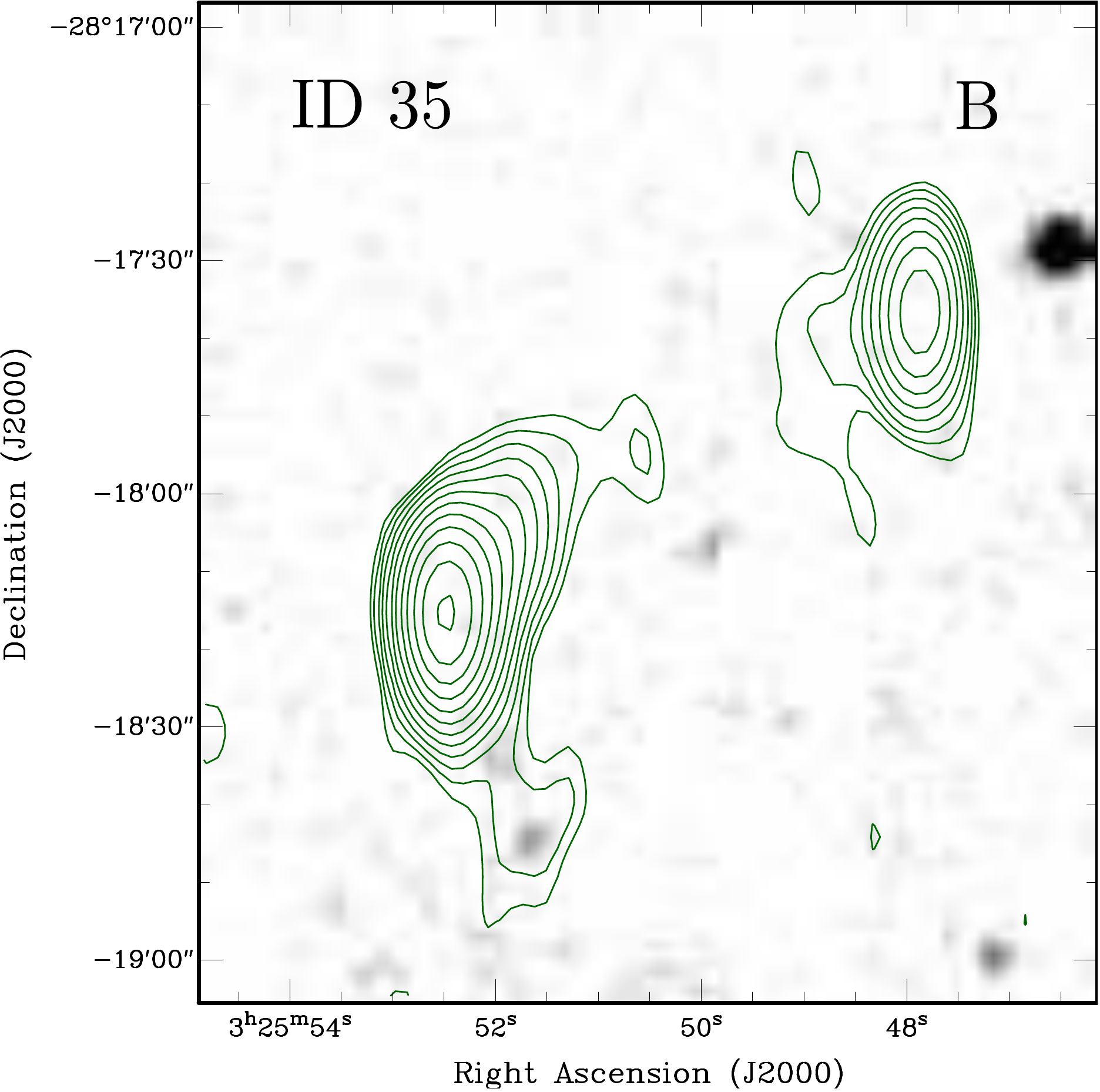}}
{\includegraphics[width=1.624in, trim= 30 20 0 0, clip=true]{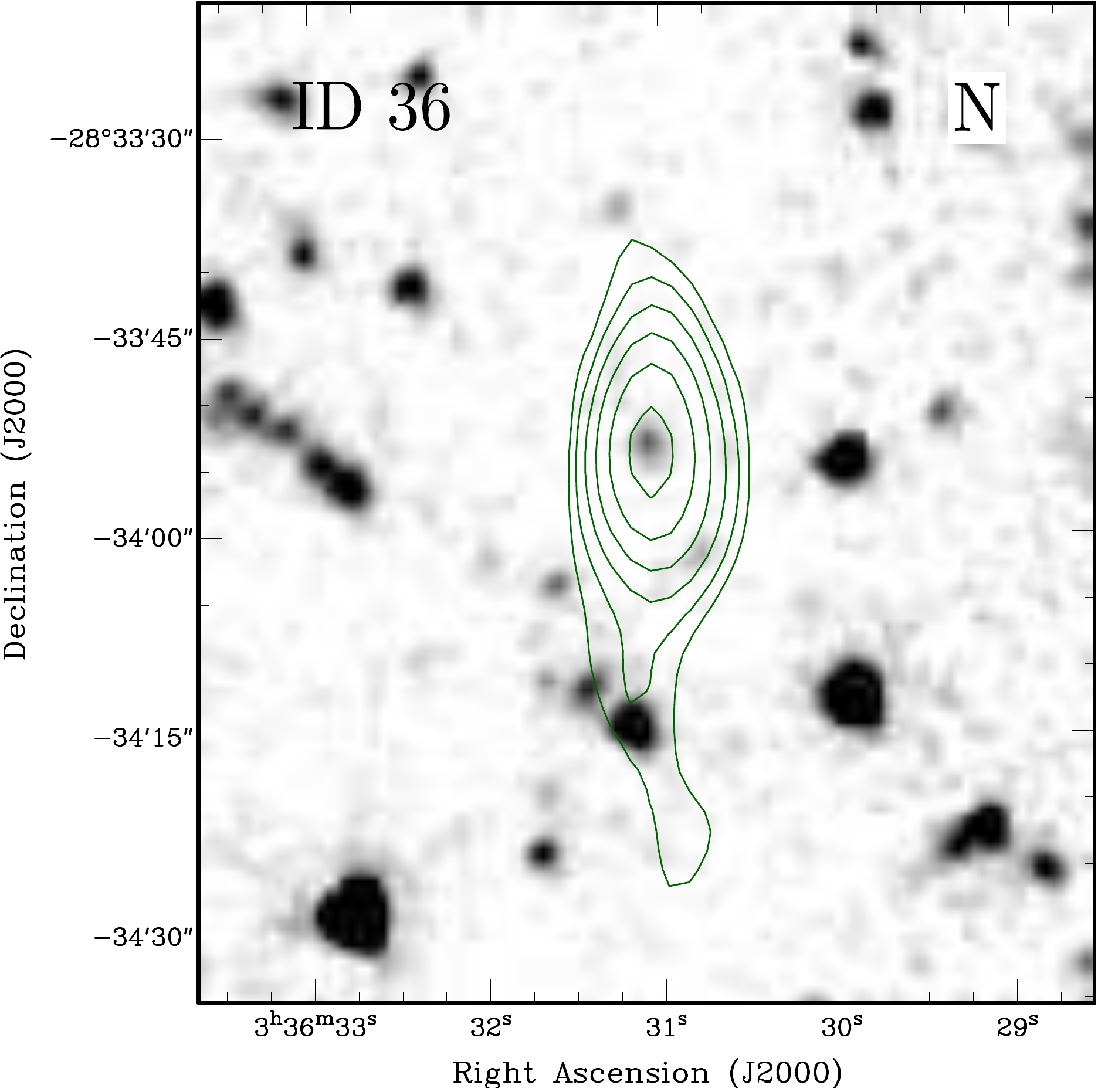}\vspace*{.02cm}}
{\includegraphics[width=1.72in, trim= 0 0 0 0, clip=true]{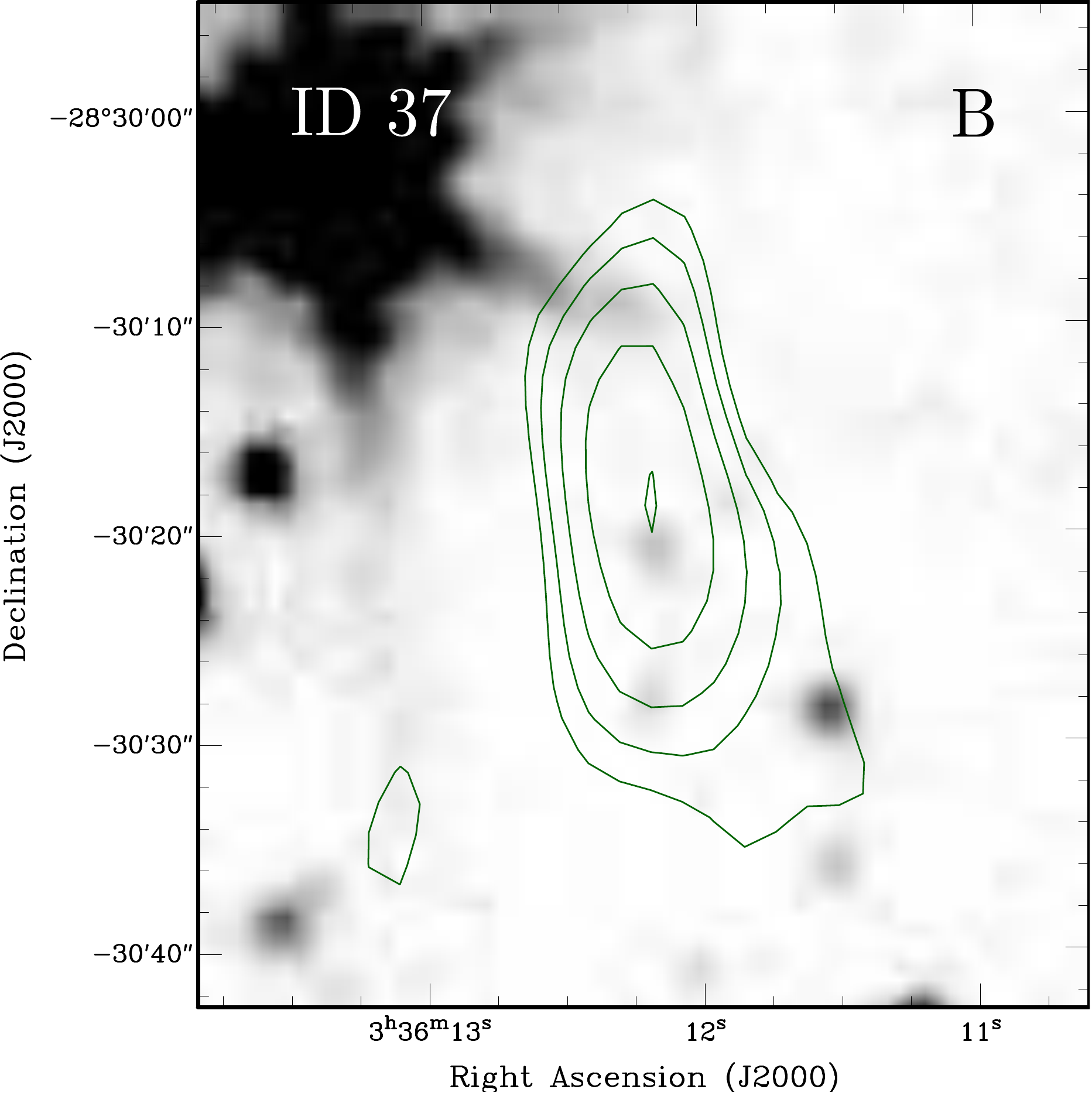}}
{\includegraphics[width=1.624in, trim= 30 0 0 0, clip=true]{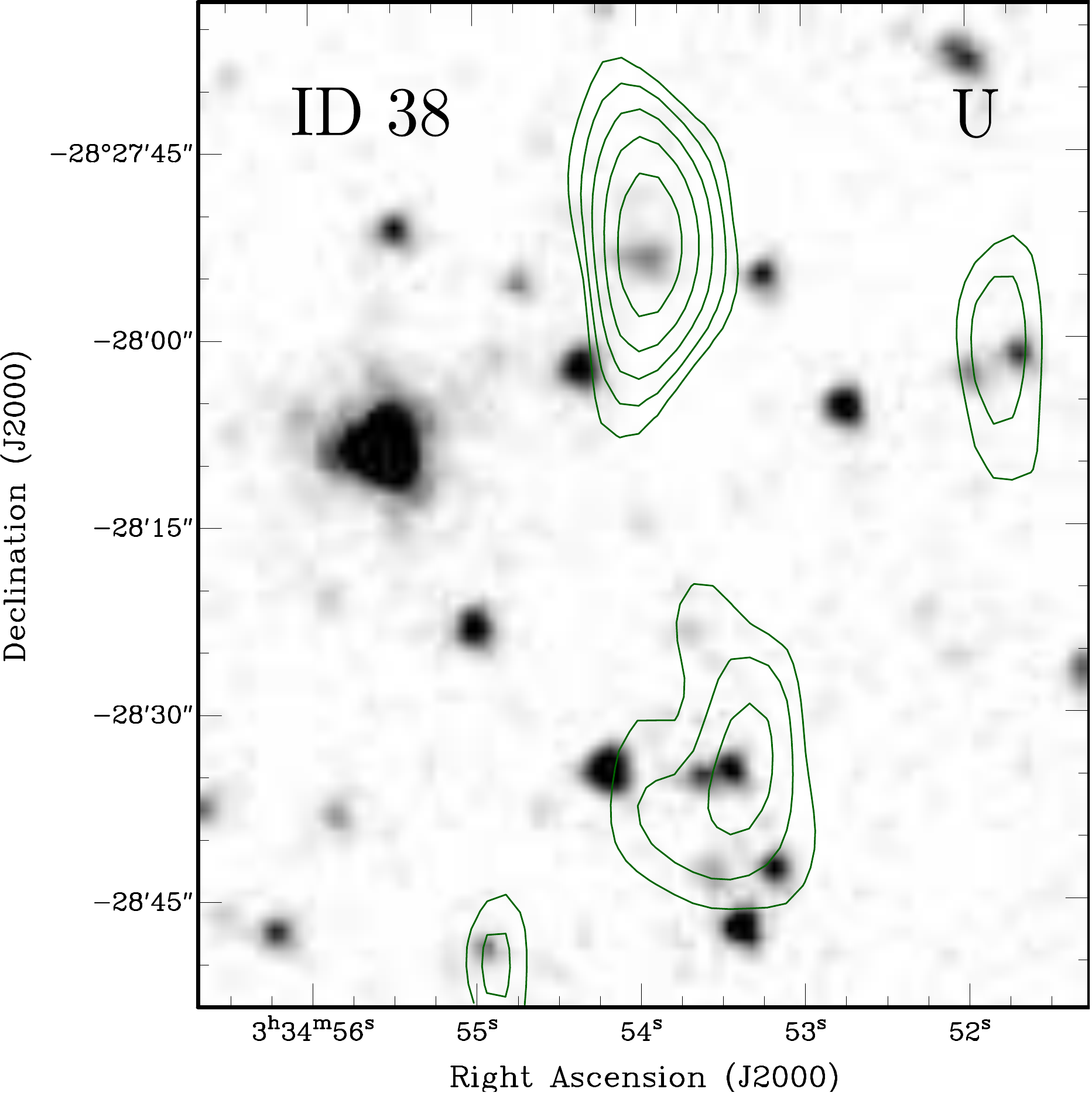}}
{\includegraphics[width=1.624in, trim= 30 0 0 0, clip=true]{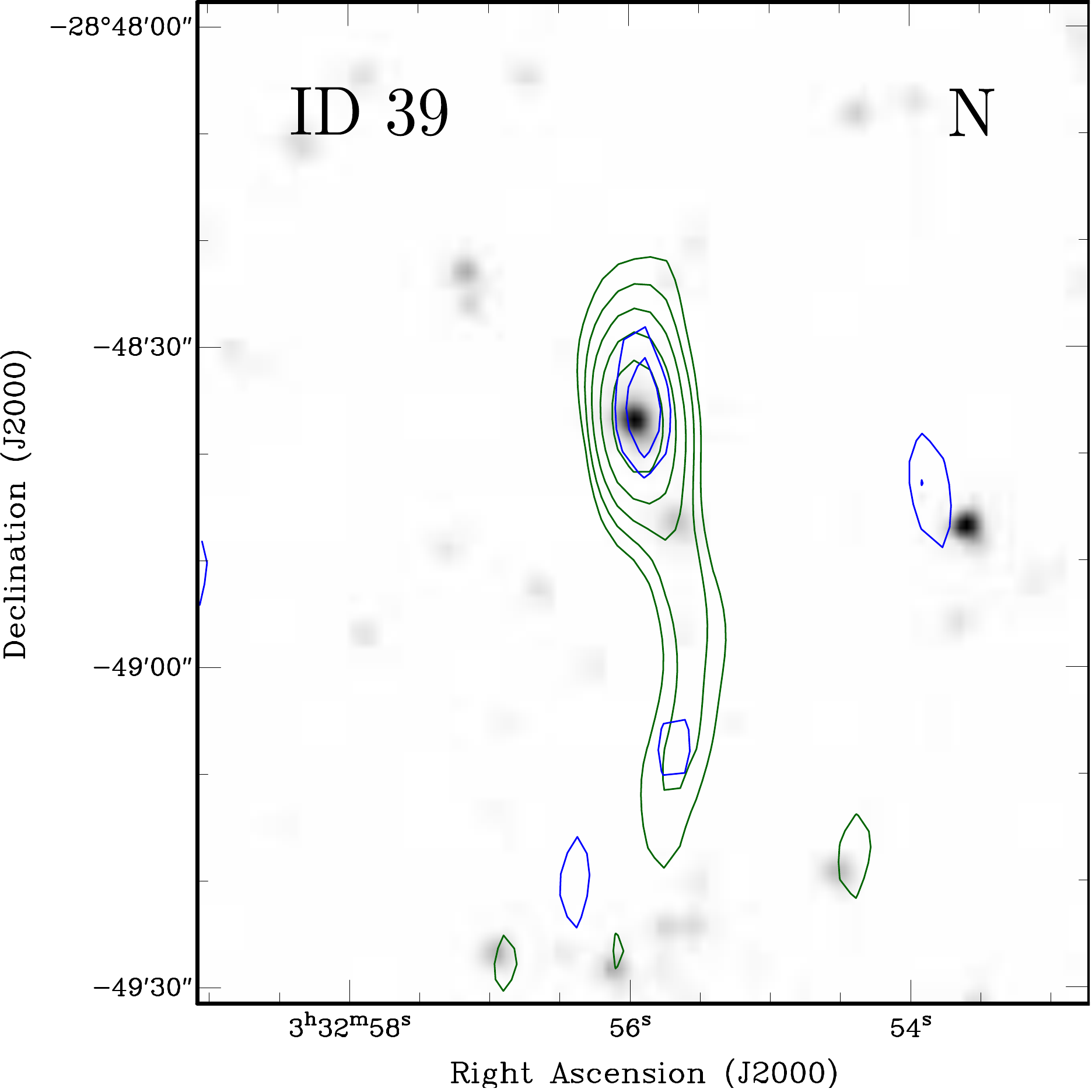}}
{\includegraphics[width=1.624in, trim= 30 0 0 0, clip=true]{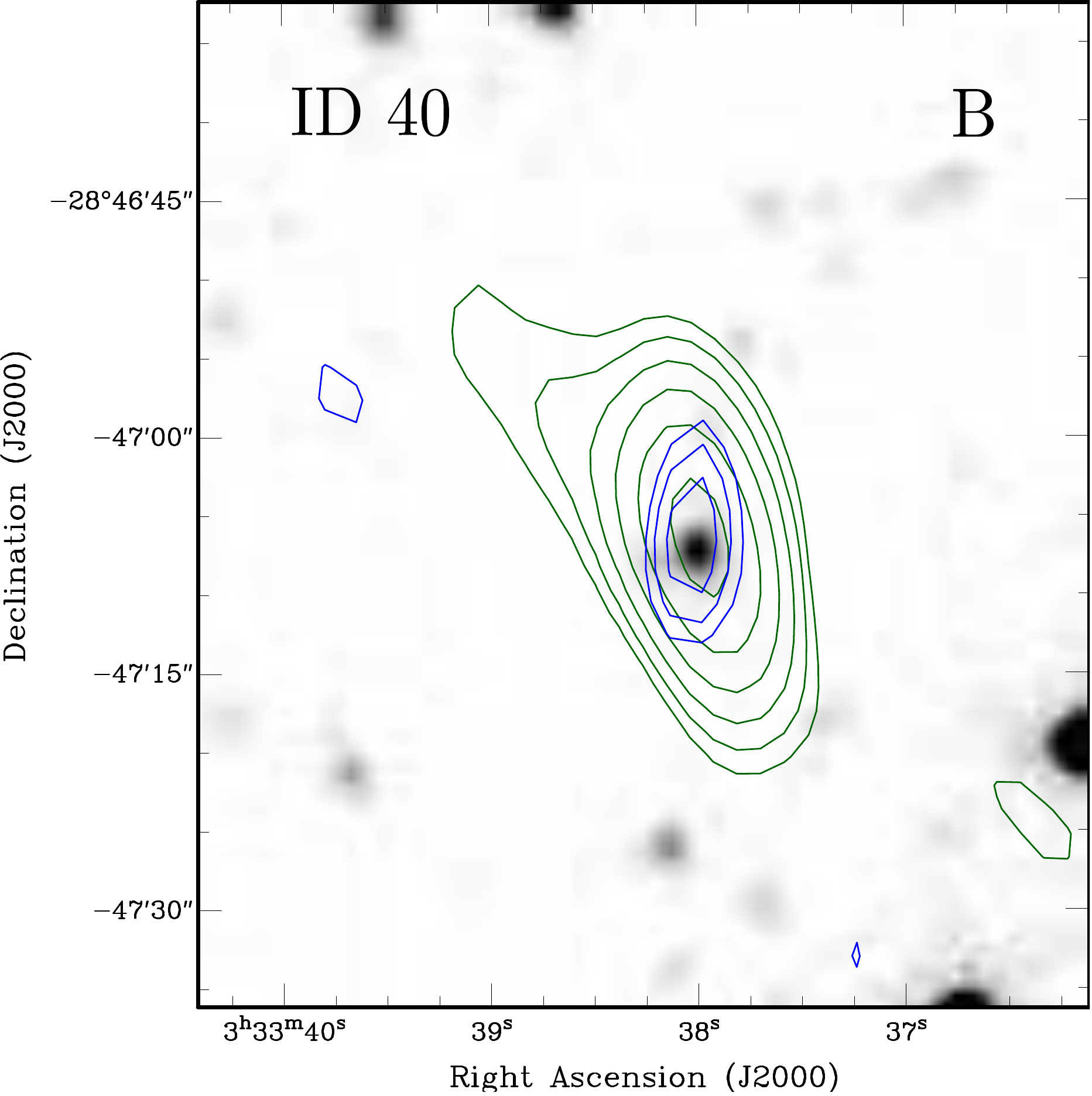}}
\caption{Detected sources in the ATLAS-CDFS field \em -- Continued}
\end{figure*}

\begin{figure*}
\figurenum{2}
\centering
{\includegraphics[width=1.72in, trim= 0 20 0 0, clip=true]{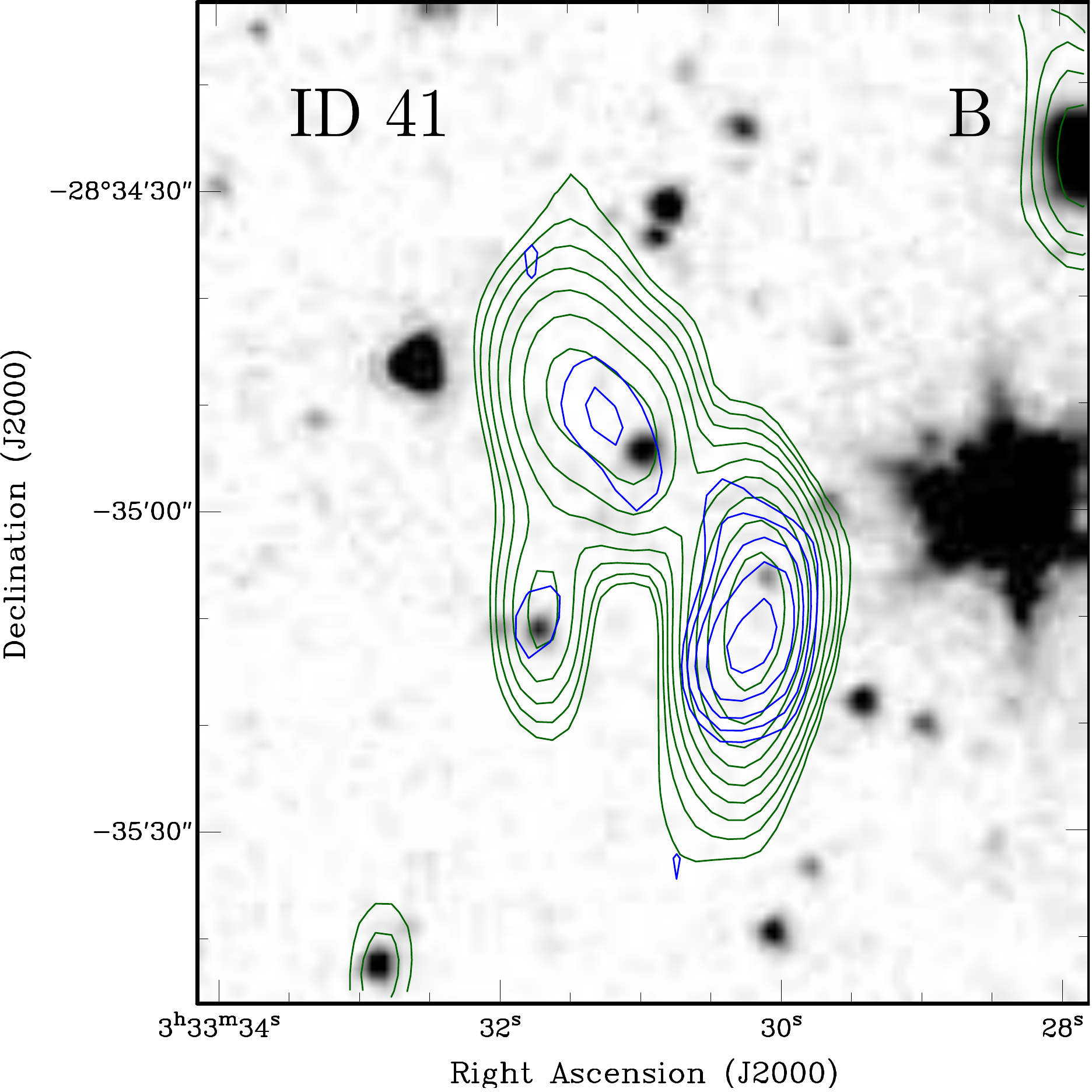}}
{\includegraphics[width=1.624in, trim= 30 20 0 0, clip=true]{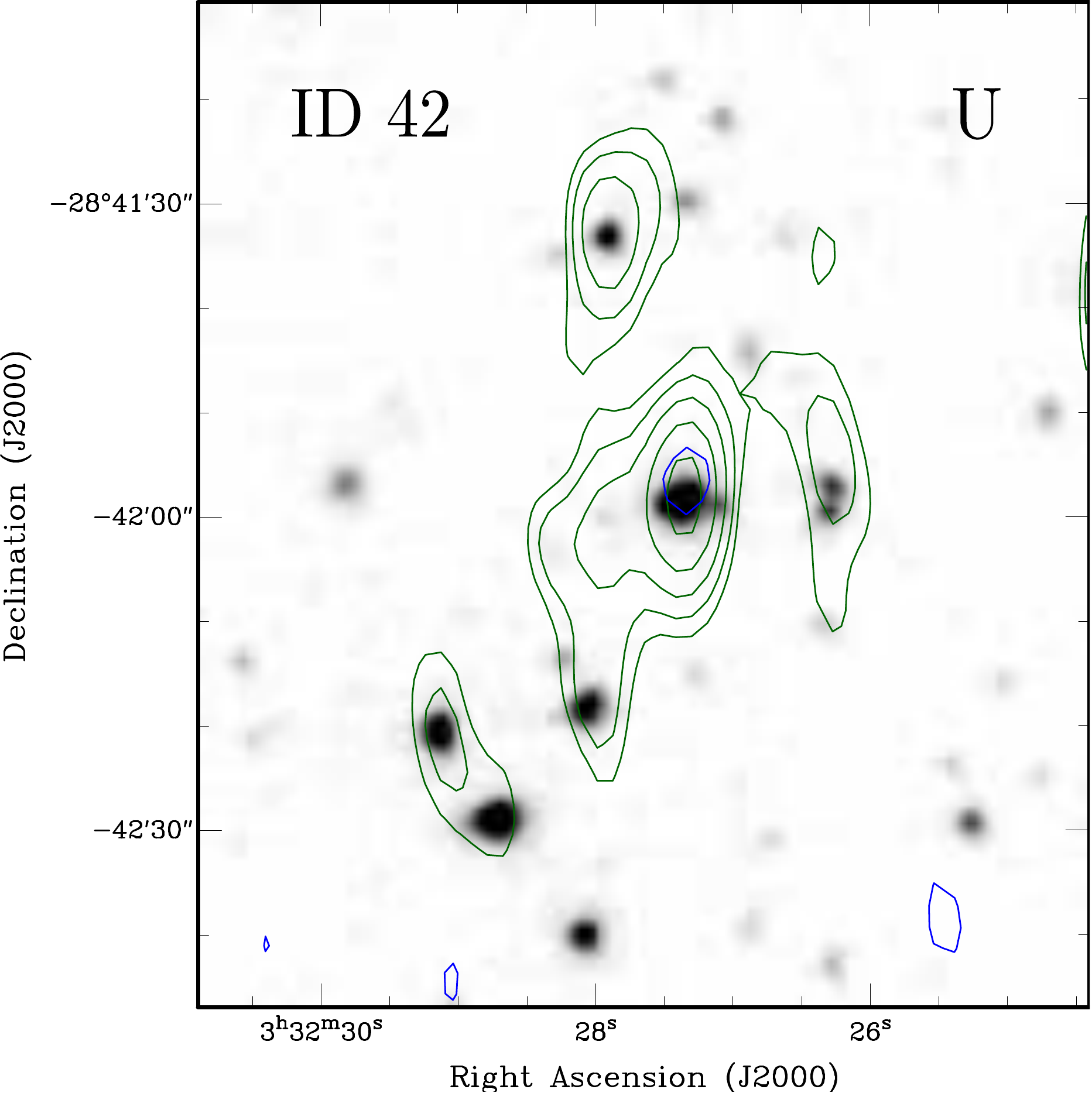}}
{\includegraphics[width=1.624in, trim= 30 20 0 0, clip=true]{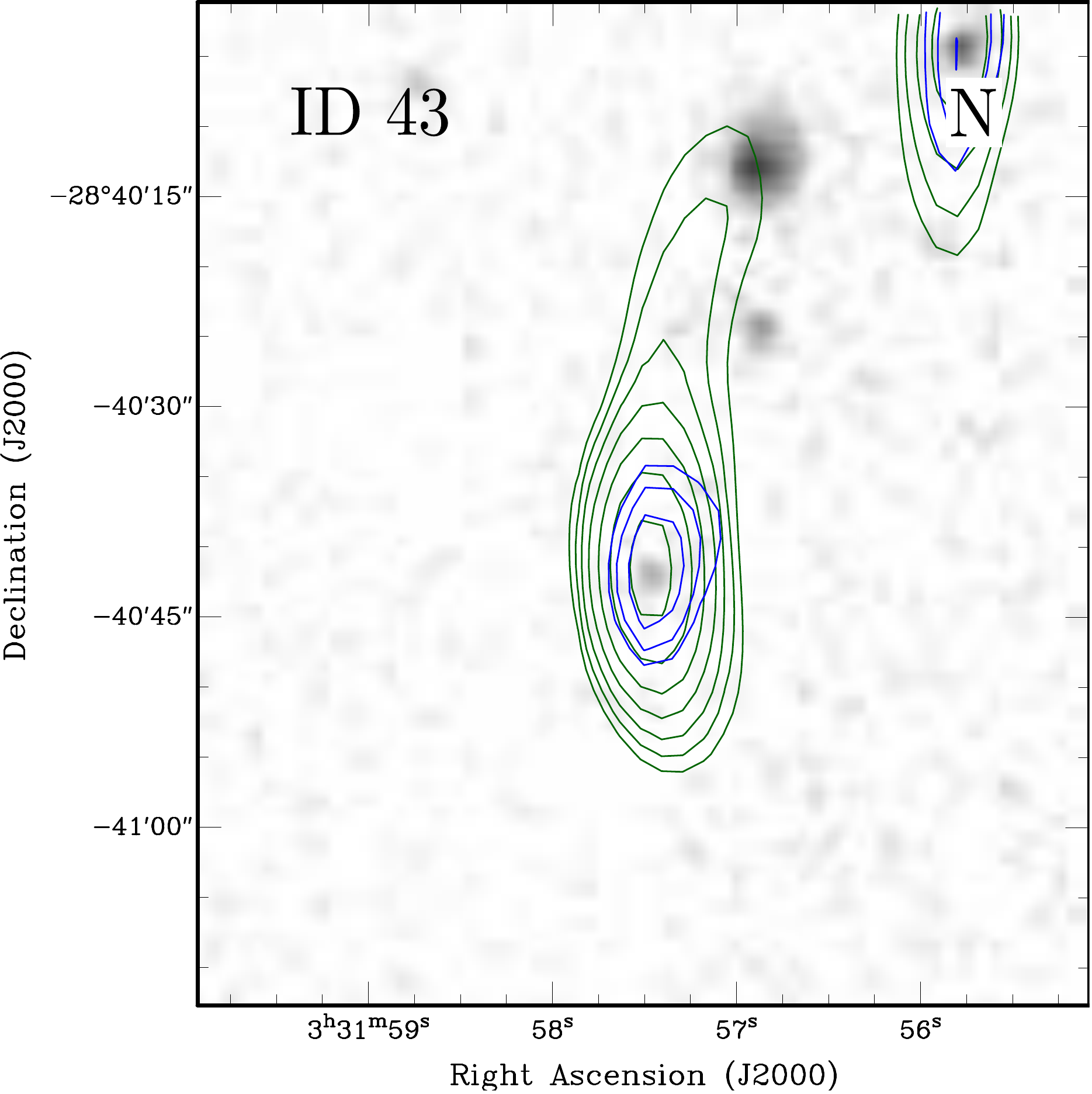}}
{\includegraphics[width=1.624in, trim= 30 20 0 0, clip=true]{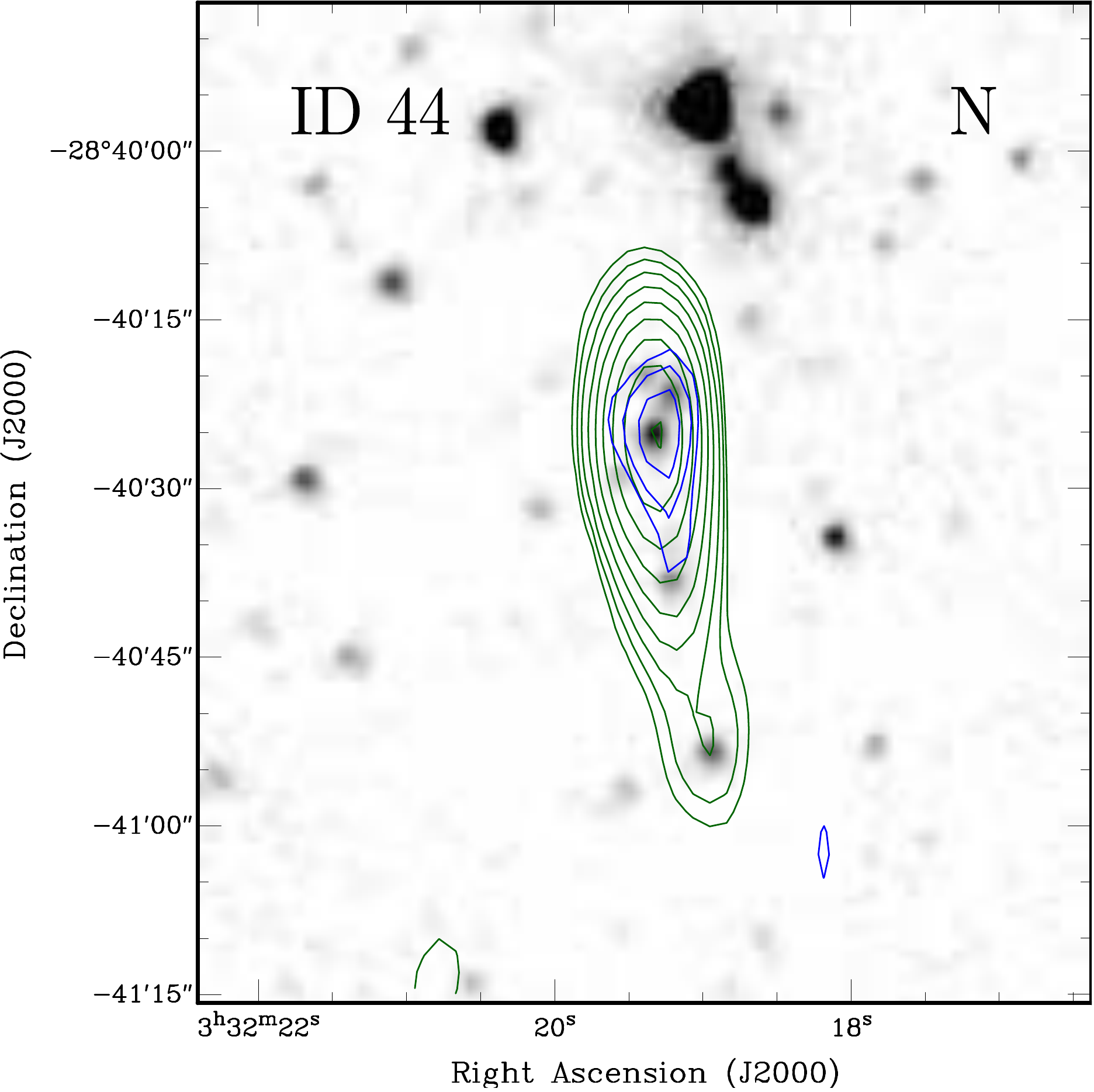}}
{\includegraphics[width=1.72in, trim= 0 20 0 0, clip=true]{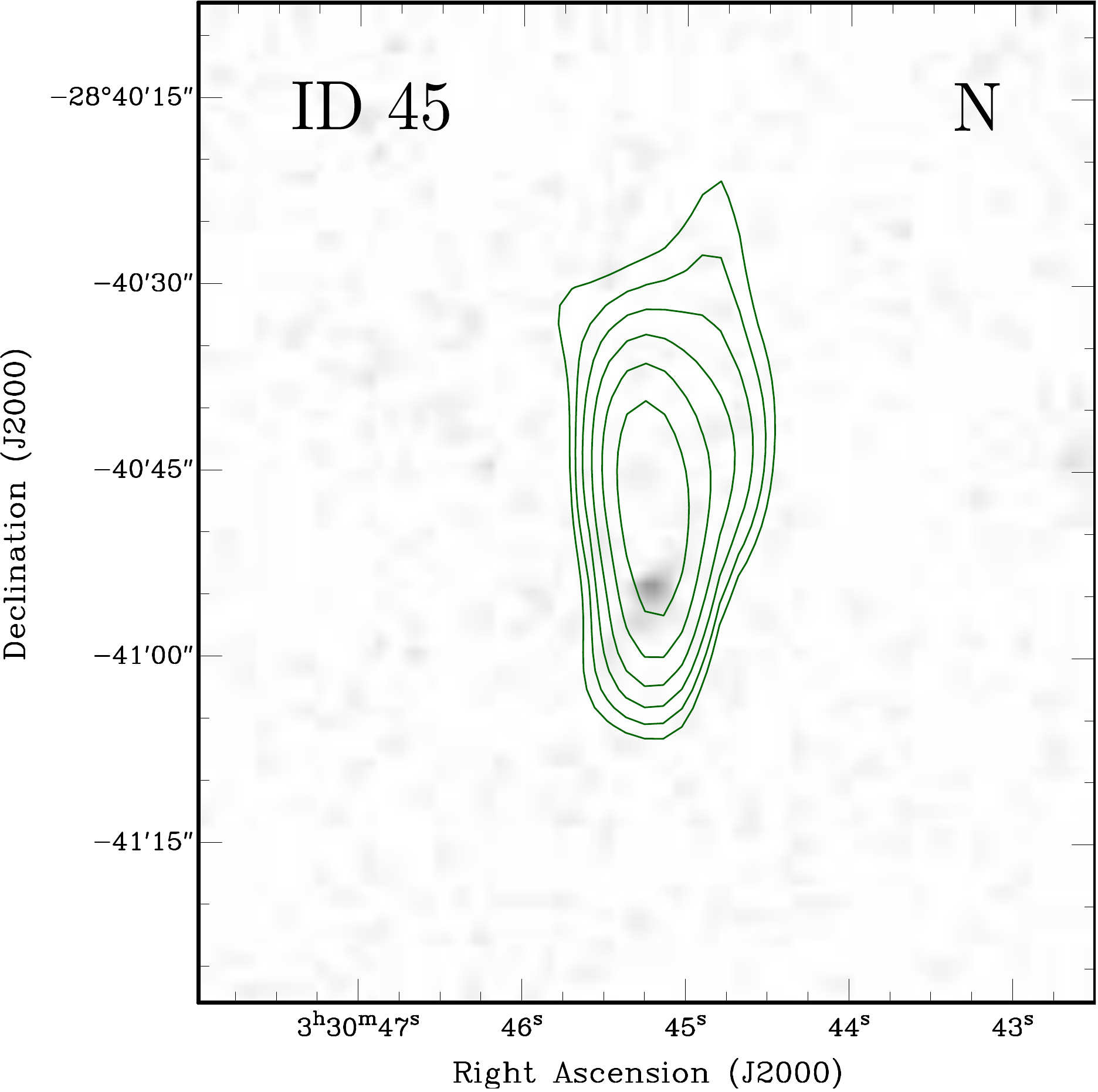}}
{\includegraphics[width=1.624in, trim= 30 20 0 0, clip=true]{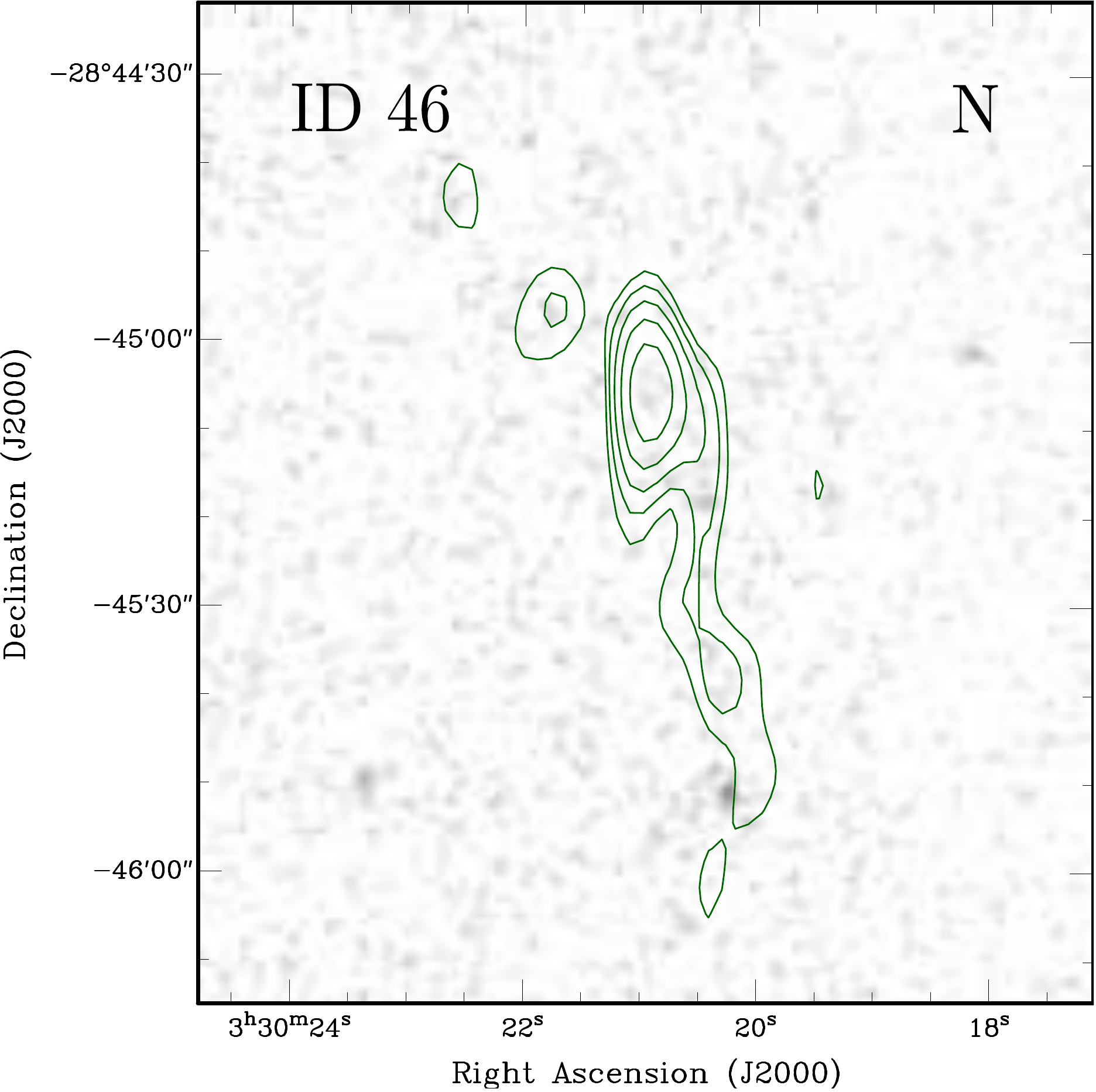}}
{\includegraphics[width=1.624in, trim= 30 20 0 0, clip=true]{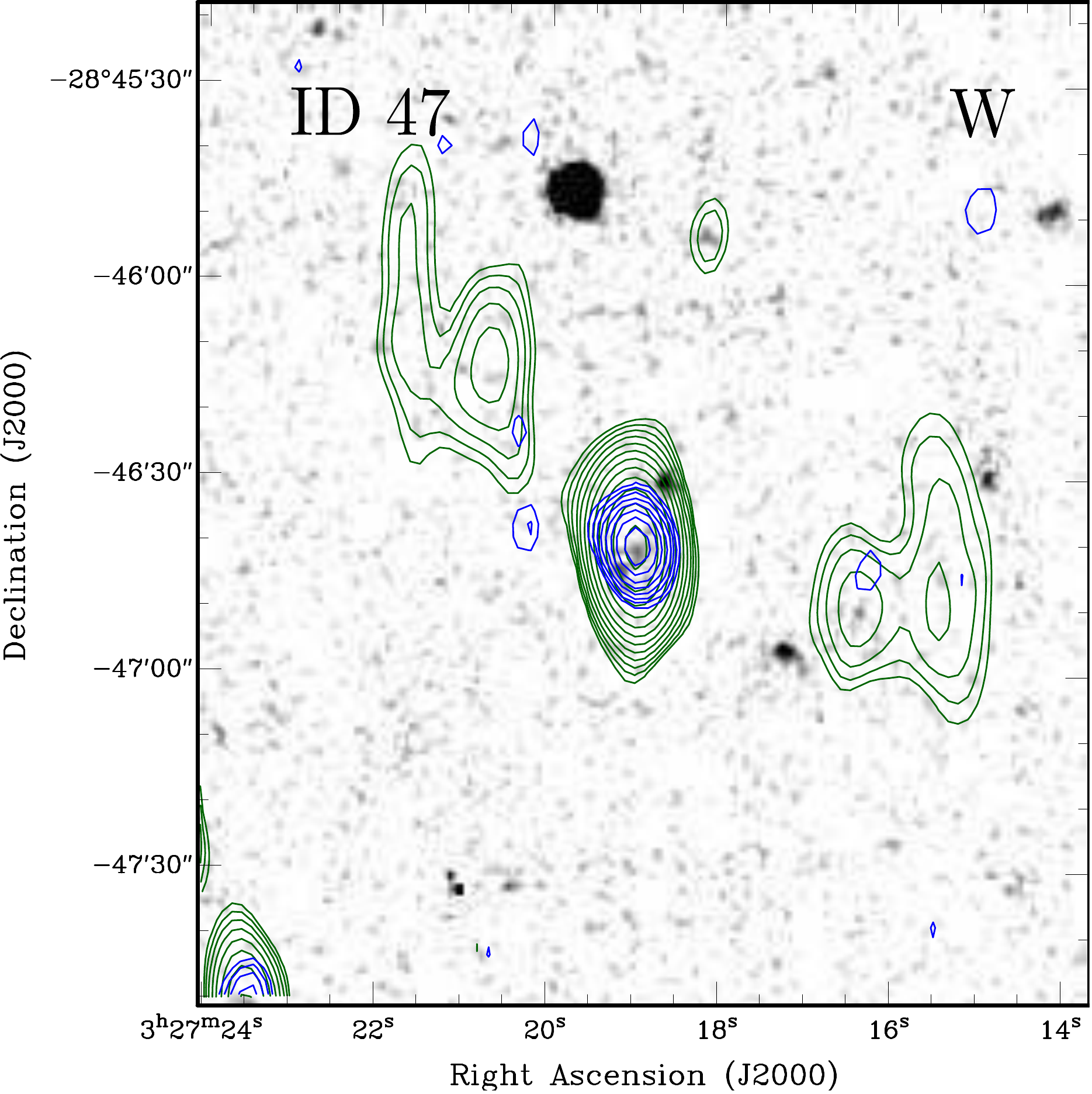}}
{\includegraphics[width=1.624in, trim= 30 20 0 0, clip=true]{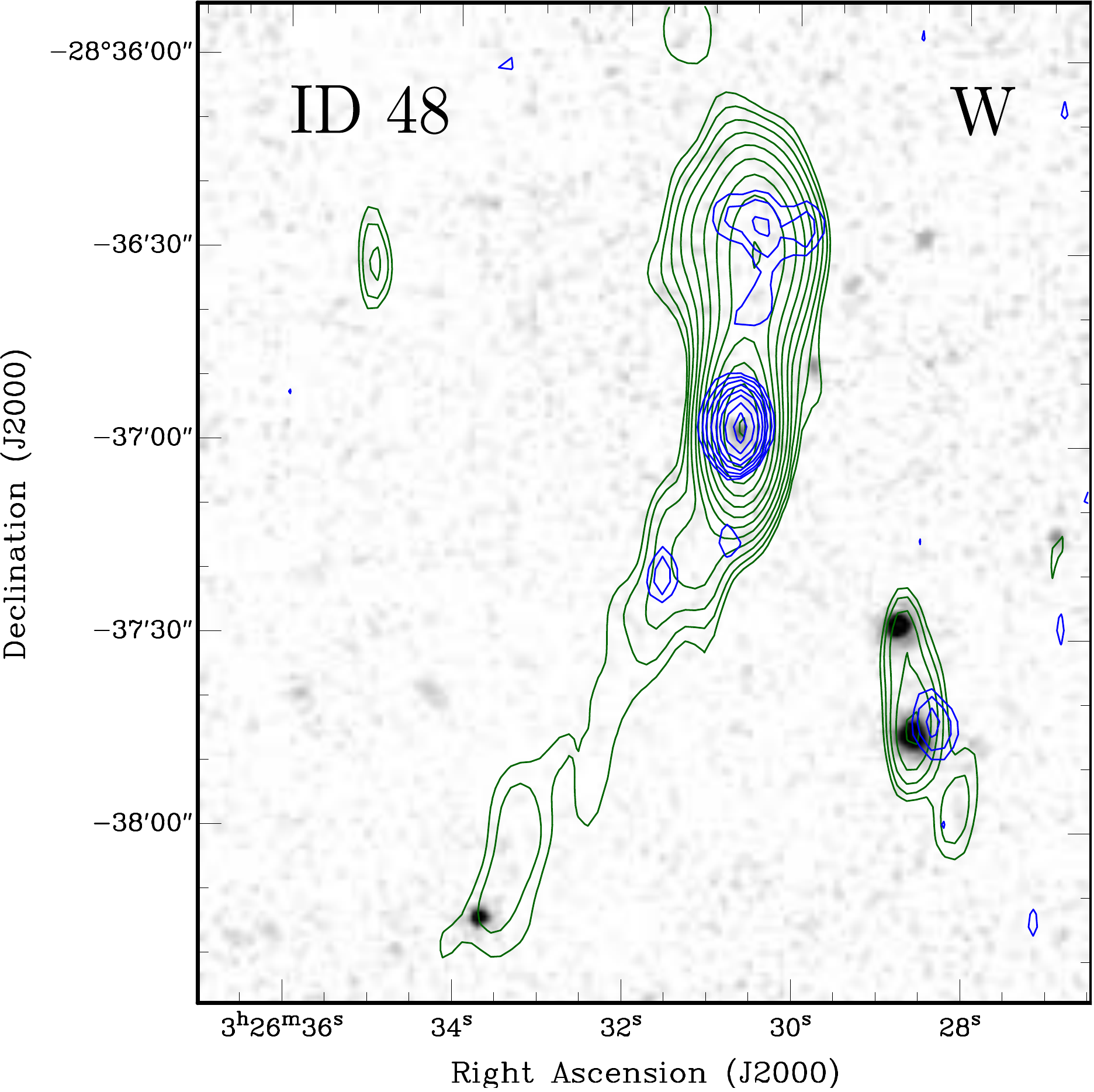}}
{\includegraphics[width=1.72in, trim= 0 20 0 0, clip=true]{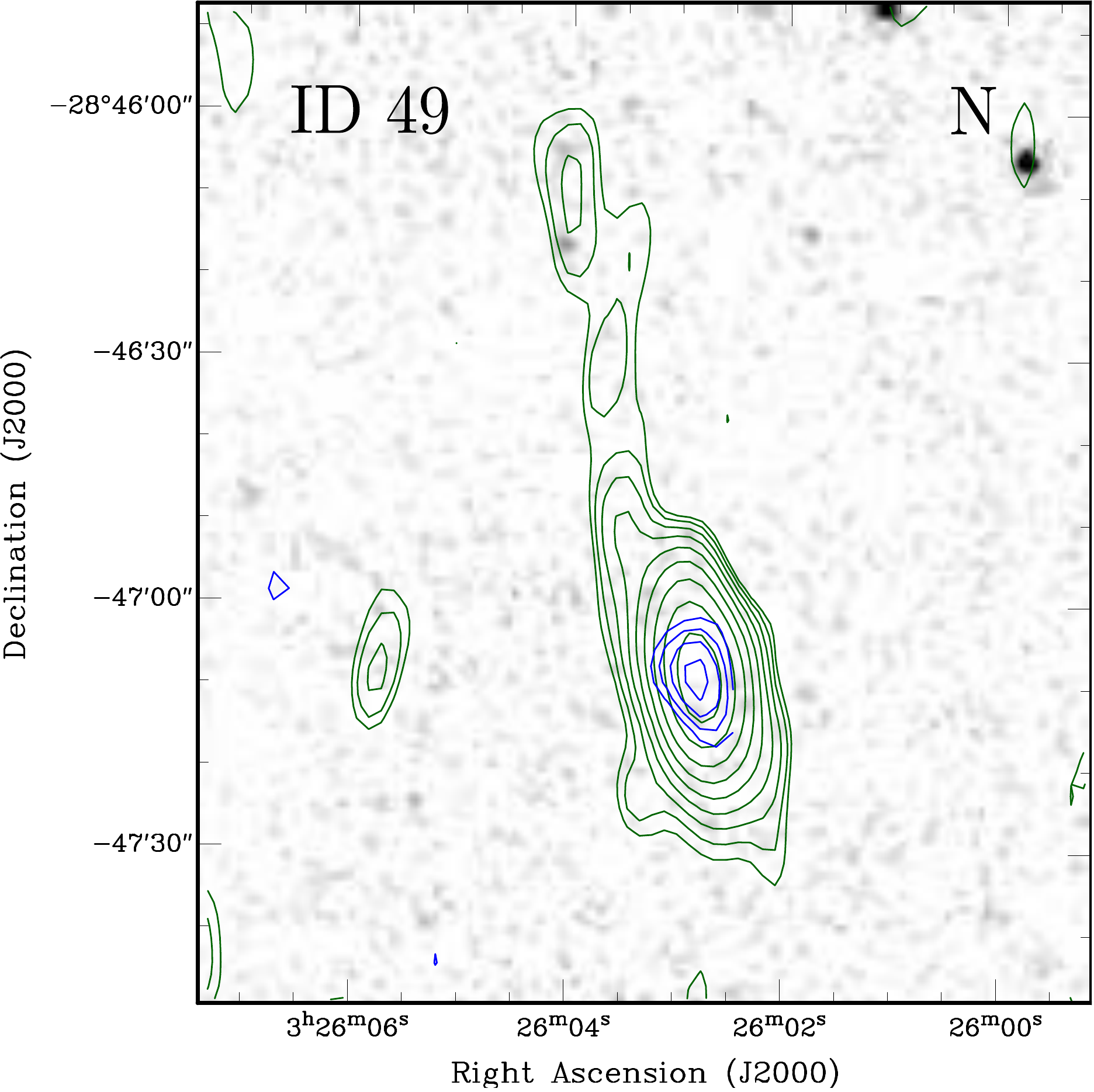}}
{\includegraphics[width=1.624in, trim= 30 20 0 0, clip=true]{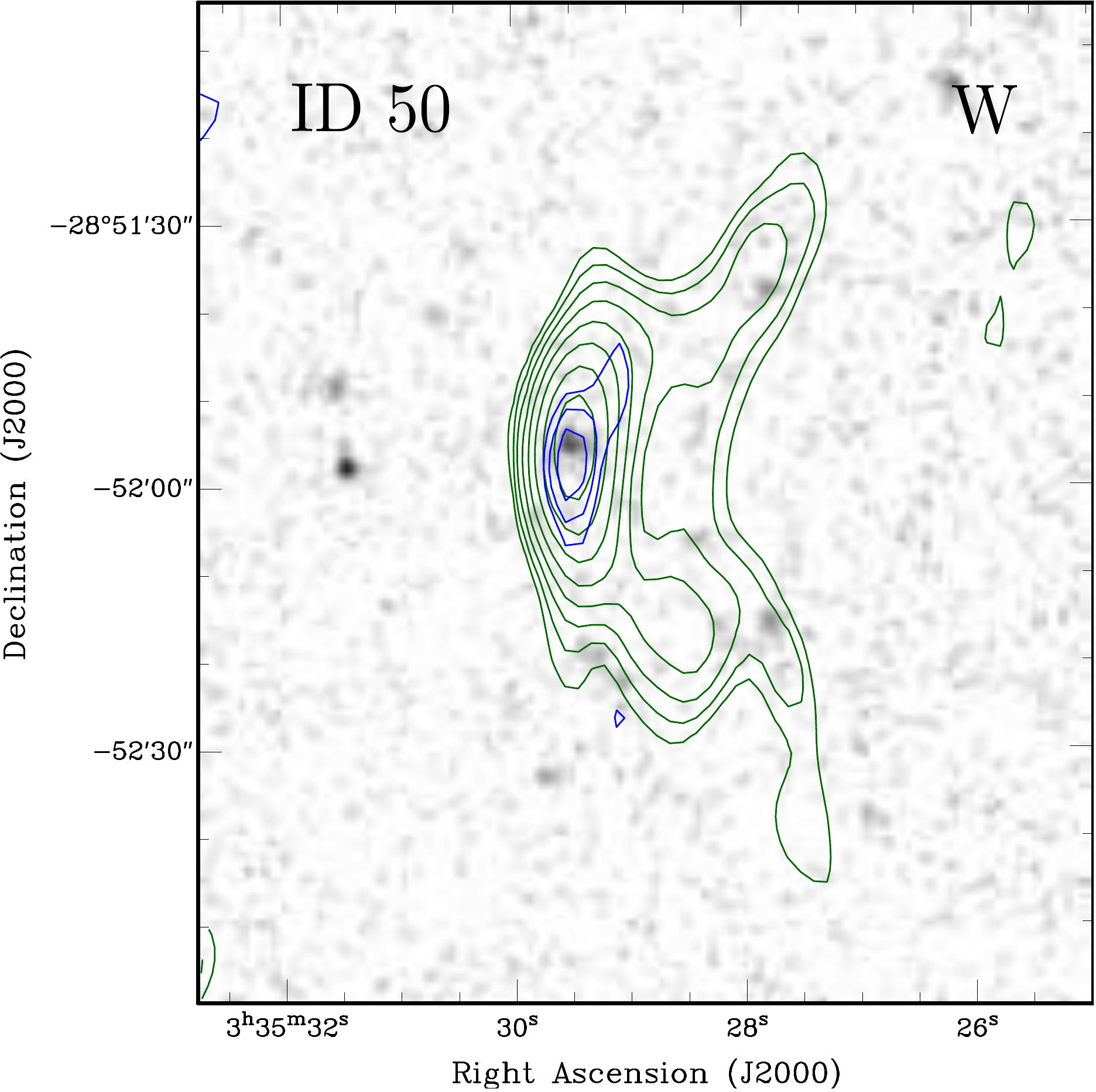}}
{\includegraphics[width=1.624in, trim= 30 20 0 0, clip=true]{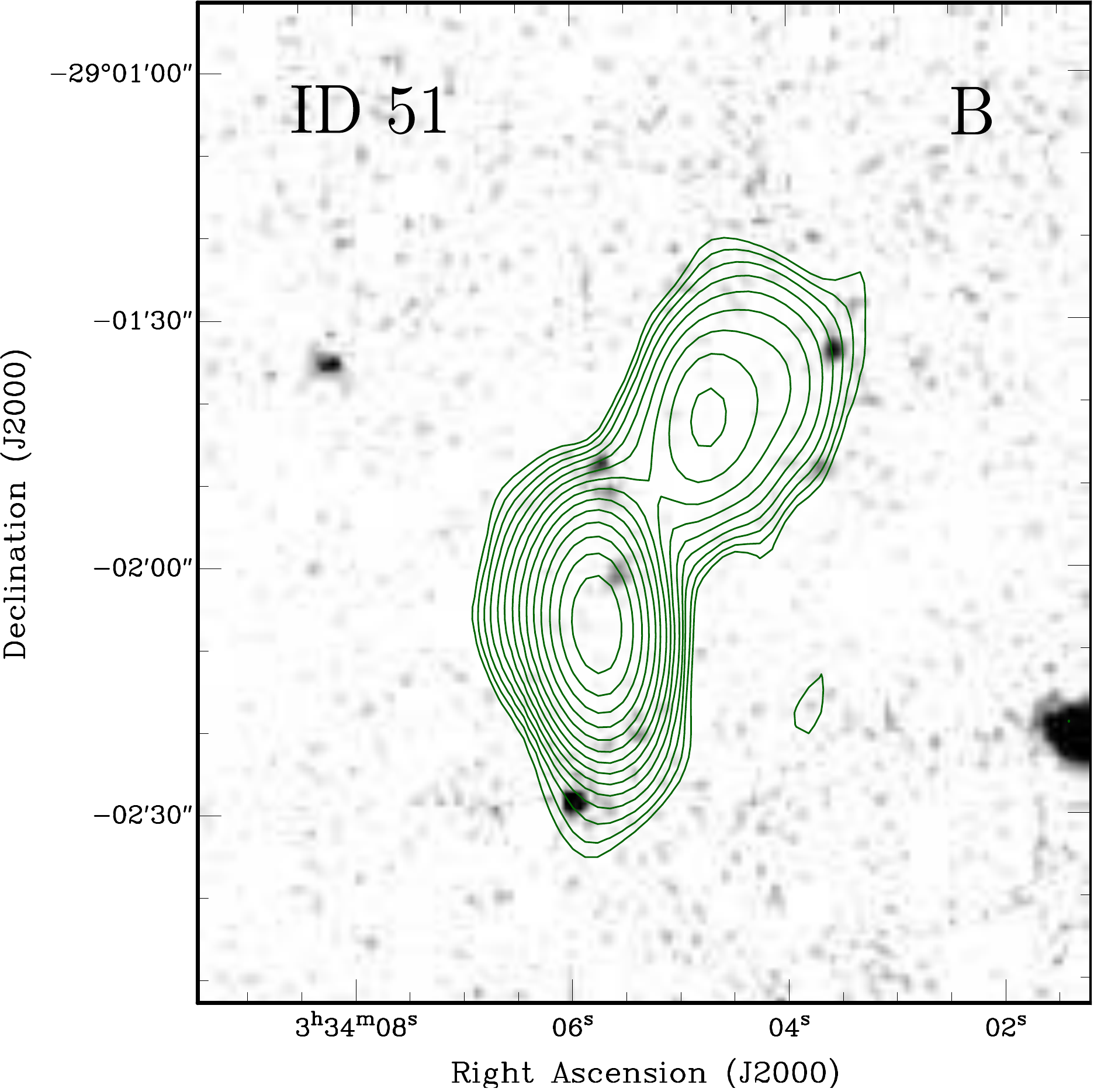}}
{\includegraphics[width=1.624in, trim= 30 20 0 0, clip=true]{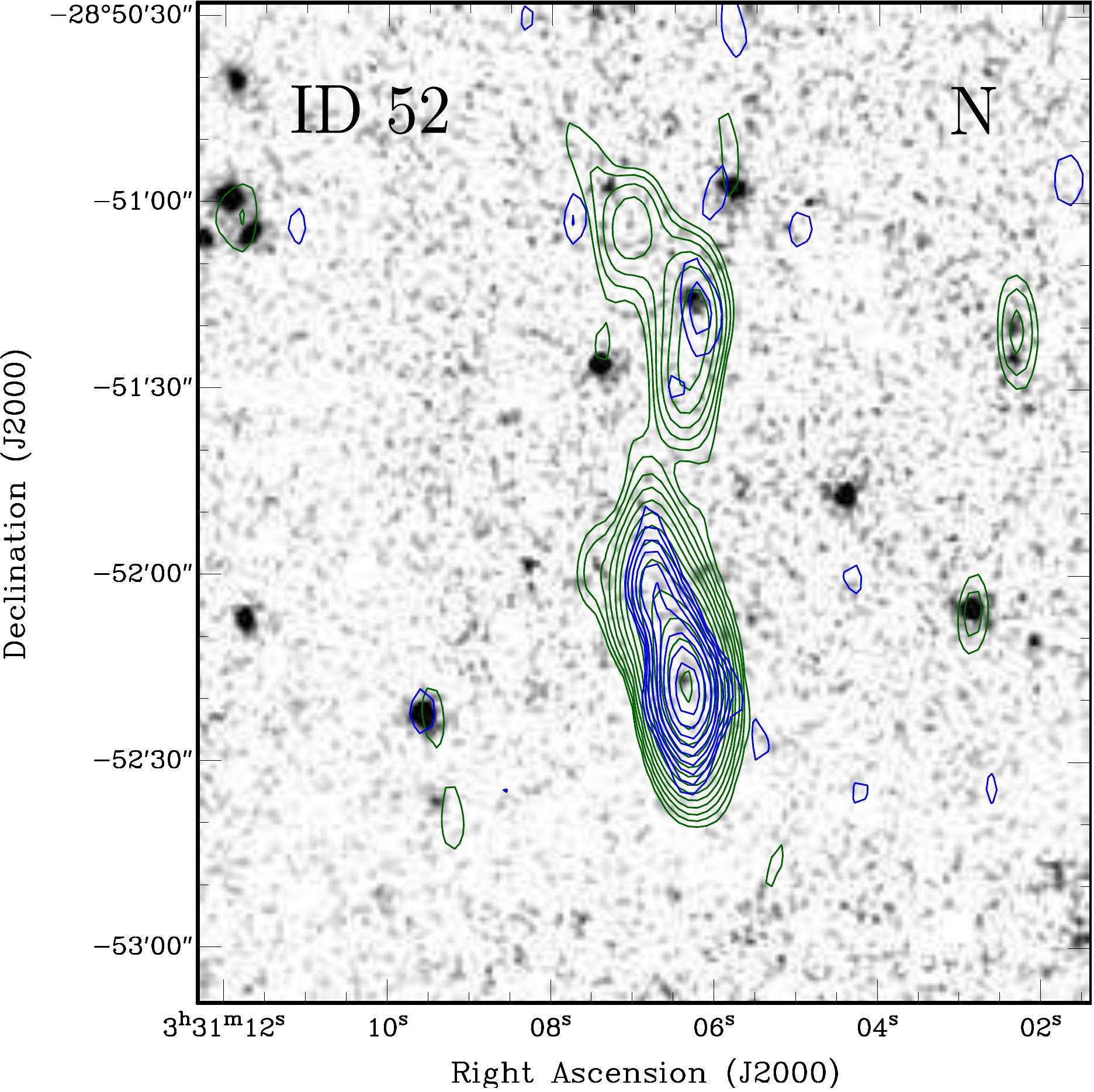}\vspace*{.02cm}}
{\includegraphics[width=1.72in, trim= 0 20 0 0, clip=true]{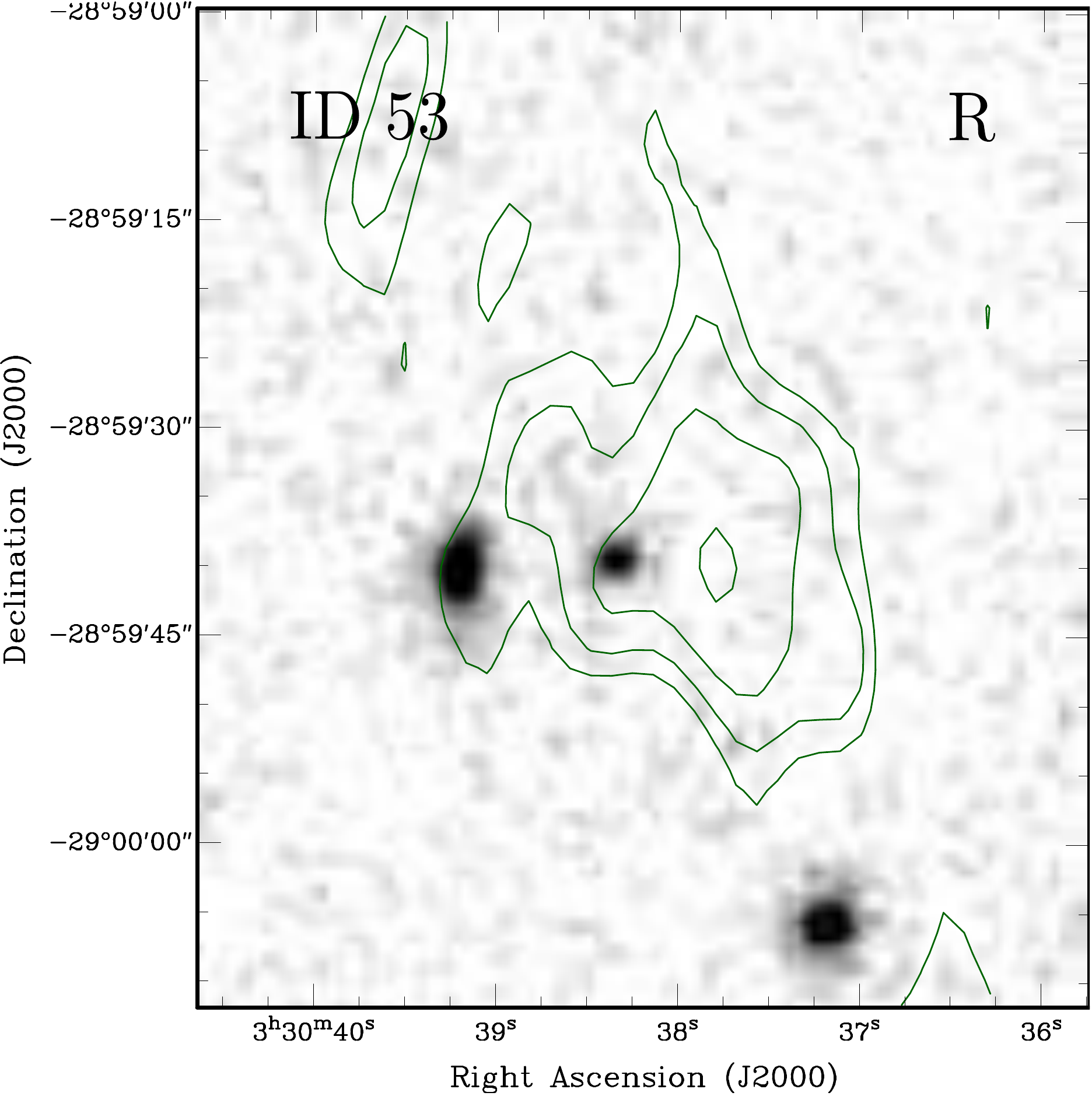}}
{\includegraphics[width=1.624in, trim= 30 20 0 0, clip=true]{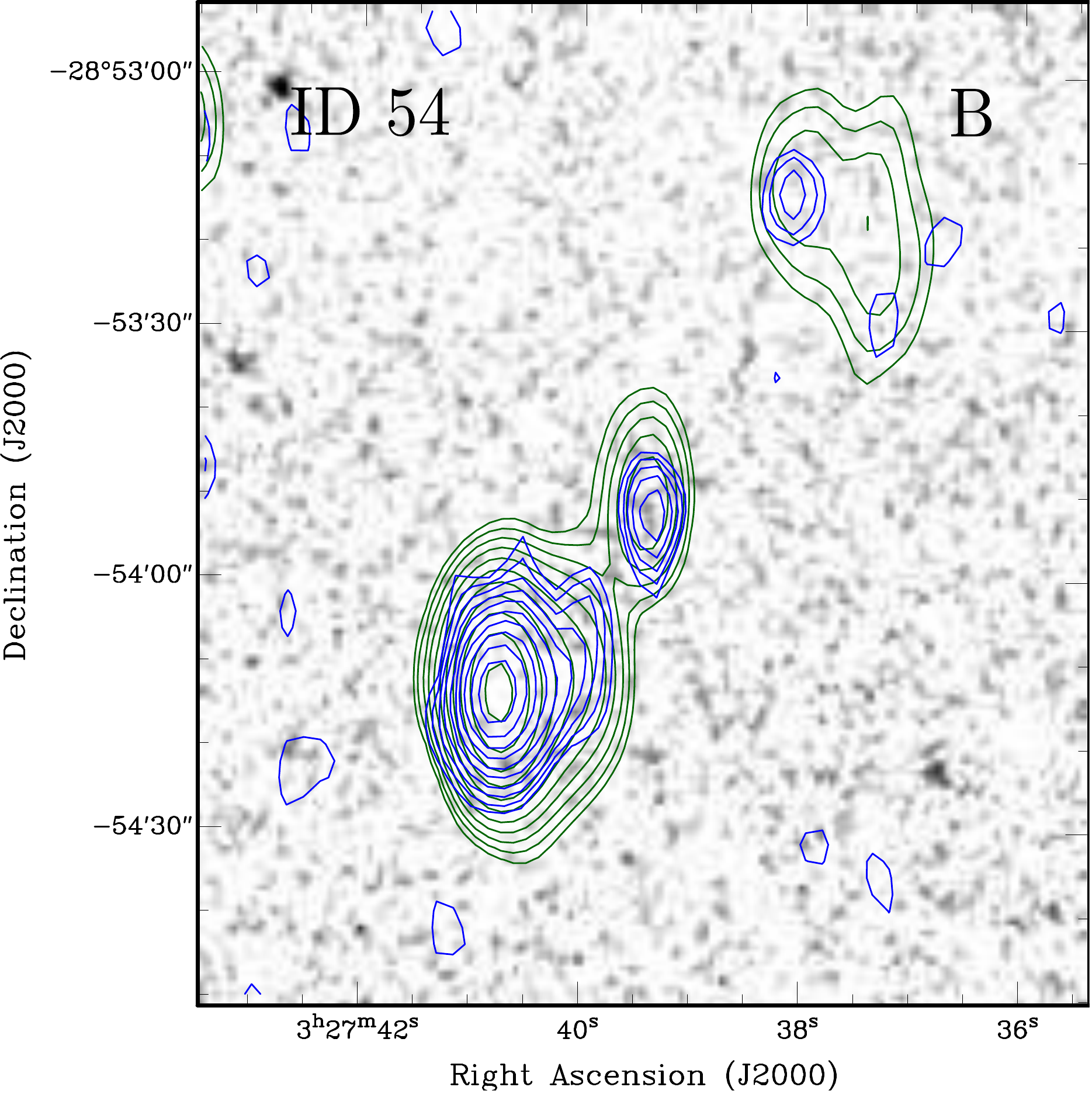}}
{\includegraphics[width=1.624in, trim= 30 20 0 0, clip=true]{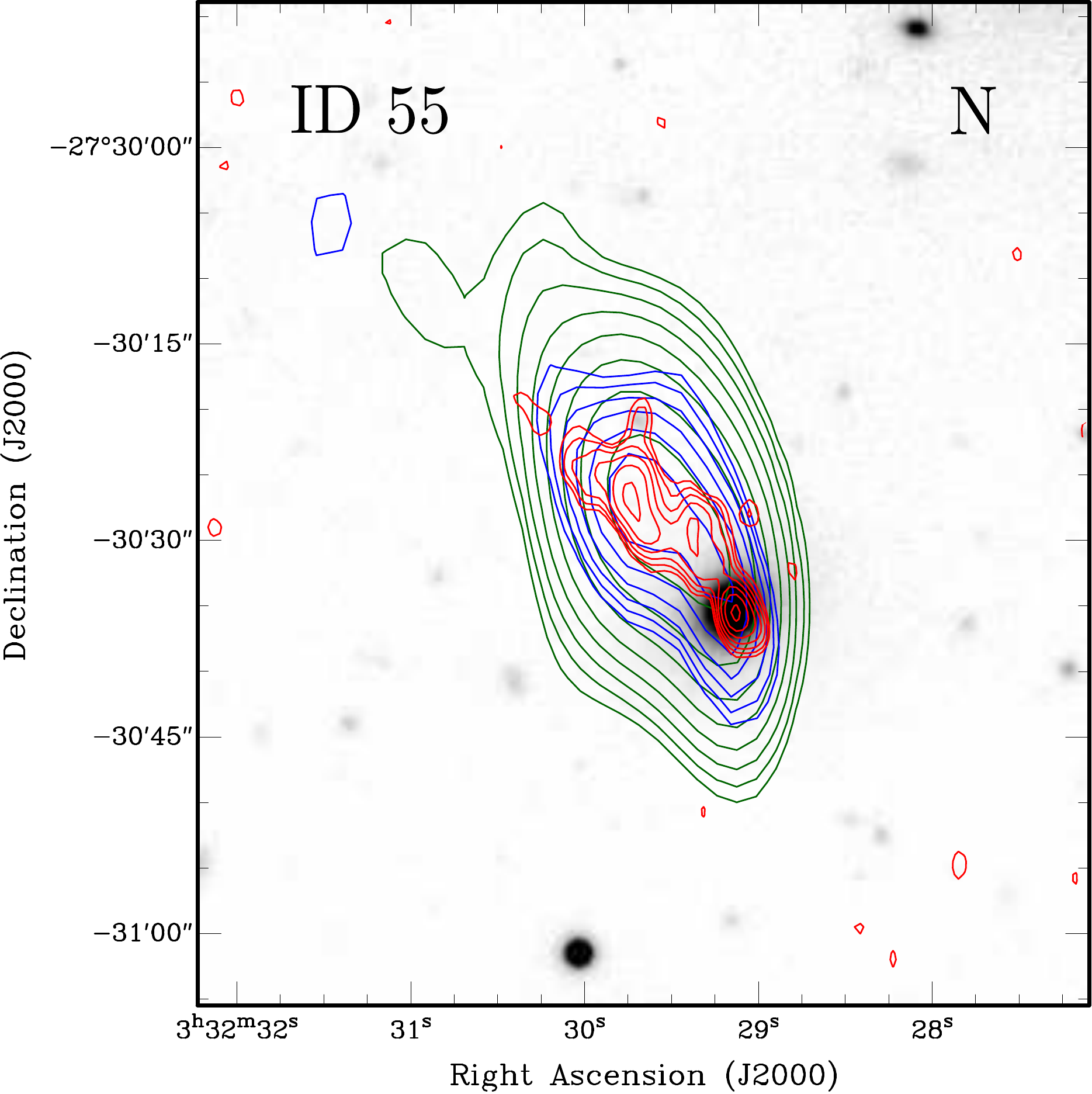}}
{\includegraphics[width=1.624in, trim= 30 20 0 0, clip=true]{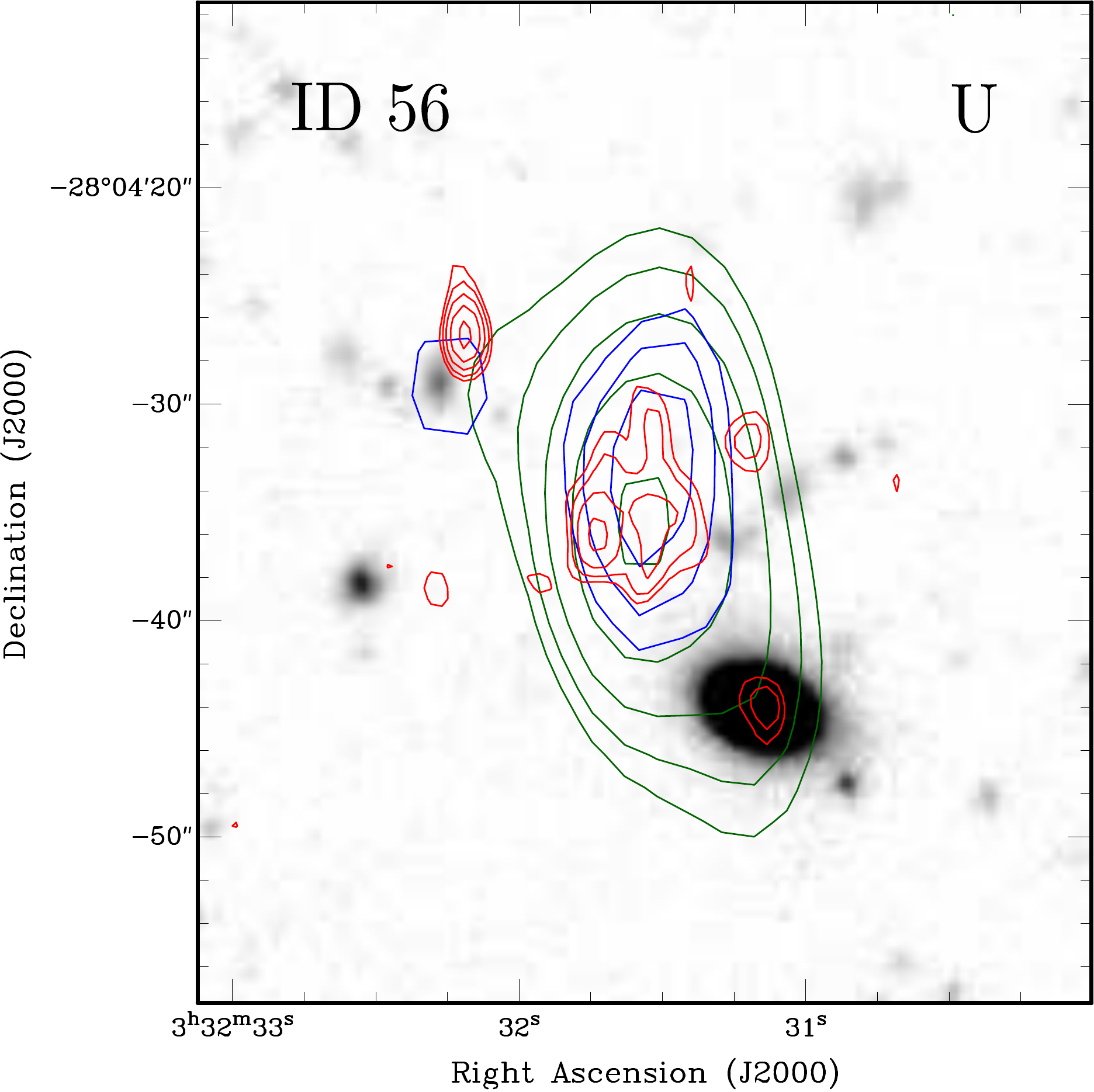}}

\caption{Detected sources in the ATLAS-CDFS field \em -- Continued}
\end{figure*}

Here we present a catalog of 56 extended radio sources detected by examining the 4 deg$^2$ area of the Australia Telescope Large Area Survey (ATLAS) associated with Chandra Deep Field-South (CDFS). The radio image has a resolution of about 17\as $\times 7^{\prime\prime}$ \citep{naa06,f14}, making it similar to FIRST, but reaches a depth of 15 $\mu$Jy, allowing the exploration of a significantly fainter population. The sample includes 45 BT radio galaxies, of which 12 were previously identified by \citet{mss10} and \citet{djm11}. The ATLAS data were supplemented over part of the field covering the Extended CDFS (ECDFS) with $2^{\prime\prime}$ resolution Very Large Array (VLA) data \citetext{second data release, \citealp{mbf13}}. The CDFS field is located in the South Galactic Hole, and a large number of deep multi-wavelength observations cover this field, including optical, near infrared, and Chandra/XMM-Newton X-ray observations. These multi-wavelength data along with this sample of extended radio sources provide a unique opportunity for future studies on the evolution and morphology of BT radio galaxies and other extragalactic extended radio sources, within a wide range of redshifts.

The paper is laid out as follows: section 2 discusses the radio, optical, and spectroscopic data used in this work. In Section 3, we present the results, including the detected sources and their properties. We summarize our work and discuss implications for future surveys in Section 4. Throughout this paper, we assumed a standard Lambda-CDM model, with $H_{0}=71$ \kms Mpc$^{-1}$, $\Omega_{m}=0.27$, and $\Omega_{\Lambda}=0.73$.

\section{Data}

Radio data were provided by the first and third ATLAS data release, with about 17\as $\times 7^{\prime\prime}$ resolution at 1.4GHz over the 4 deg$^2$ field surrounding the ECDFS \citep{naa06,f14}. The rms noise in the first release ATLAS image varies from $\sim$ 20 $\mu$Jy in the low noise regions to $\sim$ 1 mJy near the edges of the field, whereas the rms level in the third release is considerably improved ($\sim$ 15 $\mu$Jy), allowing 42 new BT detections in the field as compared to \citet{djm11}, who only assessed the first data release. Additionally, we used 2 arcsecond resolution VLA data with depth of 10 $\mu$Jy/beam over the ECDFS, which covers a small portion of the ATLAS field with $\sim$ 0.3 deg$^2$ area \citep{mbf13}. This was supplemented with the extended VLA image, which covers the regions outside the final VLA-ECDFS field.

The ATLAS field was chosen to cover the \textit{Spitzer} Wide-Area Infrared Extragalactic \citetext{SWIRE, \citealp{lsr03}} regions, in order to provide the optical identification \citep{naa06}. In addition to SWIRE survey data, we used the $495 - 584$ nm image of the Deep2c field of the Garching-Bonn Deep Survey \citetext{GaBoDS, WFI Data Release: Version 1.0, \citealp{hed06,esd05}}\footnote{Based on data obtained from the ESO Science Archive Facility under request number SDEHGHAN173380.}.

Spectroscopic redshifts of all radio objects of interest were extracted from the following sources: the ESO Nearby Abell Cluster Survey \citetext{ENACS, \citealp{kmd98}}, the 2dF Galaxy Redshift Survey \citetext{2dFGRS, \citealp{cpj03}}, the Balloon-borne Large Aperture Submillimeter Telescope \citetext{BLAST, \citealp{ecd11}} survey, and \citet{msn12}. In addition, where spectroscopic redshifts were not available in the literature, we used photometric redshifts extracted from the Multiwavelength Survey by Yale-Chile \citetext{MUSYC, \citealp{cdm10}} and the Classifying Objects by Medium-Band Observations in 17 filters \citetext{COMBO-17, \citealp{wmk04}}.

\section{Results}

There are $\sim$ 3079 radio sources above the $5\sigma$ level in the ATLAS-CDFS catalog of detected sources, including both Active Galactic Nuclei (AGN) and star forming galaxies \citep{f14,b14}. We detected 56 extended radio sources by visually inspecting over $\sim$ 4 deg$^{2}$ area of the field.

Our criteria for detection included:

\begin{enumerate*}
\item Sources with an extent greater than 1.5 times the synthesized beam.
\item Sources either with any misalignment throughout the radio structure and the host AGN, or with non-symmetrical radio structures with respect to the host galaxy.
\item Additionally, in order to include possible radio halo and relic candidates, we included all the low-surface-brightness radio sources without distinctive radio lobes or optical counterparts in our catalog.
\end{enumerate*}

The detected sources include 45 BTs and a radio relic candidate. A further 9 sources could not be unambiguously classified, and have a variety of possible interpretations including two possible radio halos. The distribution of the detected sources in the field are shown in Figure \ref{fig:map}, along with the ATLAS 1$^{\textrm{st}}$ \& 3$^{\textrm{rd}}$ data release and the VLA-ECDFS areas. VLA postage stamps for four sources just outside the published ECDFS were also included (IDs 7, 8, 13, and 14). Figure \ref{fig:bts} shows the radio structure of the detected sources based on the VLA data in red, ATLAS first data release in blue, and the ATLAS third data release in green contours overlaid on the SWIRE or GaBoDS optical images. The detected BTs were matched with optical counterparts and their redshifts were obtained from the literature, where available. We present the properties of the detected sources including the ATLAS ID, name, J2000 coordinates, redshift of the core galaxy, 1.4 GHz flux and rest-frame power, extent, and classification, in Table \ref{tab:prop}. In order to calculate the rest-frame power, we adopted a spectral index, $\alpha$, of -0.8 and employed the following equation\footnote{We use the convention S $\propto \alpha^{\nu}$.}:

\begin{displaymath}
L_{\textrm{1.4 GHz}}=\frac{4\pi D^2_L(z)}{\left(1+z\right)^{1-\alpha}}S_{\textrm{1.4 GHz}}.
\end{displaymath}

\subsection{Source Morphology}
\label{morph}

Sources displayed a variety of morphologies within the BT range. In particular, 9 were classified as WATs, 18 as NATs, and 10 have more complex morphologies. As expected, the radio powers (down to $10^{22}$ \whz) for the sources presented here probe a lower regime than previous works, and the detected redshift range of the sources extends up to z $\sim$ 2.

Here we discuss the morphologies of some of the more intriguing sources:

\textbf{ID 05} or S707 in the ATLAS catalog, is the largest detected source in our sample, with an extent of over 877 kpc. The optical identification of this BT radio galaxy shows a possible close companion at $\sim$ 5\as, which corresponds to a projected distance of $\sim$ 37 kpc at z=0.7406. The radio structure represents roughly a $90^{\circ}$ bend in the southern plume, while the northern plume appears to be featureless aside from two hotspot regions lined up with the host galaxy. Note that the detection of the low-surface-brightness regions toward the southern lobe was only made possible by the significantly improved sensitivity of the third data release of the ATLAS.

\textbf {ID 08} or S409 in the ATLAS catalog, already detected by \citet{mss10}, has a remarkable radio structure with prominent, highly bent radio tails. The morphology of the tails in this HT galaxy may indicate a chaotic and turbulent intra-cluster medium, which is interacting with the radio lobes. The high resolution VLA data represents a similar complex structure of hotspots, kinks, and curvatures within both lobes.

\begin{figure}
\centering
{\hspace*{-.31cm}\includegraphics[width=3.261in, trim= 0 0 0 0, clip=true]{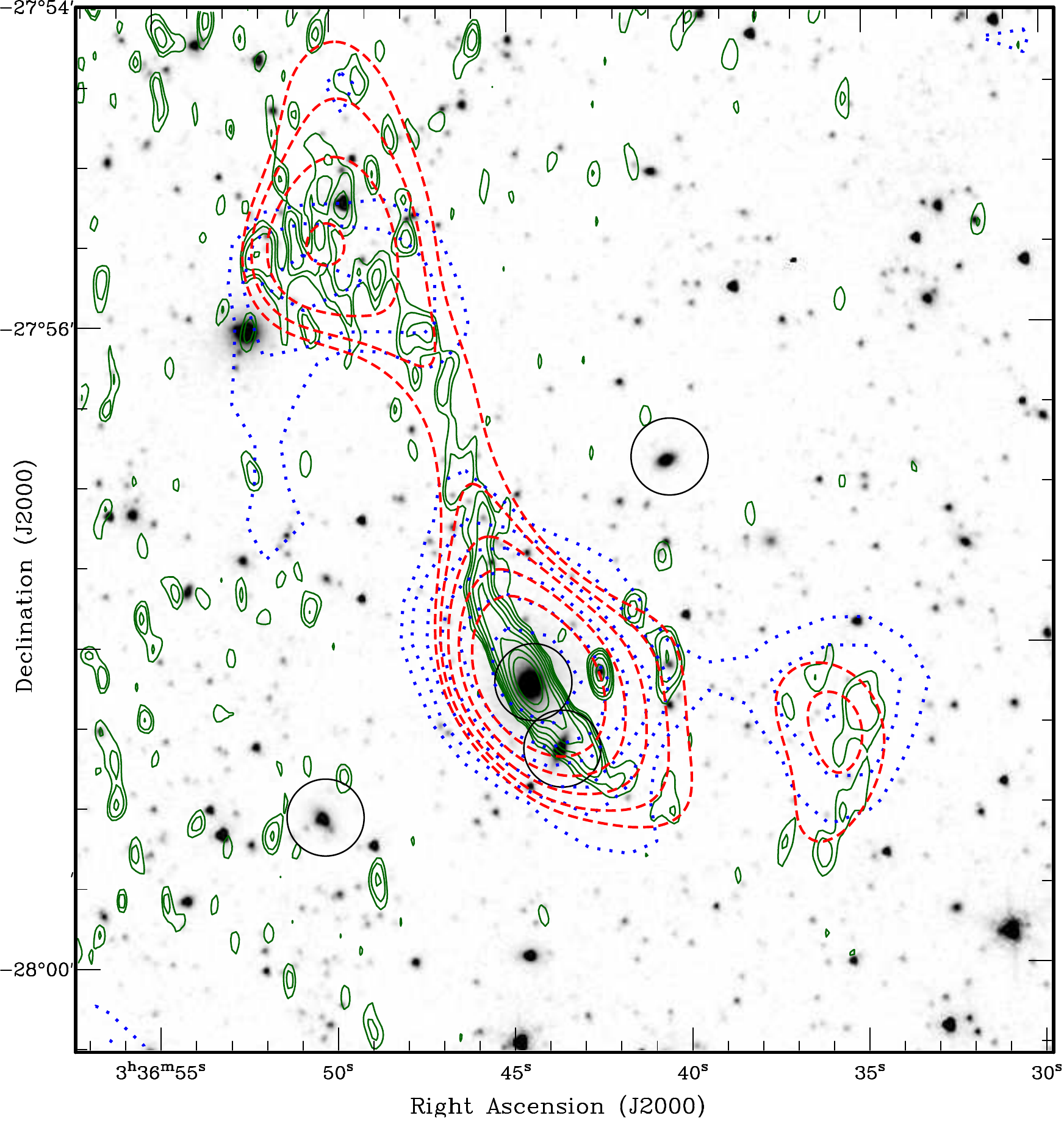}}
{\includegraphics[width=3.5in, trim= 10 0 0 0, clip=true]{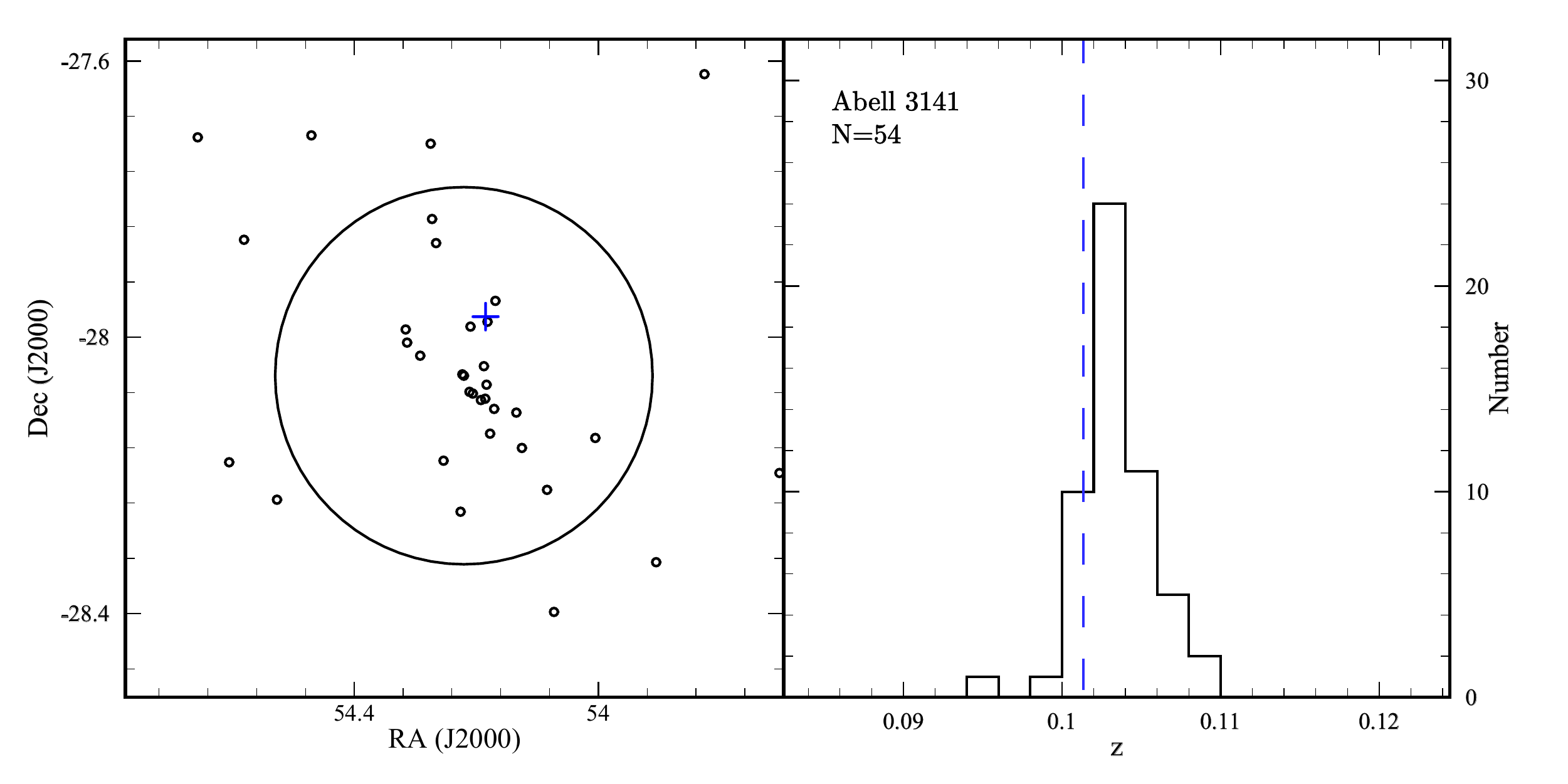}}
\caption{Top panel: the comparative radio structure of ID 11 detected via the ATLAS data and the NVSS data. The radio contours based on the ATLAS, convolved ATLAS, and VLA data are shown with the green solid, red dashed, and blue dotted contours, respectively. The black circles represent the galaxy members of the Abell 3141 cluster. Bottom left panel: the spatial distribution of galaxies in the galaxy cluster Abell 3141. The approximate extent of Abell 3141 is shown by a circle with the Abell radius ($\sim1.87$ Mpc). The location of ID 11 is shown by a blue cross. Bottom right panel: the redshift distribution of galaxies within $\sim$ 15$^{\prime}$ of ID 11. The blue dashed line represents the redshift location of the BT in the Abell 3141 cluster. The data were extracted from the 2dFGRS \citep{cpj03} and the Southern Abell Redshift Survey \citetext{SARS, \citealp{wqi05}}.}
\label{fig:a3141}
\end{figure}

\textbf {ID 11} with an overall S-shaped radio morphology is a classic BT radio galaxy. The features of the jets and lobes in ID 11 are relatively resolved out, since it is located on the noisy verge of the ATLAS field. In order to verify the detection and morphology of ID 11, we employed the data from the NRAO VLA Sky Survey \citetext{NVSS, \citealp{ccg98}}. For a better comparison between the ATLAS and NVSS images we convolved the ATLAS image with a Gaussian with full width at half maximum similar to the resolution of the NVSS image (45\as). The top panel of Figure \ref{fig:a3141} shows the radio structure of ID 11 detected via the ATLAS data with green, NVSS data with blue, and convolved ATLAS data with red contours.

The source, with z=0.1013, is accompanied by a satellite galaxy at z=0.1009 corresponding to a projected distance of $\sim$ 50 kpc. The binary system lies within the Abell radius of the galaxy cluster Abell 3141 which has a mean redshift of 0.1058 \citep{sr99}. The off-center position of ID 11 in both the spatial and redshift distributions of the cluster, along with the disturbed and elongated spatial distribution of Abell 3141 may be indicative of a possible merger process in the cluster (see the bottom panel of Figure \ref{fig:a3141}). However, further spectroscopic data is required for a decent structure analysis. It is worthwhile to note that the S-shape structure of the lobes in ID 11 may be the result of the host galaxy's orbital motion due to its companion along with its relative movement through the ICM. In fact, previous studies of BTs have suggested a close companion is typically present \citep{r82,mjs09} and recent detailed modelling of a similar large BT shows how such a morphology can be produced \citep{pjd13}.

\textbf {ID 12} or S376 in the ATLAS catalog is a BT radio galaxy with about a 90$^\circ$ opening angle (a classical WAT). ID 12 with 2$1^{\prime\prime}$ extension, which corresponds to about 170 kpc at photometric redshift of 1.04, superimposed on a compact population of galaxies in the optical image. Note that the detection of ID 12 was only made possible by employing the $2^{\prime\prime}$ resolution VLA data.

\textbf{ID 13} or S291 in the ATLAS survey is an asymmetric BT with a redshift of 0.3382 and an extent of 130 kpc, which is smaller than the average length of BTs. It appears to be a weak radio galaxy in the early stage of evolution to a BT. ID 13 represents a straight and featureless structure with a single bend towards the tail-end of the eastern jet.

\textbf{ID 21} has a complicated multiple-component radio structure, including four major segments, of which two are associated with optical identifications. Current radio data are insufficient to support a plausible scenario in order to classify ID 21.

\textbf{ID 24} or S447 has a peculiar radio structure consisting of four radio components with an overall 1.4 GHz rest-frame power of 5.98 $\times 10 ^{26}$ \whz. As stated in \citet{mbf13}, the north-eastern component coincides with an optical ID and appears to be a single separate source in the high resolution VLA image. Although ID 24 may not be unambiguously classified as a BT, it appears to consist of a core galaxy with a pair of lobes to the north and south, of which the latter undergoes a diffuse 90$^{\circ}$ bend. ID 24, which was reported the farthest such source ever detected in \citet{djm11}, has a central component identified with an optical counterpart located at z=1.9552. The discovery of ID 24 along with ID 52, at z=2.1688, extends the existence of distorted radio sources as far back as the formation of galaxy clusters, which adds a valuable clue to the mutual evolution hypothesis.

\textbf{ID 32} or S024, as reported in previous works \citep{mss10,djm11}, exhibits a clear emission cut-off between the host galaxy and two radio lobes. In addition to the significant radio emission from the lobes, a faint radio component is evident surrounding the core galaxy. This emission along with a putative companion within the host galaxy's halo, may be indicative of an AGN with recurrent jet activity, considering that close companions are believed to be the principal trigger for the AGN and radio activity \citep{kmv10}.

\textbf{ID 38} includes two significant radio components. Although ID 38 may be classified as a single BT radio galaxy with an optically faint core galaxy located at the midpoint of the two intensity peaks, a more likely scenario is that it consists of two separate sources, considering that the northern peak has an optical counterpart at its center. Furthermore, the southern component with a diffuse structure is superimposed on a compact population of galaxies, making it a conceivable candidate for a radio halo at the center of a presumptive cluster. Deeper radio observations and spectroscopic data are required to clarify whether the southern component of ID 38 is indeed a radio halo.

\begin{figure}
\centering
{\hspace{-.8cm}\includegraphics[width=2.5in, trim= 0 0 0 0, clip=true]{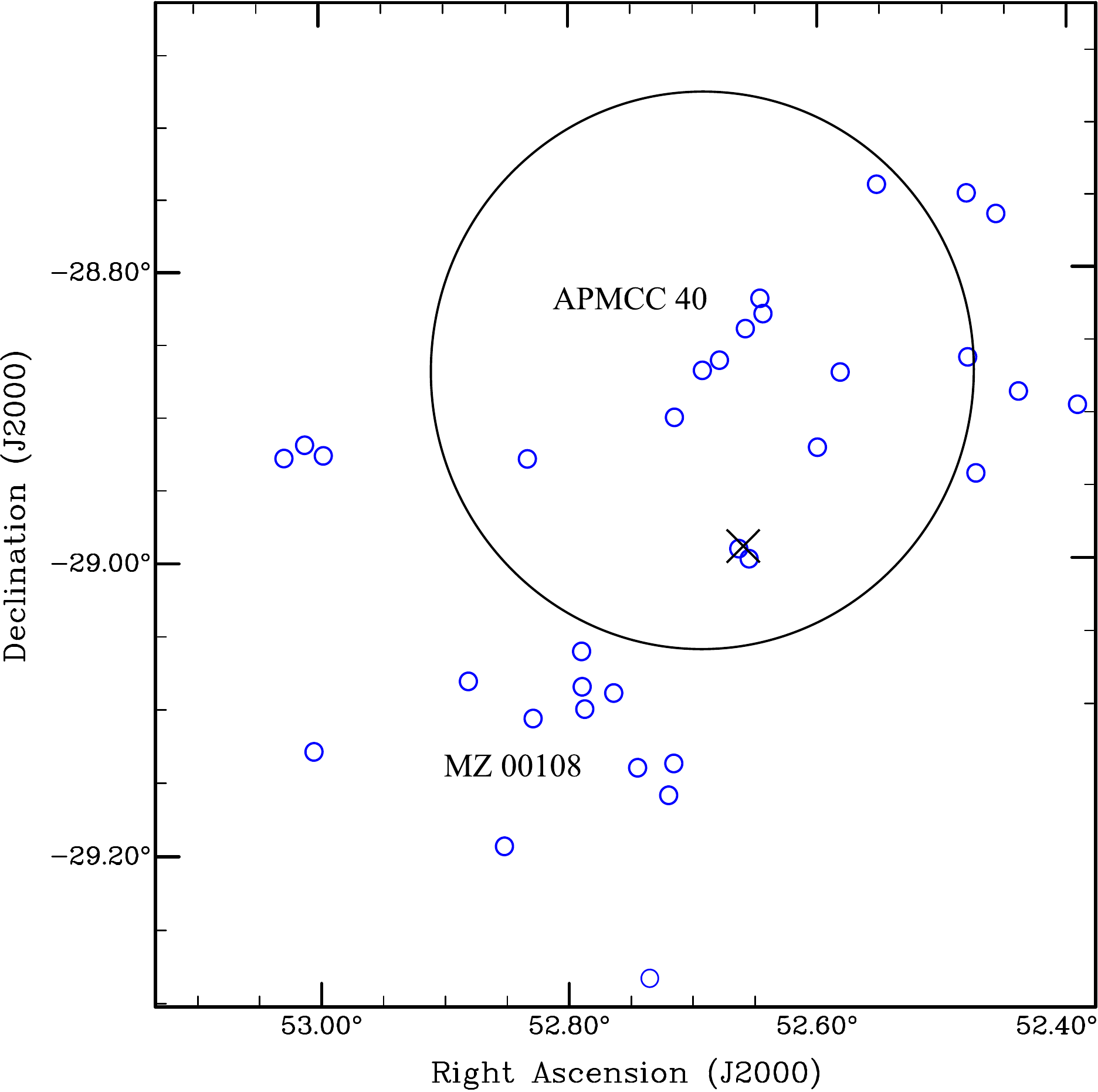}}
\caption{Spatial distribution of galaxies surrounding ID 53, including the galaxy cluster APMCC 40 and galaxy group MZ 00108. Blue circles represent galaxies within redshift range $0.145 \leq z \leq 0.165$. The approximate extent of the APMCC 40 cluster is shown by a circle with the Abell radius. The cross marks the location of ID 53.}
\label{fig:53}
\end{figure}

\textbf{ID 42} consists of four major radio components, each of which coincides with at least one optical counterpart. The central radio component may be ambiguously classified as a BT galaxy with one significant right-angle twist towards the south, where it is concomitant with another galaxy. The remaining radio components may be separate radio galaxies, irrelevant to the central BT source, or in an alternative scenario, all or part of the radio emission may correspond to a radio halo within a presumptive galaxy cluster. Supplementary spectroscopic data along with deep radio observations are required to verify whether a radio halo is subject in this multiple-component radio source.

\textbf{ID 52} or S306 in the ATLAS survey is a BT radio galaxy with an optical counterpart located at z=2.1688, making it the farthest BT galaxy discovered to date. The tail of ID 52 with three hotspot regions extends to $\sim$ 104 arcseconds, which corresponds to over 870 kpc at the host galaxy's redshift. Note that the ID 52 continuous tail is only visible in the third data release of ATLAS given the improved signal to noise ratio. Alternatively, ID 52 may consist of two unique sources, considering that one of the intensity peaks towards the north is associated with an optical identification (MRSS 418-056517). Nevertheless, whether or not the latter scenario is the case, the southern section is still classified as a BT radio galaxy.

\textbf{ID 53} is a very faint radio source with an irregular radio morphology, which is not optically coincident with any galaxy. However, there are two galaxies in the vicinity of ID 53, at z=0.1518 and 0.1529, which are within the Abell radius of the APMCC 401 galaxy cluster with a mean redshift of 0.1500, located at the north-west of the radio source (see Figure \ref{fig:53}). In addition to APMCC 401, the galaxy group MZ 00108, with a mean redshift of 0.1511, is located towards the south-east of ID 53. ID 53 appears to be positioned at a location in the middle of the cluster and group, where possible merger shocks are expected \citep{mrk01}. This overall picture makes ID 53 a plausible candidate for a radio relic. Similar size relics are known in Abell 4038, Abell 85 and Abell 133 \citep{srm01}, but with powers greater than $2.5 \times 10^{24}$ \whz. Further radio observations are required to confirm this possibility, since ID 53 is located at the edge of the ATLAS frame, where the rms value is significantly higher (50 $\mu$Jy). If ID 53 were confirmed as a relic, it would be the faintest yet detected with P$_{\textrm{1.4 GHz}} \simeq 9\times 10^{22}$ \whz.

\textbf{ID 55} or S441 in the ATLAS catalog, is a low power ($\textrm{P}_{1.4 \textrm{ GHz}}= 10^{23}$ \whz) BT galaxy with angular size of 56\as, which corresponds to $\sim$ 142 kpc at z=0.1466. This radio galaxy coincides with a local density peak known as Structure 6 at z=0.141 \citep{dj14}. Structure 6 is located at the north of a large arc-shaped structure, which appears to extend further north, where the HT galaxy ID 08 resides at redshift of 0.1469 (see Figure 2 of \citealp{dj14}). The coincidence of ID 08 \& 55 with a large-scale structure may indicate that HT galaxies are reliable probes of over-dense regions.

\textbf{ID 56} or S446 in the ATLAS catalog, is a diffuse low-surface-brightness radio source located at $85^{\prime\prime}$ south of ID 24 (the distant source S447). ID 56 is not identified with any optical counterpart, thus, reliable classification is not possible. However, assuming that ID 24 is located at the center of a putative cluster at z=1.95, which contains both ID 24 \& 56, ID 56 may be a radio relic at a projected distance of $\sim$ 720 kpc from the center of the hypothetical cluster. Alternatively, the source may be classified as a radio halo, though there is no galaxy population observed in the GaBoDS optical image. The total 1.4 GHz flux of the sources is only 0.38 mJy. Further low frequency observation are required to properly classify ID 56.

\begin{figure*}
\centering
{\hspace*{-0.5cm}\includegraphics[width=3.481in, trim= 0 -70 100 70, clip=true]{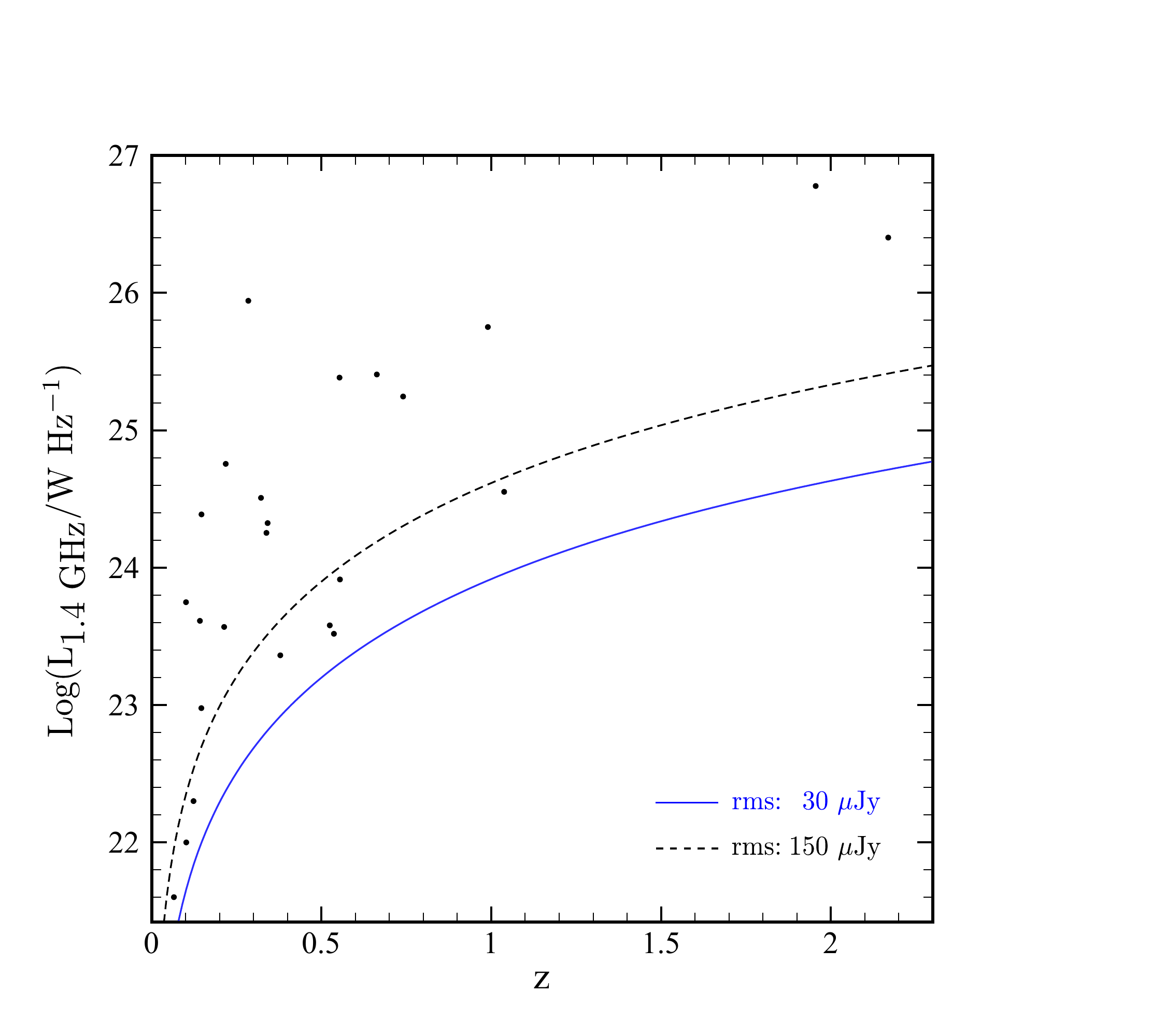}}
{\hspace*{-.5cm}\includegraphics[width=3.481in, trim= 0 -70 100 70, clip=true]{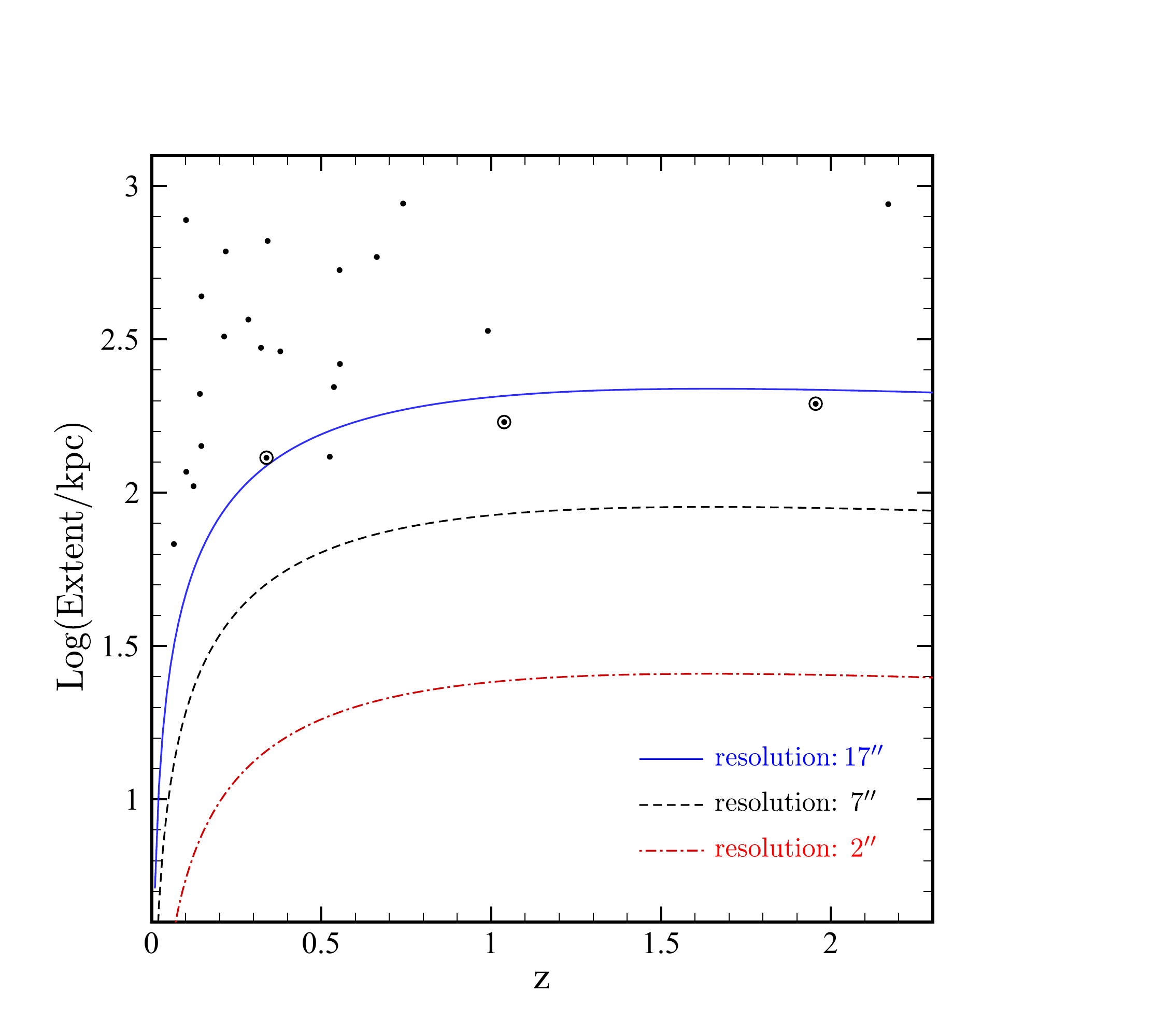}}
\vspace{-1.5cm}\caption{Left panel: the rest-frame 1.4 GHz luminosity-redshift relation for the detected BTs. The blue line and black dashed lines represent the $6\sigma$ source detection limit with the average rms of 30 and 150 $\mu$Jy, respectively. Black dots represent the detected BTs in the ATLAS-CDFS. Right panel: the extent-redshift relation for the detected BTs. The blue line, black dashed line, and red dash-dot line represent the detection limit of the sources with an extent of 1.5 beams at 17, 7, \& 2 arcseconds resolution, respectively. Black dots and circles represent detected BTs, and sources which could not be classified as BTs without employing $2^{\prime\prime}$ resolution VLA data, respectively.}
\label{fig:limit}
\end{figure*}

\subsection{Comparison with other studies}

It is worthwhile to note that the detection of $\sim 80$\% (44/56) of the sources would not be possible without obtaining the third data release of ATLAS with an improved rms value of a factor of 1.5-2 over the majority of the field, depending on location. The left panel of Figure \ref{fig:limit} plots the BT sources detected here as a function of redshift and power with a comparison between detection limits of the ATLAS (third data release) and the FIRST survey used in \citet{wb11}. Detection ($6\sigma$) of at least 33\% (8/24) of the sources is beyond the limits of the FIRST survey. The ratio will likely be increased, considering that FRI radio galaxies are often associated with low-surface-brightness lobes with respect to the host galaxy. Figure \ref{fig:limit} also demonstrates that $46\%$ (11/24) of the sources presented here are below $\textrm{P}_{1.4 \textrm{ GHz}} = 10^{24}$ WHz$^{-1}$ and thus probe a fainter population than other studies. Finally, both panels of Figure \ref{fig:limit} show that there are a pair of powerful sources at redshifts $\sim$ 2. These sources (ID 24 and 52) are discussed in Section \ref{morph} and represent the most distant BTs ever detected. ID 24 in particular (S447 in the ATLAS catalog) is a particularly powerful object, worthy of further study. 

In a comparative study of radio galaxies from the NVSS in the southern hemisphere with the 6dFGS redshift survey, \citet{ms07} examined the properties of 2864 AGNs. They found rest-frame radio powers up to $2.5 \times 10^{26}$ W Hz$^{-1}$ with an average redshift of 0.073. For this work they assumed a spectral index of $-0.54\pm0.07$ derived from comparison of 288 radio loud AGNs in the NVSS and Sydney University Molonglo Sky Survey (SUMSS). Using the spectral index value of -0.54, we measure the radio power of ID 24 and 52 to be 4.32 and 1.78 $\times 10^{26}$ \whz, respectively, which lie within the high-end of the radio power range reported by \citet{ms07}, albeit for a lower average redshift population.

The right panel of Figure \ref{fig:limit} shows a comparison between the detection limit of the ATLAS and VLA data due to resolution. Detection of 50\% (4/8) of the sources in the ECDFS, three shown with circles in the right panel of Figure \ref{fig:limit}, were only feasible by employing the $2^{\prime\prime}$ resolution VLA data. It is worth noting that a further increase in resolution will likely result in some of these sources being shown to be blended emission from multiple radio galaxies. Additionally, two of the sources here (ID 3 \& 19) rely on extensions made visible in only one $3\sigma$ contour. While it is difficult to precisely estimate the fraction of sources which will turn out to not be BTs in future, experience from previous data suggests this will be less than $1\%$ \citep{tss13}.

While ID 24 and 52 represent the extremes in terms of power and size, this sample also extends down to sources with sizes of the order of tens of kpc with 1.4 GHz rest frame powers $\sim 10^{22}$ WHz$^{-1}$. BT sources with similar powers and linear extents have previously been observed to be associated with central dominant galaxies in lower redshift ($z<0.1$) clusters \citep{gvm07}. Our sample may therefore be probing such objects in the higher z Universe in as yet undetected clusters at both ends of the power distribution.

\section{Implications and Conclusions}

We present 56 diffuse radio sources in $\sim$ 4 deg$^2$ of the ATLAS-CDFS field, including 45 BT radio galaxies and a radio relic candidate. Ten sources could not be unambiguously classified mostly due to a complex, multi-component presentation. Three of the ambiguous sources may be radio halos surrounded by other objects. Of the sources detected, two are at a redshift of $\sim$ 2 and are thus extremely powerful sources in the distant Universe. These represent the most distant BTs ever detected. Additionally, two sources appear associated with known clusters including the BT (ID 11) found in Abell 3141 and a relic candidate (ID 53) between a cluster and a group. The relic, if confirmed, will be the least powerful example yet seen. Based on extrapolating from these results, future all sky 1.4 GHz continuum surveys, comparable to the ATLAS survey in terms of resolution and sensitivity such as EMU \citep{nha11}, will detect up to $\sim$ 560,000 extended low-surface-brightness radio sources, including up to about 440,000 BT radio galaxies. This sample will span rest frame powers in the range $10^{22} \leq \textrm{P}_{1.4 \textrm{ GHz}} \leq 10^{26}$ WHz$^{-1}$, linear extents over two orders of magnitude and redshifts up to 2. Such extensive samples of tailed radio sources will significantly expand our knowledge of the origins and evolution of BTs and the mutual correlation between these sources and the environment in which they reside. By using these as tracers of overdensities, this means that EMU will detect up to $\sim$ 560,000 clusters or groups, which is more than ten times the number of clusters currently known, and comparable or exceeding the number to be detected with eROSITA \citep{mpb12}. Assuming the same clusters are detected by both eROSITA and EMU, this will result in a massive increase in the knowledge of the multiwavelength properties of clusters.

\begin{acknowledgements}

This work was supported by strategic funding from the Vice-Chancellor \& the Faculty of Science at Victoria University of Wellington. MJ-H is supported by the Marsden Fund Council from Government funding, administered by the Royal Society of New Zealand. TMOF acknowledges support from an ARC Super Science Fellowship.

This research has made use of the NASA/IPAC Extragalactic Database (NED) which is operated by the Jet Propulsion Laboratory, California Institute of Technology, under contract with the National Aeronautics and Space Administration.

\end{acknowledgements}

\clearpage
\begin{turnpage}
\begin{deluxetable}{lccccccccccc}
\tablecolumns{5} \tablewidth{0pt} \tablecaption{List of detected sources and their properties\label{tab:nmad}}
\tablehead{\multirow{2}{*}{ID} & \multirow{2}{*}{ATLAS ID \tablenotemark{a}} & \multirow{2}{*}{Name} & \colhead{RA \tablenotemark{b}} & \colhead{Dec \tablenotemark{b}} & \multirow{2}{*}{z} & \colhead{Flux} & \colhead{Power \tablenotemark{c}} & \multicolumn{2}{c}{Size} & \multirow{2}{*}{Type \tablenotemark{d}} & \multirow{2}{*}{z ref.\ \tablenotemark{e}}\\
\colhead{} & \colhead{} & \colhead{} & \colhead{(J2000)} & \colhead{(J2000)} & \colhead{} & \colhead{mJy} & \colhead{$\times 10 ^{24}$ \whz} & \colhead{\as} & \colhead{kpc}}
\startdata
1 & S619 &  ATCDFS\_J033430.10-271201.9 & $03^{\text{h}}34^{\text{m}}30.15^{\text{s}}$ & $-27 ^{\text{d}}12^{\text{m}}05.0^{\text{s}}$ & \nodata & 1.93 & \nodata & 45 & \nodata & 3 & \nodata \\ 
2 & \nodata & NVSS J033432-270853 & $03^{\text{h}}34^{\text{m}}32.60^{\text{s}}$ & $-27 ^{\text{d}}08^{\text{m}}52.9^{\text{s}}$ & \nodata & 9.47 & \nodata & 52 & \nodata & 1 & \nodata \\ 
3 & \nodata & \nodata & $03^{\text{h}}29^{\text{m}}44.99^{\text{s}}$ & $-27 ^{\text{d}}10^{\text{m}}11.1^{\text{s}}$ & \nodata & 2.22 & \nodata & 53 & \nodata & 3 & \nodata \\ 
4 & \nodata & \nodata & $03^{\text{h}}36^{\text{m}}25.50^{\text{s}}$ & $-27 ^{\text{d}}35^{\text{m}}29.9^{\text{s}}$ & \nodata & 2.07 & \nodata & 80 & \nodata & 2 & \nodata \\ 
5 & S707 &  ATCDFS\_J033533.96-273313.4 & $03^{\text{h}}35^{\text{m}}33.98^{\text{s}}$ & $-27 ^{\text{d}}33^{\text{m}}12.7^{\text{s}}$ & 0.7406 & 7.88 & 17.60 & 120 & 877 & 1 & 4 \\ 
6 & S691-S687-S694 &  ATCDFS\_J033523.04-273330.6 & $03^{\text{h}}35^{\text{m}}23.04^{\text{s}}$ & $-27 ^{\text{d}}33^{\text{m}}30.6^{\text{s}}$ & \nodata & 21.40 & \nodata & \nodata & \nodata & \nodata & \nodata \\ 
7 & S577 &  ATCDFS\_J033356.58-272444.5 & $03^{\text{h}}33^{\text{m}}56.59^{\text{s}}$ & $-27 ^{\text{d}}24^{\text{m}}44.3^{\text{s}}$ & 0.5533 & 21.62 & 24.14 & 83 & 532 & 1 & 4 \\ 
8 & S409 &  ATCDFS\_J033210.70-272635.5 & $03^{\text{h}}32^{\text{m}}10.73^{\text{s}}$ & $-27 ^{\text{d}}26^{\text{m}}35.4^{\text{s}}$ & 0.1469 & 43.99 & 2.44 & 172 & 437 & 3 & 4 \\ 
9 & S349 &  ATCDFS\_J033131.04-273815.7 & $03^{\text{h}}31^{\text{m}}30.33^{\text{s}}$ & $-27 ^{\text{d}}38^{\text{m}}15.2^{\text{s}}$ & 0.99 & 12.59 & 56.35 & 42 & 337 & 2 & 5 \\ 
10 & S119 &  ATCDFS\_J032806.04-272635.3 & $03^{\text{h}}28^{\text{m}}06.06^{\text{s}}$ & $-27 ^{\text{d}}26^{\text{m}}35.3^{\text{s}}$ & \nodata & 1.96 & \nodata & 47 & \nodata & 1 & \nodata \\ 
11 & \nodata & 2MASX J03364443-2758140 & $03^{\text{h}}36^{\text{m}}44.42^{\text{s}}$ & $-27 ^{\text{d}}58^{\text{m}}14.7^{\text{s}}$ & 0.1013 & 22.49 & 0.56 & 420 & 775 & 2 & 1 \\ 
12 & S376 &  ATCDFS\_J033150.07-273947.1 & $03^{\text{h}}31^{\text{m}}50.03^{\text{s}}$ & $-27 ^{\text{d}}39^{\text{m}}47.6^{\text{s}}$ & 1.04 & 0.71 & 3.56 & 21 & 170 & 2 & 6 \\ 
13 & S291 &  ATCDFS\_J033055.63-275201.7 & $03^{\text{h}}30^{\text{m}}55.69^{\text{s}}$ & $-27 ^{\text{d}}52^{\text{m}}02.0^{\text{s}}$ & 0.3382 & 5.04 & 1.79 & 27 & 130 & 1 & 4 \\ 
14 & S279 &  ATCDFS\_J033046.26-275517.5 & $03^{\text{h}}30^{\text{m}}46.24^{\text{s}}$ & $-27 ^{\text{d}}55^{\text{m}}20.8^{\text{s}}$ & 0.5247 & 0.39 & 0.38 & 21 & 131 & 1,3,6 & 4 \\ 
15 & \nodata & APMUKS(BJ) B032810.16-280248.8 & $03^{\text{h}}30^{\text{m}}15.72^{\text{s}}$ & $-27 ^{\text{d}}52^{\text{m}}35.5^{\text{s}}$ & 0.3789 & 0.50 & 0.23 & 56 & 289 & 2,6 & 4 \\ 
16 & \nodata & \nodata & $03^{\text{h}}29^{\text{m}}22.88^{\text{s}}$ & $-27 ^{\text{d}}52^{\text{m}}48.4^{\text{s}}$ & \nodata & 0.56 & \nodata & 47 & \nodata & 3 & \nodata \\ 
17 & \nodata & APMUKS(BJ) B032707.57-280520.8 & $03^{\text{h}}29^{\text{m}}13.26^{\text{s}}$ & $-27 ^{\text{d}}55^{\text{m}}01.8^{\text{s}}$ & 0.1021 & 0.47 & 0.01 & 63 & 117 & 2 & 2 \\ 
18 & \nodata & MRSS 418-067201 & $03^{\text{h}}29^{\text{m}}12.71^{\text{s}}$ & $-28 ^{\text{d}}02^{\text{m}}16.1^{\text{s}}$ & 0.5368 & 0.32 & 0.33 & 35 & 221 & 1,6 & 4 \\ 
19 & \nodata & APMUKS(BJ) B032701.01-281239.1 & $03^{\text{h}}29^{\text{m}}06.63^{\text{s}}$ & $-28 ^{\text{d}}02^{\text{m}}27.7^{\text{s}}$ & 0.0657 & 0.44 & 0.004 & 55 & 68 & 3 & 2 \\ 
20 & \nodata & 2MASX J03253652-2741071 & $03^{\text{h}}25^{\text{m}}36.41^{\text{s}}$ & $-27 ^{\text{d}}41^{\text{m}}06.4^{\text{s}}$ & 0.1423 & 7.86 & 0.41 & 85 & 210 & 3 & 2 \\ 
21 & \nodata & 2MASX J03253004-2739122 & $03^{\text{h}}25^{\text{m}}29.95^{\text{s}}$ & $-27 ^{\text{d}}39^{\text{m}}11.9^{\text{s}}$ & 0.1415 & 6.78 & 0.35 & \nodata & \nodata & \nodata & 2 \\ 
22 & S585 &  ATCDFS\_J033402.86-282405.6 & $03^{\text{h}}34^{\text{m}}02.87^{\text{s}}$ & $-28 ^{\text{d}}24^{\text{m}}05.8^{\text{s}}$ & 0.6632 & 14.83 & 25.45 & 84 & 587 & 1 & 4 \\ 
23 & S505 &  ATCDFS\_J033310.87-281854.7 & $03^{\text{h}}33^{\text{m}}10.87^{\text{s}}$ & $-28 ^{\text{d}}18^{\text{m}}54.8^{\text{s}}$ & \nodata & 1.21 & \nodata & 35 & \nodata & 1 & \nodata \\ 
24 & S447 &  ATCDFS\_J033232.04-280310.2 & $03^{\text{h}}32^{\text{m}}32.18^{\text{s}}$ & $-28 ^{\text{d}}03^{\text{m}}08.8^{\text{s}}$ & 1.9552 & 26.53 & 598.40 & 23 & 195 & 3 & 4 \\ 
25 & S336 &  ATCDFS\_J033127.00-281811.2 & $03^{\text{h}}31^{\text{m}}26.96^{\text{s}}$ & $-28 ^{\text{d}}18^{\text{m}}11.2^{\text{s}}$ & 0.2849 & 365.10 & 87.52 & 86 & 367 & 1 & 4 \\ 
26 & S224 &  ATCDFS\_J032956.13-280956.8 & $03^{\text{h}}29^{\text{m}}56.08^{\text{s}}$ & $-28 ^{\text{d}}09^{\text{m}}53.3^{\text{s}}$ & 0.1235 & 0.59 & 0.02 & 48 & 105 & 3 & 4 \\ 
27 & S156 &  ATCDFS\_J032851.61-280544.6 & $03^{\text{h}}28^{\text{m}}51.59^{\text{s}}$ & $-28 ^{\text{d}}05^{\text{m}}44.0^{\text{s}}$ & \nodata & 4.48 & \nodata & 52 & \nodata & 1,3 & \nodata \\ 
28 & \nodata & \nodata & $03^{\text{h}}28^{\text{m}}09.39^{\text{s}}$ & $-28 ^{\text{d}}04^{\text{m}}36.8^{\text{s}}$ & \nodata & 0.16 & \nodata & 21 & \nodata & 1,3 & \nodata \\ 
29 & \nodata & \nodata & $03^{\text{h}}28^{\text{m}}21.92^{\text{s}}$ & $-28 ^{\text{d}}10^{\text{m}}07.1^{\text{s}}$ & \nodata & 0.24 & \nodata & 27 & \nodata & 1 & \nodata \\ 
30 & \nodata & \nodata & $03^{\text{h}}28^{\text{m}}38.23^{\text{s}}$ & $-28 ^{\text{d}}19^{\text{m}}40.9^{\text{s}}$ & \nodata & 0.35 & \nodata & 43 & \nodata & 3 & \nodata \\ 
31 & \nodata & \nodata & $03^{\text{h}}27^{\text{m}}54.70^{\text{s}}$ & $-28 ^{\text{d}}25^{\text{m}}40.2^{\text{s}}$ & \nodata & 0.52 & \nodata & 28 & \nodata & 1 & \nodata \\ 
32 & S031 &  ATCDFS\_J032639.15-280800.6 & $03^{\text{h}}26^{\text{m}}39.18^{\text{s}}$ & $-28 ^{\text{d}}07^{\text{m}}59.7^{\text{s}}$ & 0.2183 & 43.11 & 5.69 & 175 & 612 & 2 & 4 \\ 
33 & S038 &  ATCDFS\_J032643.61-282209.7 & $03^{\text{h}}26^{\text{m}}43.38^{\text{s}}$ & $-28 ^{\text{d}}22^{\text{m}}10.4^{\text{s}}$ & 0.3222 & 10.15 & 3.22 & 64 & 297 & 3 & 4 \\ 
34 & \nodata & \nodata & $03^{\text{h}}26^{\text{m}}08.95^{\text{s}}$ & $-28 ^{\text{d}}09^{\text{m}}16.8^{\text{s}}$ & \nodata & 1.79 & \nodata & 109 & \nodata & 1 & \nodata \\ 
35 & \nodata & \nodata & $03^{\text{h}}25^{\text{m}}49.73^{\text{s}}$ & $-28 ^{\text{d}}17^{\text{m}}53.8^{\text{s}}$ & \nodata & 15.19 & \nodata & 125 & \nodata & 1 & \nodata \\ 
36 & \nodata & \nodata & $03^{\text{h}}36^{\text{m}}31.05^{\text{s}}$ & $-28 ^{\text{d}}33^{\text{m}}52.7^{\text{s}}$ & \nodata & 1.70 & \nodata & 40 & \nodata & 3 & \nodata \\ 
37 & \nodata & \nodata & $03^{\text{h}}36^{\text{m}}12.11^{\text{s}}$ & $-28 ^{\text{d}}30^{\text{m}}20.1^{\text{s}}$ & \nodata & 0.76 & \nodata & 30 & \nodata & 1 & \nodata \\ 
38 & \nodata & \nodata & $03^{\text{h}}34^{\text{m}}53.49^{\text{s}}$ & $-28 ^{\text{d}}28^{\text{m}}35.0^{\text{s}}$ & \nodata & 0.21 & \nodata & 17-26 & \nodata & 1,4,6 & \nodata \\ 
39 & \nodata & MRSS 418-045976 & $03^{\text{h}}32^{\text{m}}55.94^{\text{s}}$ & $-28 ^{\text{d}}48^{\text{m}}36.4^{\text{s}}$ & \nodata & 0.46 & \nodata & 52 & \nodata & 3 & \nodata \\ 
40 & S556 &  ATCDFS\_J033338.01-284706.4 & $03^{\text{h}}33^{\text{m}}38.07^{\text{s}}$ & $-28 ^{\text{d}}47^{\text{m}}05.5^{\text{s}}$ & \nodata & 0.47 & \nodata & 38 & \nodata & 1 & \nodata \\ 
41 & S538 &  ATCDFS\_J033330.20-283511.1 & $03^{\text{h}}33^{\text{m}}30.92^{\text{s}}$ & $-28 ^{\text{d}}34^{\text{m}}54.1^{\text{s}}$ & \nodata & 1.70 & \nodata & 54 & \nodata & 1 & \nodata \\ 
42 & \nodata & GALEX J033227.3-284157 & $03^{\text{h}}32^{\text{m}}27.35^{\text{s}}$ & $-28 ^{\text{d}}41^{\text{m}}58.4^{\text{s}}$ & 0.1715 & 0.38 & 0.03 & 33 & 95 & 3,4 & 2 \\ 
43 & S387 &  ATCDFS\_J033157.42-284041.0 & $03^{\text{h}}31^{\text{m}}57.41^{\text{s}}$ & $-28 ^{\text{d}}40^{\text{m}}38.2^{\text{s}}$ & \nodata & 0.46 & \nodata & 42 & \nodata & 3 & \nodata \\ 
44 & S426 &  ATCDFS\_J033219.28-284025.6 & $03^{\text{h}}32^{\text{m}}19.31^{\text{s}}$ & $-28 ^{\text{d}}40^{\text{m}}28.5^{\text{s}}$ & \nodata & 0.74 & \nodata & 45 & \nodata & 3 & \nodata \\ 
45 & \nodata & J052.68808-28.68144 & $03^{\text{h}}30^{\text{m}}45.17^{\text{s}}$ & $-28 ^{\text{d}}40^{\text{m}}54.4^{\text{s}}$ & 0.5546 & 0.73 & 0.82 & 41 & 263 & 3 & 3 \\ 
46 & \nodata & \nodata & $03^{\text{h}}30^{\text{m}}20.74^{\text{s}}$ & $-28 ^{\text{d}}45^{\text{m}}13.6^{\text{s}}$ & \nodata & 0.50 & \nodata & 59 & \nodata & 3 & \nodata \\ 
47 & S072 &  ATCDFS\_J032718.98-284640.5 & $03^{\text{h}}27^{\text{m}}19.00^{\text{s}}$ & $-28 ^{\text{d}}46^{\text{m}}40.6^{\text{s}}$ & 0.3416 & 5.82 & 2.11 & 137 & 662 & 2 & 4 \\ 
48 & S021 S020 &  ATCDFS\_J032630.68-283657.4 & $03^{\text{h}}26^{\text{m}}30.66^{\text{s}}$ & $-28 ^{\text{d}}36^{\text{m}}57.3^{\text{s}}$ & \nodata & 6.01 & \nodata & 155 & \nodata & 2 & \nodata \\ 
\enddata
\label{tab:prop}
\end{deluxetable}
\clearpage
\end{turnpage}

\begin{turnpage}
\begin{deluxetable}{lccccccccccc}[htbp]
\tablenum{1}\tablecolumns{5} \tablewidth{0pt} \tablecaption{---Continued\label{tab:nmad}}
\tablehead{\multirow{2}{*}{ID} & \multirow{2}{*}{ATLAS ID \tablenotemark{a}} & \multirow{2}{*}{Name} & \colhead{RA \tablenotemark{b}} & \colhead{Dec \tablenotemark{b}} & \multirow{2}{*}{z} & \colhead{Flux} & \colhead{Power \tablenotemark{c}} & \multicolumn{2}{c}{Size} & \multirow{2}{*}{Type \tablenotemark{d}} & \multirow{2}{*}{z ref.\ \tablenotemark{e}}\\
\colhead{} & \colhead{} & \colhead{} & \colhead{(J2000)} & \colhead{(J2000)} & \colhead{} & \colhead{mJy} & \colhead{$\times 10 ^{24}$ \whz} & \colhead{\as} & \colhead{kpc}}
\startdata
49 & S001 &  ATCDFS\_J032602.78-284709.0 & $03^{\text{h}}26^{\text{m}}02.82^{\text{s}}$ & $-28 ^{\text{d}}47^{\text{m}}08.1^{\text{s}}$ & \nodata & 2.03 & \nodata & 83 & \nodata & 3 & \nodata \\ 
50 & S701 &  ATCDFS\_J032718.79-273132.7 & $03^{\text{h}}35^{\text{m}}29.45^{\text{s}}$ & $-28 ^{\text{d}}51^{\text{m}}55.2^{\text{s}}$ & 0.2137 & 2.91 & 0.37 & 94 & 323 & 2 & 4 \\ 
51 & \nodata & NVSS J033405-290202 & $03^{\text{h}}34^{\text{m}}05.52^{\text{s}}$ & $-29 ^{\text{d}}02^{\text{m}}00.8^{\text{s}}$ & \nodata & 44.51 & \nodata & 68 & \nodata & 1 & \nodata \\ 
52 & S306 &  ATCDFS\_J033106.35-285217.5 & $03^{\text{h}}31^{\text{m}}06.34^{\text{s}}$ & $-28 ^{\text{d}}52^{\text{m}}18.1^{\text{s}}$ & 2.1688 & 8.78 & 252.22 & 104 & 873 & 3 & 4 \\ 
53 & \nodata & \nodata & $03^{\text{h}}30^{\text{m}}37.78^{\text{s}}$ & $-28 ^{\text{d}}59^{\text{m}}40.1^{\text{s}}$ & 0.152\tablenotemark{$\star$} & 1.52 & 0.09 & 34-50 & 89-130 & 5 & 2 \\ 
54 & S093 S094 &  ATCDFS\_J032739.36-285352.4 & $03^{\text{h}}27^{\text{m}}39.41^{\text{s}}$ & $-28 ^{\text{d}}53^{\text{m}}50.4^{\text{s}}$ & \nodata & 16.05 & \nodata & 108 & \nodata & 1 & \nodata \\ 
55 & S441 & ATCDFS\_J033229.52-273028.8 & $03^{\text{h}}32^{\text{m}}29.13^{\text{s}}$ & $-27 ^{\text{d}}30^{\text{m}}35.4^{\text{s}}$ & 0.1466 & 1.72 & 0.10 & 56 & 142 & 3 & 4 \\ 
56 & S446 & ATCDFS\_J033231.54-280433.5 & $03^{\text{h}}32^{\text{m}}31.54^{\text{s}}$ & $-28 ^{\text{d}}04^{\text{m}}33.5^{\text{s}}$ & \nodata & 0.38 & \nodata & 18-35 & \nodata & 4,5 & \nodata
\enddata
\tablenotetext{a}{IDs based on the ATLAS survey by \citet{naa06}.}
\tablenotetext{b}{Coordinates of the host galaxy as seen in the optical or infrared images. In cases in which the host galaxies were not recognized, the ATLAS coordinates are reported.}
\tablenotetext{c}{The rest-frame 1.4 GHz luminosity.}
\tablenotetext{d}{Numbers 1-6 respectively correspond to BT, WAT, NAT, radio halo, radio relic, and separated sources. Note that some sources could not be unambiguously classified, due to poor sensitivity or resolution, and therefore are represented by multiple type numbers. We note that the classification of BTs should encompass the historical classifications of WATs and NATs as a subset, and although the WAT$\slash$NAT distinction is somewhat arbitrary we include them here explicitly to allow further comparison with the literature.}
\tablenotetext{e}{Numbers 1-6 correspond to the spectroscopic redshifts derived from the ENACS, 2dFGRS, BLAST, and \citet{msn12}, and the photometric redshifts obtained from COMBO-17 and MUSYC surveys, respectively.}
\tablenotetext{$\star$}{The redshift is estimated by averaging the redshifts of two nearby galaxies.}
\label{tab:prop}
\end{deluxetable}
\clearpage
\end{turnpage}


\begin{thebibliography}{}
\bibitem[Banfield et al.(in prep.)]{b14}
  Banfield, J. K., et al., (in prep).

\bibitem[Becker et al.(1995)]{bwh95}
  Becker, R. H., White, R. L. \& Helfand, D. J., 1995, \aj, 450, 559

\bibitem[Blanton et al.(2001)]{bgh01}
  Blanton, E. L., Gregg, M. D., Helfand, D. J., Becker, R. H. \& Leighly, K. M., 2001, \aj, 121, 2915

\bibitem[Blanton et al.(2003)]{bgh03}
  Blanton, E. L., Gregg, M. D., Helfand, D. J., Becker, R. H. \& White, R. L., 2003, \aj, 125, 1635

\bibitem[Cardamone et al.(2010)]{cdm10}
  Cardamone, C. N., van Dokkum, P. G., Urry, C. M., et al., 2010, \apjs, 189, 270

\bibitem[Colless et al.(2003)]{cpj03}
  Colless, M., Peterson, B. A., Jackson, C., et al., 2003, astro-ph/0306581

\bibitem[Condon et al.(1998)]{ccg98}
  Condon, J. J., Cotton, W. D., Greisen, E. W., et al., 1998, \aj, 115, 1693

\bibitem[Cowie \& McKee(1975)]{cm75}
  Cowie, L. L. \& McKee, C. F., 1975, \aap, 43, 337

\bibitem[Dehghan et al.(2011)]{djm11}
  Dehghan, S., Johnston-Hollitt, M., Mao, M., et al., 2011, JApA, 32, 491

\bibitem[Dehghan \& Johnston-Hollitt(2014)]{dj14}
  Dehghan, S. \& Johnston-Hollitt, M., 2014, \aj, 147, 52

\bibitem[Eales et al.(2009)]{ecd11}
  Eales, S., Chapin, E. L., Devlin, M. J., et al., 2009, \apj, 707, 1779

\bibitem[Erben et al.(2005)]{esd05}
  Erben, T., Schirmer, M., Dietrich, J. P., et al., 2005, Astronomische Nachrichten, 326, 432

\bibitem[Fanaroff \& Riley(1974)]{fr74}
  Fanaroff, B. L., Riley, J. M., 1974, \mnras, 167, 31

\bibitem[Franzen et al.(in prep.)]{f14}
  Franzen, T. M. O., et al., (in prep).
  
\bibitem[Giacintucci et al.(2007)]{gvm07}
  Giacintucci, S., Venturi, T., Murgia, M., et al., 2007, \aap, 476, 99

\bibitem[Gunn \& Gott(1972)]{gg72}
  Gunn, J. E. \& Gott, J. R., 1972, \aj, 176, 1

\bibitem[Hildebrandt et al.(2006)]{hed06}
  Hildebrandt, H., Erben, T., Dietrich, J. P., et al., 2006, \aap, 452, 1121

\bibitem[Katgert et al.(1998)]{kmd98}
  Katgert, P., Mazure, A., den Hartog, R., et al., 1998, \aaps, 129, 399

\bibitem[Koss et al.(2010)]{kmv10}
  Koss, M., Mushotzky, R., Veilleux, S.,\& Winter, L., 2010, \apj, 716, 125

\bibitem[Ledlow \& Owen(1996)]{lo96}
  Ledlow, M. J. \& Owen, F. N., 1996, \aj, 112, 9

\bibitem[Lonsdale et al.(2003)]{lsr03}
  Lonsdale, C. J., Smith, H. E., Rowan-Robinson, M., et al., 2003, \pasp, 115, 897

\bibitem[Mao et al.(2009)]{mjs09}
  Mao, M. Y., Johnston-Hollitt, M., Stevens, J. B., \& Wotherspoon, S. J., 2009, \mnras, 392, 1070

\bibitem[Mao et al.(2010)]{mss10}
  Mao, M. Y., Sharp, R., Saikia, D. J., et al., 2010, \mnras, 406, 2578

\bibitem[Mao et al.(2012)]{msn12}
  Mao, M. Y., Sharp, R., Norris, R. P., et al., 2012, \mnras, 426, 3334

\bibitem[Mauch \& Sadler(2007)]{ms07}
  Mauch, T., Sadler, E. M., 2007, \mnras, 375, 931

\bibitem[Merloni et al.(2012)]{mpb12}
  Merloni, A.,  Predehl, P., Becker, W., et al., 2012, eprint arXiv:1209.3114

\bibitem[Middelberg et al.(2008)]{mnc08}
  Middelberg, E., Norris, R. P., Cornwell, T. J., et al., 2008, \aj, 135, 1276

\bibitem[Miniati et al.(2001)]{mrk01}
  Miniati, F., Ryu, D., Kang, H. \& Jones, T. W., 2001, \apj, 559, 59

\bibitem[Miley et al.(1972)]{mpk72}
  Miley, G. K., Perola, G. C., van der Kruit, P. C. \& van der Laan, H., 1972, \nat, 237, 269

\bibitem[Miller et al.(2013)]{mbf13}
  Miller, N. A., Bonzini, M., Fomalont, E. B., et al., 2013, \apjs, 205, 13

\bibitem[Norris et al.(2006)]{naa06}
  Norris, R. P., Afonso, J., Appleton, P. N., et al., 2006, \aj, 132, 2409

\bibitem[Norris et al.(2011)]{nha11}
  Norris, Ray P., Hopkins, A. M., Afonso, J., et al., 2011, PASA, 28, 215

\bibitem[Norris et al.(2013)]{nab13}
  Norris, R. P., Afonso, J., Bacon, D., et al., 2013, PASA, 30, 20

\bibitem[Pratley et al.(2013)]{pjd13}
  Pratley, L., Johnston-Hollitt, M., Dehghan, S., \& Sun, M., 2013, \mnras, 432, 243

\bibitem[Rose(1982)]{r82}
  Rose, J. A., 1982, \mnras, 201, 1015

\bibitem[Rudnick \& Owen(1976)]{ro76}
  Rudnick, L. \& Owen, F. N., 1976, \aj, 203, 107

\bibitem[Slee et al.(2001)]{srm01}
  Slee, O. B., Roy, A. L., Murgia, M., Andernach, H., \& Ehle, M., 2001, \aj, 122, 1172

\bibitem[Struble \& Rood(1999)]{sr99}
  Struble, M. F. \& Rood, H. J., 1999, \apjs, 125, 35

\bibitem[Thorat et al.(2013)]{tss13}
  Thorat, K., Subrahmanyan, R., Saripalli, L., \& Ekers, R. D., 2013, \apj, 762, 16  

\bibitem[Way et al.(2005)]{wqi05}
  Way, M. J., Quintana, H., Infante, L., Lambas, D. G., \& Murie, H., 2005, \aj, 130, 2012

\bibitem[Wing \& Blanton(2011)]{wb11}
  Wing, J. D. \& Blanton, E. L. 2011, \aj, 141, 88

\bibitem[Wolf et al.(2004)]{wmk04}
  Wolf, C.,  Meisenheimer, K., Kleinheinrich, M., et al., 2004, \aap, 421, 913
\end{thebibliography}
\end{document}